\documentclass[a4paper,11pt]{article}
\pdfoutput=1 

\usepackage{jcappub} 

\usepackage[T1]{fontenc} 
\usepackage[nameinlink,noabbrev]{cleveref}
\usepackage{xcolor}
\usepackage[normalem]{ulem}
\usepackage{mathrsfs}
\usepackage{comment}
\usepackage{makecell, multirow}
\usepackage{orcidlink}
\usepackage{aas_macros}
\usepackage{lineno}

\title{Model-independent cosmology with joint observations of gravitational waves and $\gamma$-ray bursts}

\author[1,2]{Andrea Cozzumbo\orcidlink{0000-0002-7332-9806},}
\author[1,2]{Ulyana Dupletsa\orcidlink{0000-0003-2766-247X},}
\author[3,4]{Rodrigo Calder{\'o}n\orcidlink{0000-0002-8215-7292},}
\author[5,6,1]{Riccardo Murgia\orcidlink{0000-0002-2224-7704},}
\author[1,2]{Gor Oganesyan\orcidlink{0000-0001-9765-1552},}
\author[1,2]{Marica Branchesi\orcidlink{0000-0003-1643-0526}}

\affiliation[1]{Gran Sasso Science Institute (GSSI), Viale F. Crispi 7, L'Aquila (AQ), I-67100, Italy}
\affiliation[2]{INFN - Laboratori Nazionali del Gran Sasso (LNGS), L'Aquila (AQ), I-67100, Italy}
\affiliation[3]{Korea Astronomy and Space Science Institute, Daejeon 34055, Korea}
\affiliation[4]{CEICO, Institute of Physics of the Czech Academy of Sciences,\\
Na Slovance 1999/2, 182 21, Prague, Czech Republic}
\affiliation[5]{Dipartimento di Fisica, Universit\`a degli Studi di Cagliari, Cittadella Universitaria, 09042 Monserrato (CA), Italy}
\affiliation[6]{INFN, Sezione di Cagliari, Cittadella Universitaria, 09042 Monserrato (CA), Italy}

\emailAdd{andrea.cozzumbo@gssi.it}
\emailAdd{ulyana.dupletsa@gssi.it}
\emailAdd{calderon@fzu.cz}
\emailAdd{riccardo.murgia89@unica.it}

\usepackage{fontawesome5}
\newcommand{\gitlink}[1]{\href{#1}{\textcolor{black}{\faGithub}}}

\abstract{Multi-messenger (MM) observations of binary neutron star (BNS) mergers provide a promising
approach to trace the distance-redshift relation, crucial for understanding the expansion history
of the Universe and, consequently, testing the nature of Dark Energy (DE). While the gravitational wave (GW) signal offers a direct measure of the distance to the source, 
high-energy observatories can detect the electromagnetic counterpart and drive the optical follow-up providing the redshift of the host galaxy.
In this work, we exploit up-to-date catalogs of $\gamma$-ray bursts (GRBs) supposedly coming from BNS mergers observed by the Fermi $\gamma$-ray Space
Telescope and the Neil Gehrels Swift Observatory, to construct a large set of mock MM data. 
We explore how combinations of current and future generations of GW observatories operating under various underlying cosmological models would be able to detect GW signals from these GRBs. We achieve the reconstruction of the GW parameters by means of a novel prior-informed Fisher matrix approach. 
We then use these mock data to perform an agnostic reconstruction of the DE phenomenology, thanks to a machine learning method based on forward modeling and Gaussian Processes (GP). Our study highlights the paramount importance of observatories capable of detecting GRBs and identifying their redshift. In the best-case scenario, the GP constraints are 1.5 times more precise than those produced by classical parametrizations of the DE evolution. We show that, in combination with forthcoming cosmological surveys, fewer than 40 GW-GRB detections will enable unprecedented precision on $H_\mathrm{0}$ and $\Omega_\mathrm{m}$, and accurately reconstruct the DE density evolution.} 

\begin{document}
\maketitle
\flushbottom

\definecolor{RedWine}{rgb}{0.743,0,0}
\def\rcb#1{\textcolor{RedWine}{[RC:  #1]}}

\definecolor{Copper}{rgb}{0.188,0.811,0.713}
\def\ac#1{\textcolor{Copper}{[AC:  #1]}}

\definecolor{Orange}{rgb}{1., 0.647, 0.}
\def\riki#1{\textcolor{Orange}{[RM:  #1]}}

\newcommand{\vect}[1]{\ensuremath{\boldsymbol{#1}}}
\newcommand{\tens}[1]{\ensuremath{\mathbfss{#1}}}
\newcommand{\fde }{\ensuremath{f_\mathrm{DE}} }
\newcommand{\Hnot }{\ensuremath{H_\mathrm{0}} }
\newcommand{\Omegam }{\ensuremath{\Omega_{\mathrm{m}}} }
\newcommand{\Omegade }{\ensuremath{\Omega_{\mathrm{DE}}} }
\newcommand{\lcdm }{$\Lambda$CDM }
\newcommand{\wnot }{\ensuremath{w_{\mathrm{0}}} }
\newcommand{\wa }{\ensuremath{w_{\mathrm{a}}} }
\newcommand{\panp }{Pantheon$\Plus$ }

\def\tpdf#1{\texorpdfstring{#1}{Lg}}
\def\Plus{\texttt{+}}

\crefname{equation}{Eq.}{Eqs.}
\crefname{section}{Section}{Sections}
\crefname{figure}{Fig.}{Figs.}
\crefname{table}{Table}{Tables}
\crefname{appendix}{Appendix}{Appendices}
\Crefname{figure}{Figure}{Figures}
\Crefname{equation}{Equation}{Equations}
\Crefname{section}{Section}{Sections}
\Crefname{table}{Table}{Tables}
\section{Introduction}
\label{sec:intro}

Current constraints on the late-time expansion history of the Universe mostly come from observations of luminosity distance - redshift $\{d_L, z\}$-relation of type Ia Supernovae (SNIa)~\cite{Perlmutter1998Measurements,  Riess1998Observational, Riess:2017lxs} together with measurements of Baryon Acoustic Oscillations (BAO)~\cite{Addison:2017fdm, SDSS:2005xqv, Percival:2007yw}. However, these measurements do not extend above redshift $z \approx 2$. Moreover, within this redshift range, the statistical uncertainty grows with increasing $z$, making the distinct effects of many cosmological models alternative to the $\Lambda$ Cold Dark Matter ($\Lambda$CDM) indistinguishable. Both the BAO and SNIa methods have intrinsic limitations that cannot be completely overcome. Distance measurements from BAO are related to the sound horizon scale, $r_d$. This is the maximum distance that acoustic waves could travel in the early Universe before recombination, which today is imprinted in the spatial distribution of galaxies. We can measure how the expansion rate $H(z)$ varies with redshift $z$ but the exact values depend on an uncertain scaling factor, which is the combined product of $H_0$ and $r_d$, making it impossible to determine $H_0$ or $r_d$ independently without additional data.
This implies that a calibration is needed~\cite{Poulin:2024ken}.
Similarly, (uncalibrated) SNIa  measurements can only constrain distances up to an additive constant, $\mathcal{M}=M_B+5\log_{10}{(\frac{c}{H_0})
+25}$\footnote{Taking the distance modulus $\mu(z) \equiv m_B(z) - M_B$, one gets  $m_B(z)=5\log_{10}\left(\frac{d_L(z)}{\rm Mpc}\right)+25+M_B = 5\log_{10}\left({\frac{H_0}{c}d_L(z)}\right)+\mathcal{M}$, having defined $\mathcal{M}\equiv M_B+5\log_{10}{\left(\frac{c}{H_0}\right)}
+25$}.
Thus, extracting information on \Hnot from either BAO or SNIa requires distances to be calibrated,~\textit{i.e.},~the inferred $H(z)$ and $\mu(z)$ need to be ``anchored'' at low redshift~\cite{Camarena2021Use,Bernal2016Trouble}. This is usually achieved by imposing an informative prior on the sound horizon $r_d$ or the absolute magnitude $M_B$, or by including external data sets such as Cosmic Microwave Background (CMB), or measurements of the baryon density $\omega_b=\Omega_bh^2$ from Big-Bang Nucleosynthesis (BBN)~\cite{Planck:2018vyg}. However, the latter approach often assumes a $\Lambda$CDM-like expansion history at early times, which could be an obvious source of systematic in case the latter is not the model that better describes our Universe~\cite{Knox2019Hubble}.

Recently, thanks to the detection of gravitational waves (GWs)~\cite{LIGOScientific:2016aoc} 
from the merger of binary compact objects, we have access to an alternative, self-calibrated\footnote{Luminosity distance measurements from GWs can also be affected by systematics, for instance due to the waveform modeling. However, they are completely negligible in the context of this work.} way to measure luminosity distances~\cite{Mastrogiovanni:2024mqc}. Although GWs qualify as standard sirens, they do not provide any direct redshift measurement, which has to be inferred relying either on the identification of the host galaxy~\cite{Abbott2017GW170817, 1986Natur.323..310S} or on statistical methods~\cite{Mastrogiovanni:2021wsd, Mastrogiovanni:2023emh}. There are, in fact, two primary methods to exploit GW data for cosmological inference: one involves the presence of the electromagnetic counterpart of the gravitational event (bright sirens), and the other does not (dark sirens). An example of the first category is the GW signals from binary neutron star (BNS) mergers, which, when jointly detected with their electromagnetic counterpart, provide an independent and self-calibrated tool to trace the late-time Universe expansion history~\cite{LIGOScientific:2017adf}. The electromagnetic signal allows us to estimate the redshift of the event.
In the second case, such as with binary black hole (BBH) mergers, cosmological information can still be obtained by using various statistical methods, referred to the BBH population, to estimate the redshift of the events~\cite{Scelfo:2021fqe, Moresco:2022phi, Gray:2023wgj, Chen:2017rfc, Gray:2019ksv, Muttoni:2023prw, Dalya:2018cnd, Dalya:2021ewn, LIGOScientific:2019zcs, Finke:2021aom, LIGOScientific:2021aug, DES:2019ccw, DES:2020nay, Palmese:2021mjm, Alfradique:2023giv,DESI:2023fij, Ezquiaga:2022zkx, Farr:2019twy, Afroz:2024joi, Mukherjee:2022afz, Diaz:2021pem, Mukherjee:2020hyn}.

Another intriguing opportunity lies in exploring alternative classes of standard sirens such as, for instance, mergers of binary white dwarfs~\cite{Maselli:2019mzt} and supermassive black hole binaries~\cite{Yan2020using}.
The prospects for accessing cosmological parameters using different classes of GW sources will be dramatically improved by the advent of new and innovative observatories, such as the next generation of ground-based GW observatories~\cite{Bailes:2021tot}, Einstein Telescope (ET)~\cite{Branchesi:2023mws}  and Cosmic Explorer (CE)~\cite{Evans:2021gyd}, as well as moon- and space- based observatories probing the deci-, milli-, and nano-Hertz bands, such as the Lunar Gravitational Wave Antenna (LGWA)~\cite{LGWA:2020mma,Cozzumbo:2023gzs}, the Laser Interferometer Space Antenna (LISA)~\cite{LISACosmologyWorkingGroup:2022jok} and the Pulsar Timing Array facilities~\cite{Sesana:2008xk, Wang:2016tfy}.

In this work we focus on the bright siren science case, exploring both current and future GW detector capabilities.
The joint detection of GW170817~\cite{Abbott2017GW170817}, with its associated
EM emission~\cite{Abbott2017Gravitational, Goldstein2017Ordinary, Savchenko2017INTEGRAL}, enabled
us to determine {the current expansion rate of the Universe (although still with a large level of uncertainty)}. This was done by measuring the luminosity distance to the source from the GW signal and the redshift from the host galaxy observations. The almost simultaneous detection of the GRB170817A by \textit{Fermi}-GBM after the coalescence provided the first direct evidence that BNS mergers are progenitors of short GRBs~\cite{Ascenzi:2020xqi}. In addition, GW170817 taught us that binary-driven GRBs can occur simultaneously with a kilonova (KN) emission. KNe can be detected as thermal emission peaking in the optical/near-infrared (NIR) domain, which is indicative of the rapid neutron capture processes responsible for heavy nuclei production.

GRBs are usually classified according to their $T_{\mathrm{90}}$, that is the time at which we measure 90$\%$ of the incoming signal. Short GRBs ($T_{\mathrm{90}} \lesssim 2 ~\mathrm{s}$) are associated to BNS mergers~\cite{Piran:2004ba, Bartos:2012vd, Ciolfi:2018tal}, while long GRBs ($T_{\mathrm{90}} \gtrsim 2 ~\mathrm{s}$) are associated to the collapse of massive stars~\cite{Paczynski:1997yg, Bloom:2000pq}.
However, observations have recently pointed out that such a standard assumption may result in an oversimplified classification~\cite{Levan:2023qdh, Rastinejad:2022zbg, Mei:2022ncd, Lu:2022tpu, Nuessle:2024pvq, Troja:2022yya}. That is why, in this work, we base our analyses on those GRBs (either short or long) that are most likely driven by a BNS merger~\cite{Fong:2022mkv}.
Operating in synergy with electromagnetic observatories, the next generation GW observatories like ET are expected to detect GW signals associated to almost each observed $\gamma$-ray burst (GRB) produced by a BNS merger up to $z \approx 3$~\cite{Ronchini:2022gwk,Branchesi:2023mws}. These multi-messenger (MM) events will provide a valuable tool that will complement existing and upcoming constraints from other cosmological probes.
In addition, several electromagnetic surveys, such as the Dark Energy Spectroscopic Instrument (DESI)~\cite{DESI:2016fyo}, Euclid~\cite{Euclid:2024yrr} and the James Webb Space Telescope~\cite{Rigby_2023}, in the optical band, and the Square Kilometer Array Observatory (SKAO) ~\cite{Braun:2019gdo}, in the radio band, will further provide data of unprecedented precision. 
Given the improved sensitivity of the above mentioned facilities, as well as the complementarity of scale and redshift regimes probed with respect to currently available data, establishing efficient synergies among GW and EM observations represents an important frontier for cosmology~\cite{Belgacem:2019tbw,Branchesi:2023mws}.

Over the years, a plethora of alternative theoretical models have been proposed to provide a more comprehensive fit to current cosmological probes than the standard $\Lambda$CDM scenario~\cite{Bull:2015stt,Perivolaropoulos:2021jda,DiValentino2021Realm,Abdalla:2022yfr,Schoneberg2021Olympics,Khalife:2023qbu,Verde:2023lmm}.
In fact, despite the remarkable success of the latter, the persistence of cosmic tensions seems to hint at an inconsistency between some of the cosmological parameters inferred from
early Universe observations, and data from the later stages of cosmic evolution~\cite{Knox2019Hubble, Poulin:2024ken}. 
However, given the excellent fit to Cosmic Microwave Background (CMB) data provided by the $\Lambda$CDM model, any alternative scenario should only induce limited departures in order to improve the fit to post-recombination data without spoiling the CMB {predictions}.
It is thus essential to develop agnostic methods covering the phenomenology induced by broad classes of theoretical models.
In this work we exploit a model-independent, non-parametric, data-driven approach based on Machine Learning (ML) techniques~\cite{Calderon2022}. Aiming at covering any model inducing deviations from the $\Lambda$CDM expansion history still allowed by current data -- yet expected to leave a distinct imprint in forthcoming observations -- we devised a new pipeline, interfaced with the widely used numerical Einstein-Boltzmann solver {\tt CLASS}~\cite{Lesgourgues2011CosmicI,Blas2011Cosmic}.
We stick to General Relativity (GR) in the present work, but the same method is also suitable to test modified gravity (MG) theories~\cite{Calderon2023, Matos:2021qne, Matos:2023jkn}. We leave this task for future work.

To generate the mock MM data needed for inferring the expansion history, we make use of the most updated catalogs of GRBs from BNS mergers, as measured by currently operating high-energy satellites (\textit{Fermi}-GBM and \textit{Swift}-BAT/XRT). We assume that ET, as well as other combinations of both current (dubbed 2G) and future (3G) GW interferometers, was operating simultaneously to catch the corresponding GW counterparts. We leave for a follow-up work a joint analysis with future high-energy satellites~\cite{Ronchini:2022gwk, THESEUS:2017wvz}.

{We generate our data sets assuming two cosmologies, both of which are in very good agreement with current observations. The accuracy of the $d_L$ estimate is governed by the GW detector networks that we employ. We then make use of these mock MM data in two steps. To begin with, we perform a MCMC analysis on the data to estimate the cosmological parameters in a model-dependent manner, \textit{i.e.}, by assuming a parametric model for the dark energy (DE) sector. Finally, we adopt a model-independent approach to reconstruct the DE phenomenology without assuming any specific parametric form for its Equation of State (EoS).} As a promising alternative to the standard \lcdm concordance model, to generate the mock data, we focus on the so-called Phenomenological Emergent Dark Energy (PEDE)~\cite{Li:2019yem, Li:2020ybr} scenario. This model features the same number of free parameters as \lcdm but predicts an emergent component of DE in the late Universe. Such a phenomenology, describing an emergent DE dynamics, has recently gained significant interest following the data release by the DESI collaboration~\cite{DESI:2024mwx,DESI:2024aqx, DESI:2025fii}. We compare the theoretical predictions of the $\{d_L, z\}$-relation, generated using a modified version of {\tt CLASS}, with the previously described mock data. This enhanced version of {\tt CLASS} allows us to implement non-standard cosmologies and apply Gaussian Process (GP) evolution to model DE phenomenology~\cite{Calderon2022}. 

The devised software, a \texttt{python}-based likelihood for the MCMC samplers {\tt MontePython}~\cite{Audren2013Conservative,Brinckmann2018MontePython} and {\tt cobaya}~\cite{Torrado:2020dgo}, is publicly available on GitHub\footnote{\gitlink{https://github.com/AndreaCozzumbo/bright_cosmology}~\url{https://github.com/AndreaCozzumbo/bright_cosmology}}, together with all the data catalogs used in this paper. Our likelihood can be used in joint analyses with all main complementary cosmological probes.

This paper is structured as follows: in Section~\ref{sec:methods} we present the data sets and methods used to produce the joint GRB-GW mock data. We also explain the rationale behind using both parametric and non-parametric approaches for cosmological inference. In Section~\ref{sec:results} we present the cosmological constraints obtained with the two approaches and we discuss the analyses performed with future GW and BAO surveys. Section~\ref{sec:summary} provides summary and conclusions.

\section{Data and methods}
\label{sec:methods}

As outlined in the Introduction, our study follows a three-step procedure: \textit{(i)} the selection of the most up-to-date catalogs of GRBs from BNS mergers from almost 20 years of observations by ongoing high-energy surveys,~\textit{i.e.},~from \textit{Fermi}-GBM and \textit{Swift}-BAT/XRT (Subsection~\ref{subsec:GRB-GW}); \textit{(ii)} the production of a set of mock MM data by injecting a GW signal per each GRB, thereby generating a GW signal corresponding to a BNS located at the same position and redshift of the GRB. The reconstruction of their parameters is performed using a prior-informed Fisher matrix analysis~\cite{Dupletsa:2024gfl}, assuming different combinations of current and future generations of GW detectors under various underlying cosmologies (Subsection~\ref{subsec:GWfish+}); \textit{(iii)} the use of the mock data to perform a model-dependent inference of the cosmological parameters and, finally, an agnostic, model-independent reconstruction of the DE phenomenology, thanks to GP techniques (Subsection~\ref{subsec:mock} and \ref{subsec:GPth}).

\subsection{GRB data selection: redshift measurements}{\label{subsec:GRB-GW}}

The starting point of this work is an extensive analysis of the most updated catalogs of GRB data collected by \textit{Fermi}-GBM and \textit{Swift}-BAT/XRT from 2005 to 2023. The catalog is presented in Appendix~\ref{ap:grbt_comparison} (Table~\ref{tab:GRBall}), alongside with further details on the selection procedure. The aim is to identify all events -- associated to a BNS merger with high probability --  whose host galaxy has a redshift determined with less than 10\% uncertainty. To begin with, we focused on well-localized (XRT radius $\leq 5 ''$) short GRBs ($T_{\mathrm{90}} \lesssim 2~\mathrm{s}$)~\cite{Fong:2022mkv}. Following
Ref.~\cite{Fong:2022mkv}, we also considered as merger-driven GRBs those events with $T_{\mathrm{90}} \gg 2~\mathrm{s}$ due to their extended emission\footnote{The extended emission in a GRB refers to a prolonged phase of lower-intensity $\gamma$-ray radiation that follows the initial, intense burst of energy, lasting from seconds to several minutes.}.

In addition, we included in the resulting catalog of merger-driven GRBs, the events GRB230307A and GRB191019A, as they have been associated with a KN explosion~\cite{JWST:2023jqa,  Stratta:2024kbs}, as well as the event GRB211211A, interpreted as compact object mergers ~\cite{Lazzati:2023ayh, Levan:2023qdh, Rastinejad:2022zbg}.
For each event, we took into account its probability to be associated to the actual host galaxy, by accounting for the probability of chance coincidence, $P_{cc}$, as evaluated in Ref.~\cite{Fong:2022mkv}.
The full catalog of GRB events that we have considered for our analyses, classified according to their $P_{cc}$, is reported in Appendix~\ref{ap:grbt_comparison}. Hereafter, we refer to a \textit{fiducial} catalog (containing only those events with $P_{cc} \leq 0.02$), an \textit{extended} catalog (events with $P_{cc} \leq 0.10$), and a \textit{very extended} catalog ($P_{cc} \leq 0.20$).

\begin{figure}[t]

    \centering
    \includegraphics[width=0.7\linewidth]{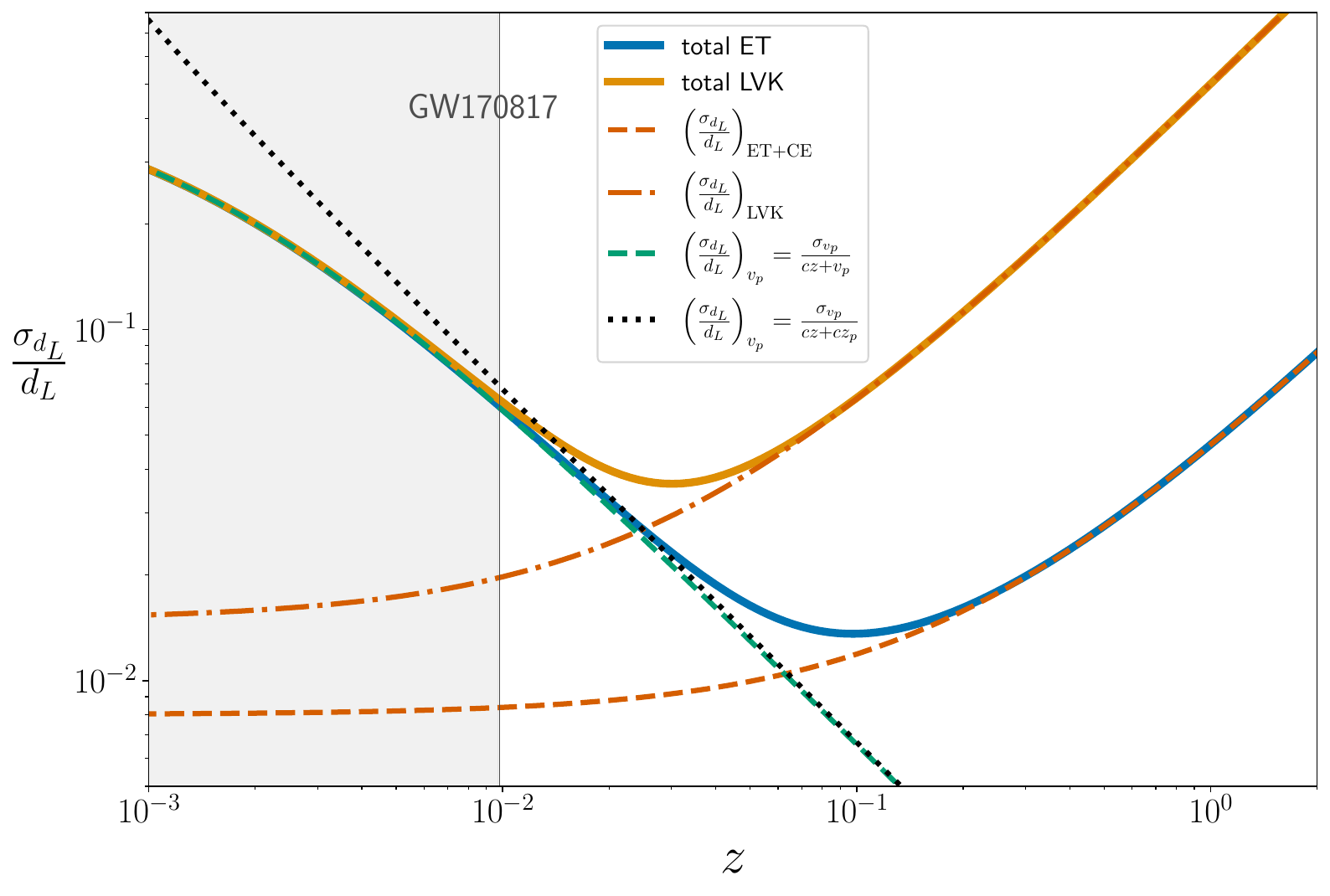}
    \caption{Impact of the peculiar velocity on the $d_L$ relative uncertainty. The orange curves show the LVK (dashed-dotted) and ET$\Delta$+CE (dashed) relative uncertainty. They are obtained by fitting the estimated uncertainty (without correcting for $v_p$) with a power-law~\cite{Belgacem:2018lbp, Zhao:2010sz}. The green dashed curve is the peculiar velocity contribution to the $d_L$ uncertainty for $v_p=400~\mathrm{km}~{\mathrm{s}}^{-1}$ and $\sigma_{v_p}=200~\mathrm{km}~{\mathrm{s}}^{-1}$. The more precise is the estimation on $d_L$ given by the GW detectors, the more the peculiar velocity contribution is important. The dotted black curve shows the peculiar velocity contribution computed adopting the prescription of Ref.~\cite{Davis:2014jwa}. The vertical line corresponds to the redshift of GW170817.}
    \label{fig:peculiar}
\end{figure}

One key difference compared to previous works is that our results do not depend on BNS population synthesis simulations~\cite{Ronchini:2022gwk, Branchesi:2023mws, Chen:2024gdn, Mukherjee:2023lqr, Belgacem:2019tbw, Iorio:2022sgz} as our GW parameter estimation code is based on actually \textit{observed} GRB data. 
Typically, the BNS merger rate density as a function of redshift of synthetic populations depends on assumptions about common envelope, core-collapse SN model, star formation and metallicity evolution \cite{Santoliquido2021}. In contrast, our approach provides a realistic redshift distribution of BNS mergers because it relies directly on observed data. This is particularly important because our data actually trace the redshift distribution of BNS, which plays a key role in constraining the evolution of DE.

It is crucial to emphasize that predicting the rate of well-localized events or those with a known redshift in MM observations is highly challenging, due to numerous uncertainties related to the instruments and the observational strategies. For this reason, we have chosen to base our work on the two decades of experience from \textit{Swift} and \textit{Fermi}. We can reasonably consider the results presented in this work as a lower bound of the constraining power from GRB-GW data provided by future high energy surveys~\cite{THESEUS:2021uox}.

\subsection{GW parameter reconstruction: luminosity distance estimates}
{\label{subsec:GWfish+}}

We used the BNS merger-driven GRB catalog presented in Subsection~\ref{subsec:GRB-GW} to produce fast and accurate forecasts on the detection capabilities of their GW counterpart, assuming different observational configurations, by means of the GW detector simulation code \texttt{GWFish}~\cite{Dupletsa2022}. \texttt{GWFish} is capable of performing parameter estimation for an individual GW event in a few seconds. However, this comes at the cost of approximating the likelihood $\mathscr{L}(d|\theta)$ with a multivariate normal distribution,~\textit{i.e.} the Fisher matrix

\begin{equation}\label{eq:fisher}
    \mathcal{F}_{ij}= \mathcal{C}^{-1}_{ij}= -\left\langle \frac{\partial^2}{\partial\theta_i\partial\theta_j}
    \mathrm{ln}~\mathscr{L}(d|\theta)\right\rangle {\Bigg|_{\theta= \theta_{\mathrm{inj}}}} =\left(\frac{\partial h}{\partial \theta_i}\Big{|}\frac{\partial h}{\partial \theta_j}\right) \equiv 4\Re{\int_0^\infty df\,\frac{1}{S_n(f)}\frac{\partial h}{\partial\theta_i}\frac{\partial h^*}{\partial\theta_j}}~,
\end{equation}  
where the data stream is defined as $d(t)=h(t) + n(t)$, $h(t)$ is the GW signal, $\theta$ is an array containing all the GW signal model parameters, $\theta_\mathrm{inj}$ are the fiducial injected values, $\mathcal{F}_{ij}$ is the Fisher information matrix, that is the inverse of the covariance matrix $\mathcal{C}_{ij}$.
The average in the above expression of Eq.~\eqref{eq:fisher} is taken over all the noise realizations. The function $S_n(f)$ is the detector power spectral density (PSD), encoding the sensitivity of the detector configuration. In our treatment we use the IMRPhenomHM waveform approximant~\cite{Kalaghatgi:2019log}. This allows us to leverage the higher modes of GW signals to more accurately constrain the parameters.
The full set of GW parameters is $ \theta = \{m_1, m_2, d_L, \alpha, \delta, \iota, \psi, t_c, \Phi_c \}$ where $m_1$ and $m_2$ are the detector frame masses of the two compact objects, $d_L$ is the luminosity distance to the merger event, $\alpha$ and $\delta$ are the right ascension (R.A.) and declination (dec.) angles, $\iota$ is the inclination angle between the line of sight vector and the orbital angular momentum of the binary, $\psi$ is the polarization angle, $t_c$ and $\Phi_c$ are the time and phase at coalescence, respectively. 

In the analyses presented in this paper, the sky localization of the event ($\alpha, \delta$) and the time of coalescence $t_c$ are fixed, as they are given by the GRB detection. The redshift $z$ is fixed too, as it is given by the host galaxy association (see Subsection~\ref{subsec:GRB-GW}).
The values of the masses and polarization angle are drawn from the following uniform intervals: $m_1, m_2 \in [1.1, 2.1~]\mathrm{M}_\odot$; $\psi \in [0, 2\pi]$~\cite{LIGOScientific:2020ibl, KAGRA:2021vkt}. The viewing angle $\iota$ is generated uniform in cosine in the range $\iota \in [0, 5/180\pi]\cup[175/180\pi, \pi]$. 
The luminosity distance is set by

\begin{equation}\label{eq:dLz}
    d_L(z, \theta_\mathrm{c})= c~(1+z)\int^{z}_0 \frac{dz'}{H(z, \theta_\mathrm{c})}
\end{equation}
as a function of the set redshift $z$ and the chosen cosmological parameters $(\theta_\mathrm{c})$.
We have verified that our results are almost insensitive to the NS mass distribution, be it Gaussian, uniform or considering BNS systems characterised by high and low total mass or mass ratio. In Appendix~\ref{ap:tables}, Figure~\ref{fig:gaussmass_ETD} we show the comparison between Gaussian and uniform mass sampling.

As already discussed, for the purposes of this work we are mostly interested in the uncertainties on the luminosity distance $d_L$. The standard Fisher matrix approach provides the uncertainties as the square root of the corresponding diagonal elements of the covariance matrix. Here we adopt an improved, novel approach capable of providing prior-informed Fisher matrix reconstruction of $d_L$ and the other GW parameters~\cite{Dupletsa:2024gfl}.

The procedure we use to produce mock data can be summarized as follows. Initially, we obtain a Gaussian posterior of $d_L$ as output of \texttt{GWFish} analysis. We sample from this distribution by first imposing truncation based on the physical bounds~\cite{Dupletsa:2024gfl}. This ensures a faster sampling process by preventing the generation of numerous samples with null priors that would need to be discarded. For instance, we avoid negative values of $d_L$ that do not have any physical meaning.
Finally, we sample again from this truncated multivariate Gaussian and for each realization of this likelihood we evaluate the prior probability. Therefore, for each event, we finally obtain a prior-informed probability density of the $d_L$, among all the GW parameters. 

We now describe how we take into account the main potential sources of uncertainty on $d_L$ estimation from GW signals, besides the systematic uncertainties associated to the GW measurement~\cite{Huang:2022rdg}.
The first source of uncertainty that we address for what concerns the association to the host galaxy is the treatment of its peculiar velocity $v_p$~\cite{Mukherjee:2019qmm, Nimonkar:2023pyt}. For nearby events the peculiar velocity correction to the pure Hubble flow contribution can be dominant. This correction accounts for the peculiar motion of the host galaxy with respect to the observer and it is mainly induced by the large-scale matter distribution, whereas in the relatively close Universe, the peculiar velocities are dominated by nearby galaxies. We provide further details on the comparison of our results by accounting or not this correction in Appendix~\ref{ap:tables}. The peculiar velocity contributes to the overall uncertainty on $d_L$ as follows:
\begin{equation}
    \frac{\sigma_{d_L}}{d_L} = \sqrt{\left(\frac{\sigma_{d_L}}{d_L}\right)_{v_p}^2 + \left(\frac{\sigma_{d_L}}{d_L}\right)_{\mathrm{GW}}^2}
\end{equation}
where the label GW refers to the uncertainty given by the interferometer network parameter estimation. The other component, due to $v_p$, can be expressed as $\sigma_{d_L} = d_L  \sigma_{v_p}/(c z + v_p)$. This treatment can be further refined according to the guidelines provided in Ref.~\cite{Davis:2014jwa}, where the authors show that the standard approach for inferring the impact of peculiar velocity can lead to significant errors.
Note that with this expression, we are performing a Gaussian approximation of the GW $d_L$ posteriors. This procedure is useful for illustrating the relative contribution of the two uncertainties. However, in the mock data set presented later, this is done in a more refined manner.
In Figure~\ref{fig:peculiar} we show the relative impact of the two contributions for the case of LVK(O5) and ET$\Delta$+CE (see Subsection~\ref{subsec:datasets} for the description of the adopted detector configurations). We notice that the more constraining is the GW detector, the more important becomes the peculiar velocity contribution. The green dashed line represents the peculiar velocity uncertainty estimation used in this work. For events at redshifts higher than the closest event in our data set (i.e., GW170817), we do not observe any significant differences between the peculiar velocity treatment from Ref.~\cite{Davis:2014jwa} (black dotted line) and the other, less refined approach.
From the considerations made using Figure~\ref{fig:peculiar}, we can safely assume the peculiar velocity contribution to be relevant only for $z \leq 0.15$ (see Figure~\ref{fig:z_vp_comaparison} in Appendix~\ref{ap:tables} for the choice of this value). For events falling into this category, we account for this source of uncertainty by generating random Gaussian-distributed values of $v_p$ and $\sigma_{v_p}$.

The last source of uncertainty that we considered is the GW weak lensing noise, that we parameterize as in Refs.~\cite{Hirata:2010ba, Tamanini2016Science, Muttoni:2023prw}. This phenomenon is particularly important at high redshift, due to the larger distance travelled by the signal. We evaluated that the effect is subdominant for all the events, even in our most constraining configuration, thus we will not report any corrections due to weak lensing.

The posterior distribution of the luminosity distance is thus characterized
either by the GW contribution alone coming from the Fisher analysis or by a combination of both GW and peculiar velocity contributions at $z \leq 0.15$.
The GW posterior is interpolated with a Kernel Density Estimator (KDE), allowing to capture all the non-Gaussian features of the distribution. The result of this process serves as the input for our newly devised likelihood for cosmological inference. Within the likelihood, if the event falls below the redshift limit for the peculiar velocity correction, we convolve the input KDE with the Gaussian distribution of the $d_L$ uncertainty due to peculiar velocity. The resulting probability density is then fit again using a KDE interpolation.

We want to stress that the pipeline we developed is versatile enough to accommodate any $d_L$ posterior distributions from any distance tracer. In our work we relied on the Fisher matrix approximation for the GW $d_L$ posterior as presently a Bayesian analysis tool, capable of processing BNS signals with the future GW detectors, is still in the developing process\footnote{One major issue in analysing the BNS signals comes from their long duration in 3G detectors band ($\sim$ hours). This implies a high computational cost, on top of the fact that it requires to consider the effects of Earth rotation. In \texttt{GWFish} these effects are implemented while they are not present in currently used GW Bayesian analysis software, such as \texttt{bilby}~\cite{Ashton:BILBY}.}.

\subsection{GW detector configurations}
{\label{subsec:datasets}}

In what follows we describe the GW detector configurations used in our analyses:

\begin{itemize}

\item LVK stands for the LIGO Hanford, LIGO Livingston, Virgo, KAGRA network~\cite{LIGOScientific:2014pky, VIRGO:2014yos, KAGRA:2020tym}. For the LVK network we use both the so-called O5 (more optimistic) PSDs
\cite{2020LRR}
~and the A$^\#$ one~\cite{Fritschel2022F}. The A$^\#$ refers to the potential upgrade of LIGO following the O5 data collection run. In the latter, we are adding to the network the LIGO India detector (LVKI). The A$^\#$ PSD is used for the LIGO detectors (Hanford, Livingston and India) alone. Virgo and KAGRA are using O5 sensitivity in both configurations.

\item ET$\Delta$ is the Einstein Telescope baseline configuration, corresponding to three nested detectors with arm length and aperture of 10 km and 60$^\circ$, respectively. Each of them is composed according to the so-called xylophone concept,~\textit{i.e.},~by two different co-located interferometers: one optimized for the low frequency (LF) and one for the high frequency (HF). We dub ET HF $\Delta$ a configuration where the cryogenic LF detector is not included~\cite{Branchesi:2023mws}. The location of the detector is assumed to be at the ET Sardinian candidate site (Sos Enathos mine).

\item ET2L is a xylophone configuration with 15 km arm length, and aperture of 90$^\circ$. We use the 45$^\circ$ misaligned configuration. It corresponds to two separated detectors, one in the Sardinian site, and one in the Meuse-Rhine candidate site (Netherlands). As in the ET$\Delta$ case, we dub as ET HF 2L a configuration where the cryogenic LF detector is not included~\cite{Branchesi:2023mws}.

\item ET X + CE configuration, with X$\in \{\Delta, \mathrm{2L}\}$, that refers to a detector network composed of the ET X and the Cosmic Explorer (CE). We consider a single CE detector positioned in the LIGO Hanford site, with an arm aperture of 90$^\circ$ and a 40 km arm length~\footnote{The PSD of CE can be found \href{https://dcc.cosmicexplorer.org/CE-T2000017/public}{here}.}.

\end{itemize}

In Table~\ref{tab:config_numb} we report the number of detected joint GRB-GW events for each GRB catalog (fiducial, extended, and very extended) by all the detector configurations that we have considered. These numbers are valid for either of the two underlying cosmologies that we have analyzed (see Subsection~\ref{subsec:mock}). We adopt a detection threshold of SNR $\geq 8$ as in Ref.~\cite{Branchesi:2023mws}. From Table~\ref{tab:config_numb} one can easily extract useful information to understand what will be presented in Section~\ref{sec:results}. One can notice that combining more than one 3G detector in a network significantly improves the number of detections, allowing to catch the GW counterpart of practically all injected events, even the ones at the highest redshift (see also Appendix~\ref{ap:det_comparison}). Note also that the duty cycle influences the number of detected events\footnote{In all our analyses we set the duty cycle of the GW detectors to $85\%$. The number of missed events decreases when combining more detectors, because it is statistically more probable that they are active together.}, while the detector sensitivity (\textit{i.e.},~the PSD) primarily affects the uncertainty in GW parameter estimation, impacting the cosmological constraints (see Section~\ref{sec:results}). Finally, we stress how properly accounting for the impact of peculiar velocities at low redshift, even at the level of a single event, significantly improves the accuracy on the inferred cosmological parameters. Having a network of N detectors increases the number of detectable events (compared to having N-1 detectors) thanks to the combined duty cycles. However, the quality of the $d_L$ posterior provided by improved detector sensitivities is more critical for an accurate cosmological analysis than the sheer number of events. This is well demonstrated by the LVK configuration, which includes three detectors. While this setup offers a more favorable duty cycle, it lacks sufficient sensitivity. Consequently, as we will discuss in Section~\ref{sec:results}, the LVK configuration provides very weak cosmological constraints.

\begin{table}[t]
\centering

\caption{Summary of the number of events observed by all GW detector configurations considered in this work. From left to right we report the detector configuration, the fiducial catalog (containing 38 GRB events), the extended catalog (containing 54 GRB events) and the very extended catalog (containing 61 GRB events). For each configuration we also report the redshift of the furthest event detected. The detection threshold is set to $\mathrm{SNR} \geq8$. The results listed in this table are the same for both the mock data catalogs that we have produced, assuming as underlying cosmology either \lcdm (MOD1) or PEDE (MOD2). See the main text for further details.\\}

    \begin{tabular}{cc|c|c|c|c|c|}
        \hline
        \multicolumn{1}{|c|}{\multirow{2}{*}{\textbf{Detector}}}    
        & \multicolumn{2}{c|}{\textbf{Fiducial (38 ev.)}}      
        & \multicolumn{2}{c|}{\textbf{Extended (54 ev.)}}  
        & \multicolumn{2}{c|}{\textbf{Very extended (61 ev.)}} \\ \cline{2-7}
        
        \multicolumn{1}{|c|}{}                                                                          & $\boldsymbol{z_{\mathrm{max}}}$ & \textbf{N. events}                            &  $\boldsymbol{z_{\mathrm{max}}}$ & \textbf{N. events}                             & $\boldsymbol{z_{\mathrm{max}}}$ & \textbf{N. events}                                   \\ \hline

        \multicolumn{1}{|c|}{\begin{tabular}[c]{@{}c@{}}LVK (O5)\end{tabular}}          & 0.248           & 8                                        & 0.248           & 10      & 0.248           & 10                        \\ \hline
        
        \multicolumn{1}{|c|}{\begin{tabular}[c]{@{}c@{}}LVKI A$^{\#}$\end{tabular}}          & 0.46           & 16                                        & 0.46           & 20      & 0.46           & 21                        \\ \hline
        
        \multicolumn{1}{|c|}{\begin{tabular}[c]{@{}c@{}} ET HF $\Delta$ \end{tabular}}          & 1.370            & 30                                        & 1.465           & 40      & 1.465           & 43                        \\ \hline
        
         \multicolumn{1}{|c|}{\begin{tabular}[c]{@{}c@{}} ET$\Delta$ \end{tabular}}          & 1.754           & 36                                        & 2.58           & 48      & 2.58           & 55                        \\ \hline
         
        \multicolumn{1}{|c|}{\begin{tabular}[c]{@{}c@{}} ET$\Delta$+CE \end{tabular}}          & 2.609           & 38                                        & 2.609           & 53      & 2.609           & 60                        \\ \hline

        \multicolumn{1}{|c|}{\begin{tabular}[c]{@{}c@{}} ET HF 2L \end{tabular}}          & 1.754           & 36                                        & 2.58           & 48      & 2.58           & 54                        \\ \hline
        \multicolumn{1}{|c|}{\begin{tabular}[c]{@{}c@{}} ET2L \end{tabular}}          & 1.754           & 37                                        & 2.58           & 50      & 2.58           & 57                        \\ \hline
        \multicolumn{1}{|c|}{\begin{tabular}[c]{@{}c@{}} ET2L + CE \end{tabular}}          & 2.609           & 38                                        & 2.609           & 53      & 2.609           & 60                        \\ \hline
    
    \end{tabular}

    \label{tab:config_numb}
\end{table}

\subsection{Mock MM data with different underlying cosmologies}
\label{subsec:mock}

Assuming an isotropic and homogeneous Universe described by a flat FLRW metric, the first Friedmann equation reads
\begin{equation}\label{eq:H2}
    \frac{H^2(z)}{H_0^2}=\Omega_{\rm m,0}(1+z)^3+\Omega_{\rm r,0}(1+z)^4+(1-\Omega_{\rm m,0}-\Omega_{\rm r,0})\;f_{\rm DE}(z)~,
\end{equation}
where $\Omega_{i, 0}=8\pi G\rho_{i,0}/3H_0^2=\rho_{i,0}/\rho_\mathrm{0,cr}$ is the fractional energy density of the $i$-th component,
the subscript ``0'' denotes its present value (at $z=0$) and $\rho_\mathrm{0,cr}$ is the critical energy density today. Under the flatness condition, we can define $\Omega_{\mathrm{DE,0}}=1-\Omega_{\rm m,0}-\Omega_{\rm r,0}$ and consequently 
\begin{equation}
    \fde(z) = 
    \rho_\mathrm{DE}(z)/\rho_\mathrm{DE,0}~.
\end{equation}
The quantity $\fde(z)$ defines the redshift evolution of the DE, such that $\Omega_{\mathrm{DE}}(z) = H_0^2/H^2(z) \allowbreak \Omega_{\mathrm{DE,0}} \fde(z)$. From the conservation of the stress-energy tensor \fde can be expressed as an integral of the DE Equation of State (EoS) $w(z) \equiv p_{\rm{DE}}/\rho_{\rm{DE}}c^2$, namely:

\begin{equation}\label{eq:fde_exp}
    f_{\mathrm{DE}}(z) =\exp{\Bigg\{3\int_0^z[1+w(z')]~d\ln{(1+z')}\Bigg\}}~.
\end{equation}
\sloppy
Within the standard \lcdm model, DE is assumed to be a cosmological constant,~\textit{i.e.}~$w(z)= -1$, that leads to $\fde = 1$. Note that $\fde(z)$ is an \emph{effective} quantity capable of capturing in a model-independent fashion a wide variety of alternatives to a cosmological constant behaviour in Eq.~\eqref{eq:H2}, such as time-evolving DE, the effective energy density of modified gravity, or any additional degree of freedom contributing to the expansion that does not evolve over time in the same way as matter.
Taking the logarithm and differentiating both sides of Eq.~\eqref{eq:fde_exp} we find the expression for the \textit{effective} EoS parameter
\begin{equation}\label{eq:weff}
      w(z)= \frac{1}{3} (1+z)\frac{d\,\textrm{ln} \fde(z)}{dz} - 1 ~ .
\end{equation}

In this work, we consider a simple beyond-\lcdm model, the PEDE model, where DE has no effective presence in the past ($z\gtrsim2$) and it phenomenologically emerges at the late times~\cite{Li:2019yem,Li:2020ybr}. The PEDE model has zero extra-degrees of freedom compared to the standard $\Lambda$CDM model. Within PEDE, the redshift evolution of the DE EoS (Eq.~\ref{eq:weff}) is given by

\begin{equation}\label{eq:wPEDE} 
     w^{\rm PEDE}(z)=-\frac{1}{3~\mathrm{ln}(10)}\left( 1+\mathrm{tanh} \left[\mathrm{log}_{\mathrm{10}}(1+z) \right] \right) -1 ~,
\end{equation}
which, using Eq.~\eqref{eq:fde_exp}, becomes
\begin{equation}\label{fde-PEDE}
     \fde(z)= 1-\mathrm{tanh}\left[\mathrm{log}_{\mathrm{10}}(1+z)\right] ~.
\end{equation}

For the purposes of this work, we generated several sets of mock MM data corresponding to various GW detector configurations, assuming two different cosmological scenarios:
\begin{itemize}
    \item the first model (MOD1) assumes \lcdm as underlying cosmology with background parameters set to the default \texttt{CLASS} values~\cite{Blas2011Cosmic},~\textit{i.e.},~$\theta_c^{\Lambda\mathrm{CDM}}=\{H_\mathrm{0}=67.81~\mathrm{km ~Mpc}^{-1}\mathrm{s}^{-1}, \, \Omega_\mathrm{m}=0.309\}$.

    \item the second model (MOD2) assumes an underlying PEDE cosmology with the best-fit cosmological parameters to \textit{Planck}2018+eBOSS+\panp data~\cite{Planck:2018vyg, eBOSS:2019qwo, Brout:2022vxf},~namely~$\theta_c^{\mathrm{PEDE}}=\{H_\mathrm{0}=70.3 ~\mathrm{km~Mpc}^{-1}\mathrm{s}^{-1}, \, \Omega_\mathrm{m}$ $=0.295\}$.
\end{itemize}

Note that both MOD1 and MOD2 are best-fit models to the same data sets, despite the underlying DE models being significantly different. This represents the most extreme case of degeneracy in light of current data. 
In light of the recent DESI results~\cite{DESI:2024mwx,DESI:2024aqx,DESI:2024kob, DESI:2025fii}, which points toward a late-time emergent DE, a PEDE-like model could in principle rule out the cosmological constant hypothesis with decisive statistical significance.

\subsection{Non-parametric modeling of DE beyond \tpdf{$\Lambda$}}
{\label{subsec:GPth}}
GP regression has been extensively used in the literature to reconstruct smooth functions from noisy data, without the need to assume a parametric form \citep[\textit{e.g.}][for a non-exhaustive list]{Holsclaw2010,Holsclaw2010PRL,Shafieloo:2012ht,Seikel_2012,Shafieloo2013,Keeley2019Implications,Hwang:2022hla,Calderon:2023obf, Ormondroyd:2025exu, Berti:2025phi, Raffaelli:2025kew,Afroz:2024lou} (see also \citep{Belgacem:2019zzu,Keeley:2019hmw,Li:2021ukd,Mukherjee:2021kcu,Ashton:2022ztk,Mukherjee:2023lqr} for some applications in the context of GW cosmology).
In this work, following Refs.~\citep{Calderon2022,Calderon2023} we reconstruct the time-evolution of \fde$(z)$\ through GPs, by modelling it as follows:
\begin{equation}\label{eq:fde-GP}
    f_{\mathrm {DE}}(z)\sim\mathcal{GP}\left(\bar f(z)=1,K=k(\sigma_f,\ell_f) \right)~,
\end{equation} 
where $\bar{f}(z)$ is the mean function of a multivariate Gaussian distribution and  $K$ is the covariance function between two points $k(x_i,x_j)$, known as a \emph{kernel} in the GP literature \cite{rasmussen2006gaussian}. We choose to work with a squared-exponential kernel of the form
\begin{equation}\label{Kernel}
     k(x_i,x_j;\sigma_f,\ell_f)=\sigma_f^2~\exp\left(-\frac{(x_i-x_j)^2}{2\ell_f^2}\right)~,
\end{equation}
where $\sigma_f$ and $\ell_f$ are hyperparameters determining the amplitude of the deviations from the mean function $\bar f$ and the typical correlation length, respectively.
As discussed in Refs.~\citep{Shafieloo2013,Calderon2023,Calderon:2023obf}, the posterior distributions of the hyperparameters $\sigma_f$ and $\ell_f$ carry valuable information on the mean function at hand (a cosmological constant, $\Lambda$, in this case). While we do not delve into the details of GP regression in this work\footnote{We refer the reader to the usual reference \cite{rasmussen2006gaussian} and \cite{Calderon2022,Hwang:2022hla,Calderon2023} for more details on the GP method.}, we briefly discuss the differences between our approach and conventional analyses (see \cite{Hwang:2022hla} for a thorough comparison). A key quantity in GP regression is the log-marginal likelihood (LML), given by 
\begin{equation}\label{eq:LML}
    \ln\mathcal{L}=-\frac12\big[\vect{r}^{T}\,\mathbf{K}^
    {-1}\,\vect{r}+\ln{|\mathbf{K}|} + N\ln{(2\pi)}\big]~, 
\end{equation}
where $\vect{r}=\vect{y}-\vect{y}^{\rm data}$ is the residual vector, $\mathbf{\Sigma}$ is the data covariance and $\mathbf{K}=\mathbf{\Sigma}+k(\sigma_f,\ell_f)$ is the full covariance. Unlike conventional GP approaches, which rely on the optimization of the hyperparameters such that the LML in Eq.~\eqref{eq:LML} is maximized, we instead do \emph{forward-modeling} and draw a sample of \fde from Eq.~\eqref{eq:fde-GP} at each point in parameter space ($\Omega_\mathrm{m}$,$H_\mathrm{0}$,$\sigma_f,\ell_f$) and compute the likelihood of the data. This effectively allows us to marginalize over the space of realizations numerically\footnote{This ``forward modeling'' approach has been shown to be equivalent to the conventional (posterior) GPR method in \cite{Hwang:2022hla}.} and obtain posterior distributions on $\fde(z)$ in a non-parametric manner.
While this approach can be computationally time-consuming, it allows us to consistently combine different probes (such as SNIa and GW distance measurements) and break the degeneracies between different cosmological parameters.

\begin{figure}[t]
    \centering
    \includegraphics[width=0.45\textwidth]{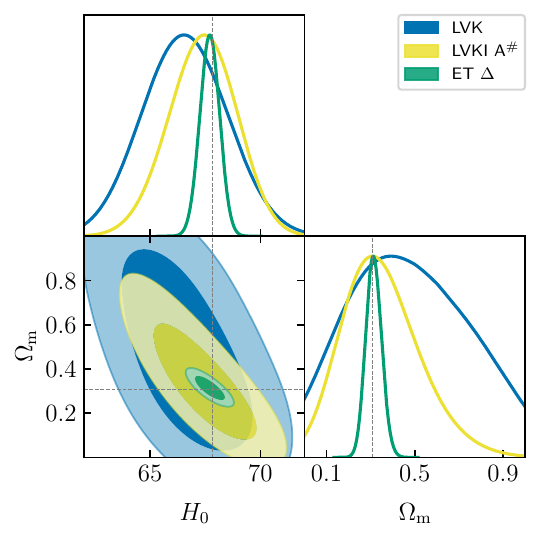}
    \includegraphics[width=0.45\textwidth]{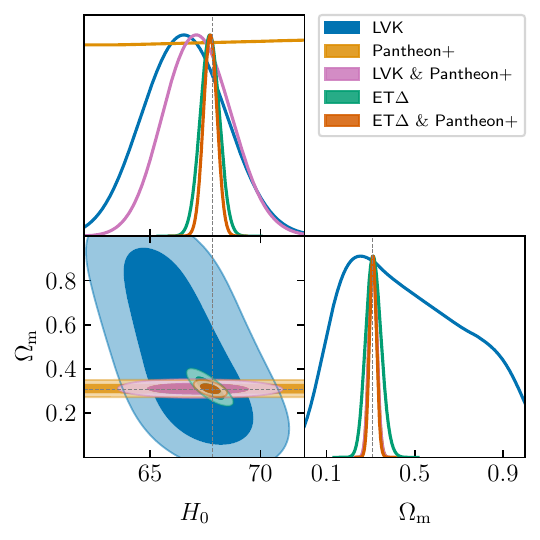}
    \caption{\lcdm constraints (68\% and 95\% C.L.) on \Hnot and \Omegam obtained by fitting an underlying \lcdm cosmology (MOD1) over mock MM and Pantheon$\Plus$ data. (\textit{Left}) Comparison between LVK (O5) (blue) with 8 events, LVKI A$^\#$ (yellow) with 16 events and ET$\Delta$ (green) with 36 events. (\textit{Right}) LVK (O5) and ET$\Delta$ as in the left panel. We over-plot the constraints obtained with the \panp data set alone (orange) and in a joint analysis with LVK (O5) (pink) and ET$\Delta$ (dark orange). Dashed lines refer to the injected cosmology used to produce the mock GRB-GW data.}
    \label{fig:LVK_ET_LL}
\end{figure}

\section{Cosmological constraints}
\label{sec:results}

In this Section we will present the constraints on the two cosmological scenarios described in Subsection~\ref{subsec:mock} -- MOD1 ($\Lambda$CDM) and MOD2 (PEDE) -- obtained by means of extensive MCMC analyses with the \texttt{MontePython-v3}\footnote{\url{https://github.com/brinckmann/montepython_public}} sampler \cite{Audren2013Conservative,Brinckmann2018MontePython} in its default Metropolis-Hastings mode and \texttt{Cobaya}\footnote{\url{https://github.com/CobayaSampler/cobaya}}~\cite{Torrado:2020dgo}, interfaced with a modified version of~\texttt{CLASS}~\cite{Blas2011Cosmic,Calderon2022}.

In both our \lcdm and PEDE analyses we scan the parameter space defined by the standard set of two background parameters, namely the CDM abundance $\Omega_{\rm{cdm}}$, and the Hubble parameter $H_0$. We fix the remaining ones -- the baryon abundance $\Omega_b$, the tilt and amplitude of the primordial power spectrum ($n_s$, $A_s$) and the optical depth at reionization $\tau_{\rm{reio}}$ -- to the best fit values from \textit{Planck}2018~\cite{Planck:2018vyg}. In addition, we perform a series of MCMC analyses exploring the Chevallier-Polarski-Linder (CPL) parameterization of the DE EoS \cite{CHEVALLIER_2001,Linder:2002et},~\textit{i.e.},~$w(a) = \wnot + \wa(1-a)$, where \wnot and \wa are the two extra free parameters with respect to \lcdm and PEDE. Finally, we scan the parameter space defined in Subsection~\ref{subsec:GPth}, where the DE sector is described in a model-independent fashion by the two GP hyperparameters $\sigma_f$ and $\ell_f$.
We have adopted the following uniform priors: $H_{\mathrm{0}}\in\left[50, 90\right]$ km~Mpc$^{-1}$~s$^{-1}$, $\Omega_{\mathrm{m}}\in\left[0, 1\right]$, $w_{\mathrm{0}}\in\left[-2,0\right]$, $w_{\mathrm{a}}\in\left[-5,2\right]$,
$\sigma_f\in\left[10^{-4},5\right]$, $\ell_f\in\left[10^{-2},10\right]$. Note that the results presented here are based on the specific priors used in our analyses and -- in some of our GW-only analyses -- the parameter \wa reaches the lower bound of its prior range. We verified that expanding the prior ranges weakens the constraints on the CPL parameters. However, this also opens up the $H_\mathrm{0}$-\Omegam parameter space in a region that is entirely ruled out by external data such as SNIa data from \panp. This ensures that all our conclusions are prior-independent. Our priors on \wnot and \wa are similar to those adopted by the DESI collaboration~\cite{DESI:2024mwx, DESI:2025fii}, and we have verified that this choice has no impact on our results.

The relation between each set of free cosmological parameters $\theta_\mathrm{c}$ and the luminosity distance is given by  Eq.~\eqref{eq:dLz}. 
Through our MCMC analyses we sample the posterior distribution of the cosmological parameters, using the likelihood $\mathscr{L}(d|\theta_\mathrm{c})$, a generalization of the $\chi^2$ function that encompasses all the non-Gaussian features of the $d_L$ posterior distributions:

\begin{equation}\label{eq:lkl_mpt}
\mathrm{ln}~\mathscr{L}(d|\theta_\mathrm{c})=\sum^{N_{\rm events}}_i \mathrm{ln}~\mathscr{K}^{i}_{_{\mathrm{GW}}}\left(d_L(\theta_\mathrm{c})\right)
\end{equation} 
where $N_{\rm events}$ depends on the choice of data set and detector configuration (see Tab.~\ref{tab:config_numb}), and $\mathscr{K}^{i}_{_{\mathrm{GW}}}$ is the KDE interpolation of the $i$-th posterior.
We consider the MCMC chains to be converged when the Gelman-Rubin criterion~\cite{Gelman1992Inference} satisfies $R-1 < 0.01$.

Besides the large set of joint GRB-GW mock data sets resulting from the detector configurations that we have just illustrated, for some of the runs we make use of the following additional complementary data:

\begin{itemize}

\item the catalog of uncalibrated SNIa from Pantheon$\Plus$, spanning redshifts $0.01 < z < 2.3$~\cite{Brout:2022vxf}. In this paper, we always use mock \panp data generated distributing the distance modulus according to the underlying cosmology that we want to test (MOD1 or MOD2) while retaining the original \panp uncertainties. This is done to ensure to not mix real data (for which the underlying cosmology is unknown) with artificially generated ones. The only case when we use the actual \panp data set is when combining the latter with the real BAO data set (for further details see Subsection~\ref{subsec:3G});

\item the BAO measurements from 6dFGS at $z=0.106$~\cite{Beutler2011Galaxy}, SDSS DR7 at $z=0.15$~\cite{Ross2014Clustering}, BOSS DR12 at $z=0.38, 0.51$ and $0.61$~\cite{Alam2016Clustering}, and the joint constraints from eBOSS DR14 Lyman-$\alpha$ auto-correlation at $z=2.34$~\cite{deSainteAgathe:2019voe} and cross-correlation at $z=2.35$~\cite{eBOSS:2019qwo}. In the following we use these data to calibrate the \panp real data set. Where it is specified, we combine BAO with a Gaussian prior on the sound horizon at the baryon drag epoch ($r_{\mathrm{s, drag}}$). This is done to have more reliable contours due to the addition of a piece of information coming from the CMB epoch;

\item the forecast for DESI constraining power is based on the use of the full data set from the 5-year mission. We assume a sky coverage of $14,000~\rm deg^2$, as described in \cite{DESI:2016fyo}. Hereafter we dub this mock data set as DESI(Y5).
\item the SH$_0$ES of SNIa at redshift $z \leq 0.01$, calibrated through direct distance ladder method based on Cepheid variables~\cite{Riess2019Large}.

\end{itemize}

\begin{figure}[t]
    \centering
    \includegraphics[width=0.9\linewidth]{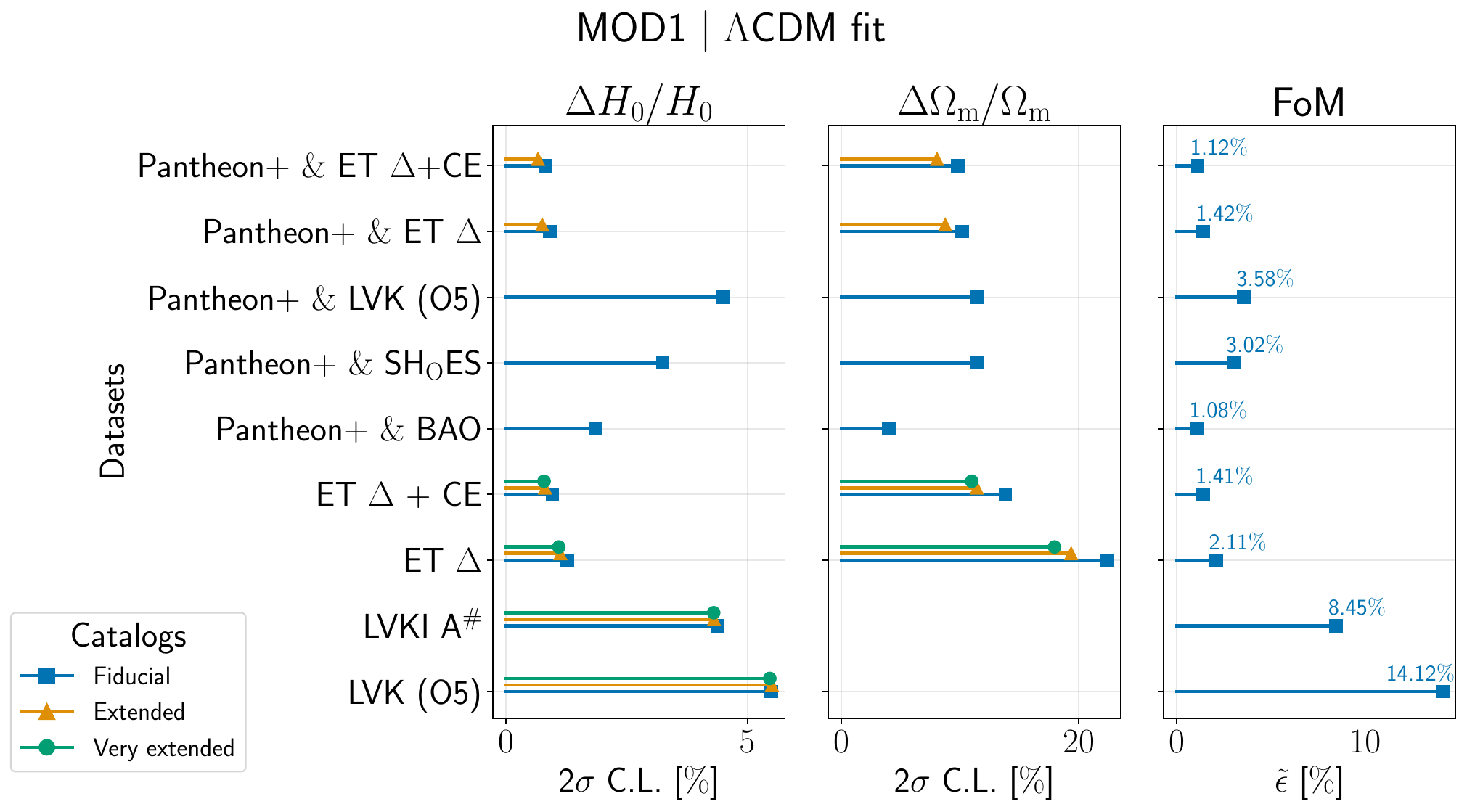}
    \caption{Whisker plot reporting the 2$\sigma$ relative uncertainties on \Hnot and $\Omega_\mathrm{m}$, as well as the corresponding FoMs, obtained by fitting an underlying MOD1 cosmology assuming $\Lambda$CDM. For some of the cases, we report the results for the three different GRB catalogs. In the main text, we refer only to the fiducial catalog.
    We do not show the error bars larger than $\gtrsim100\%$. The complete version of this plot is in Appendix~\ref{ap:tables}.}
    \label{fig:whisk_mod1_lcdm}
\end{figure}

\begin{figure}[b]
    \centering
    \includegraphics[width=0.42\textwidth]{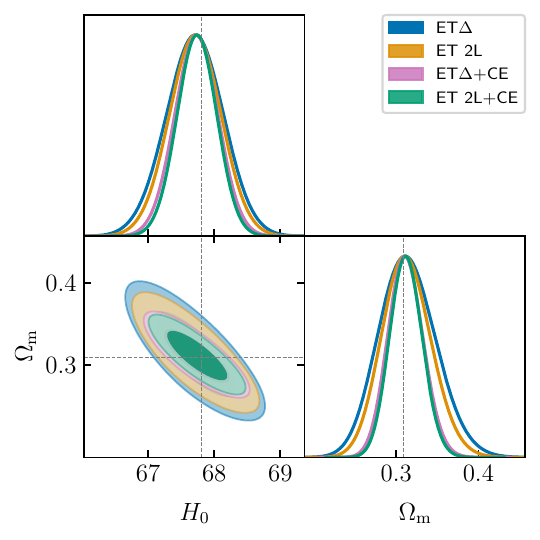}
    \includegraphics[width=0.42\textwidth]{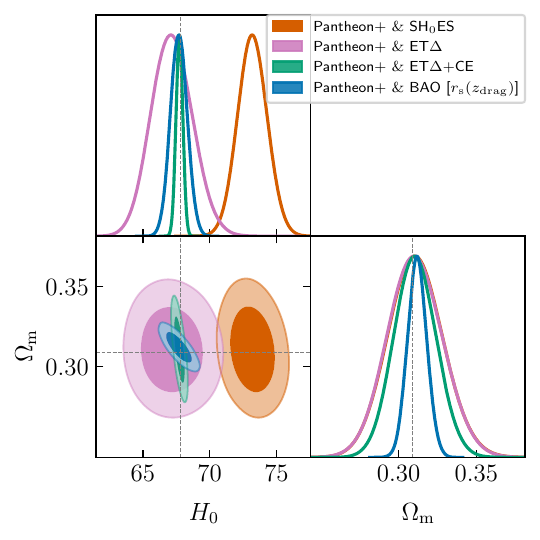}
    \caption{\lcdm constraints (68\% and 95\% C.L.) on the \Hnot and \Omegam obtained by fitting MOD1 (dashed lines) over mock MM and other complementary data sets. 
    (\textit{Left}) Comparison among the ET$\Delta$(+CE) and ET2L(+CE) configurations. (\textit{Right}) Focus on ET$\Delta$(+CE) jointly analyzed with uncalibrated SNIa data. These are compared with BAO and SH$_0$ES data sets.}
    \label{fig:triangle_mod1_lcdm}
\end{figure}

These complementary data sets are used to increase the constraining power on $\Omega_\mathrm{m}$ (Pantheon$\Plus$ or DESI) or simply for comparison purposes (BAO and SH$_0$ES). We present constraints on MOD1 and MOD2 in this fashion: in Subsection~\ref{subsec:mod1_2Gvs3G} we compare the 2G and the 3G GW detector capabilities; Subsection~\ref{subsec:3G} is dedicated to investigate the potential of 3G detector configurations in constraining a MOD1 Universe  and a MOD2 Universe; in Subsections~\ref{subsec:cpl_gp_mod1} and \ref{subsec:cpl_gp_mod2}
we discuss constraints obtained with our model-independent reconstruction method, on MOD1 and MOD2, respectively.

To quantify the constraining capabilities of the various combinations of data sets and detector combinations, we refer to a Figure of Merit (FoM) quantity~\cite{Albrecht:2006um} defined as:

\begin{equation}
    \mathrm{FoM}_X =\left[ \mathrm{det}~\mathcal{C}_X(\theta_c)\right]^{-1/2}
\end{equation}
where $X$ stands for the detector configuration (\textit{e.g.}~$X$=ET$\Delta$+CE or $X$=$\mathrm{PP}$+$\mathrm{BAO}$) and $\mathcal{C}_X(\theta_c)$ is the covariance matrix extracted from the MCMC chains. The higher is the FoM, the higher is the constraining power over the set of cosmological parameters ($\theta_c$). We now introduce a \textit{reference} FoM, $\mathrm{FoM}^{\mathrm{ref}}_{\epsilon}$, that embodies the constraining power over a set of fully uncorrelated parameter estimates, \textit{i.e.}~with a diagonal covariance matrix. Given a set of reference cosmological values $\bar{\theta}_c$, we associate $\mathrm{FoM}^{\mathrm{ref}}_{\epsilon}$ to the relative uncertainty $\epsilon \equiv \sigma_{\theta_c}/\bar{\theta}_{c}$ we want to accomplish.
In the \Hnot- \Omegam plane we can write 

\begin{equation}\label{eq:FoMref}
    \mathrm{FoM}^{\mathrm{ref}}_{\epsilon} = \left[\mathrm{det}\begin{pmatrix}
    \sigma_{H_\mathrm{0}}^2 & 0 \\
    0 & \sigma_{\Omega_\mathrm{m}}^2
    \end{pmatrix}\right]^{-1/2} = \left[\mathrm{det}\begin{pmatrix}
    \epsilon^2 \bar{H}_\mathrm{0}^2 & 0 \\
    0 & \epsilon^2 \bar{\Omega}_\mathrm{m}^2
    \end{pmatrix}\right]^{-1/2} = \frac{1}{\epsilon^2\bar{H}_\mathrm{0}~\bar{\Omega}_\mathrm{m}}
\end{equation}

This provides an intuitive method to obtain the FoM of a specific detector configuration $X$ relative to a reference FoM by setting a reference value for $\epsilon$. For example, setting $\epsilon = 1\%$ establishes the reference FoM as a fixed value for comparison. If the investigated FoM$_X$ is similar to the reference one, it indicates that the $X$ detector configuration is approaching a $1\%$ equivalent precision for both the cosmological parameters.
We can invert Eq.~\eqref{eq:FoMref} to find 
\begin{equation}\label{eq:eps}
    \epsilon = \sqrt{\frac{1}{\mathrm{FoM}^{\mathrm{ref}}_{\epsilon}\bar{H}_\mathrm{0}~\bar{\Omega}_\mathrm{m}}}
\end{equation}
We now introduce a practical way to quantify the precision achieved by a specific configuration $X$, \textit{i.e.}~the effective relative uncertainty $\tilde{\epsilon}_X$, defined as
\begin{equation}\label{eq:epstilde}
    \tilde{\epsilon}_X = \sqrt{\frac{1}{\mathrm{FoM}_X \bar{H}_\mathrm{0}~\bar{\Omega}_\mathrm{m}}}~.
\end{equation}
We stress that $\tilde{\epsilon}_X$ does not identify the true statistical uncertainty. Rather, it represents a useful proxy to compare the capabilities of different detectors. In this specific case study, this metric is straightforward and cannot be misinterpreted because the parameter degeneracies remain unchanged. This makes it an optimal method for comparing the significance of different detector configurations. We report the actual $2\sigma$ relative uncertainties on the relevant cosmological parameter in Appendix~\ref{ap:tables} and the whisker plots throughout the main text.

\begin{figure}
    \centering
    \includegraphics[width=0.42\textwidth]{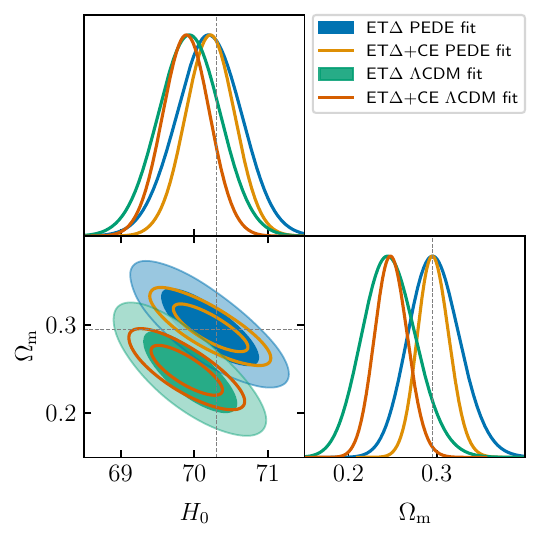}
    \includegraphics[width=0.42\textwidth]{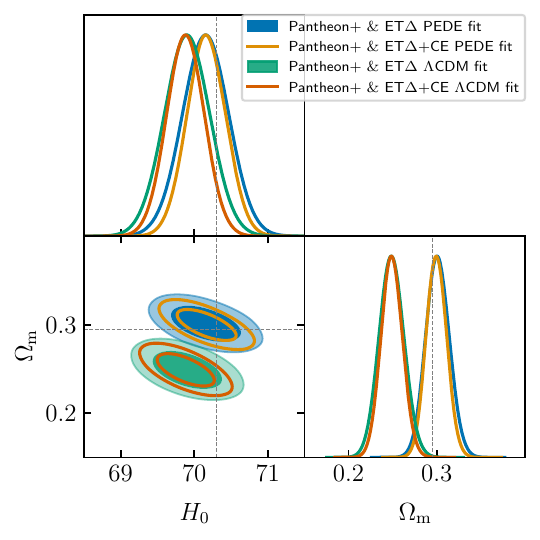}
    \caption{PEDE (blue and orange) and \lcdm (green and dark orange) constraints on \Hnot and \Omegam obtained by fitting a PEDE cosmology (MOD2) over mock MM and Pantheon$\Plus$ data. (\textit{Left}) Comparison between ET$\Delta$ and ET$\Delta$+CE. (\textit{Right}) Joint analyses of the configurations used in the left panel with uncalibrated SNIa data.}
    \label{fig:triangle_mod2}
\end{figure}

\subsection{Comparison between 2G and 3G GW detectors}\label{subsec:mod1_2Gvs3G}

In this Section we compare the performances of 2G and 3G GW detectors. Unless otherwise stated, we always assume that the analysis is performed with the fiducial GRB catalog. We show and compare results obtained with different catalogs in Appendix~\ref{ap:grbt_comparison}. 
We discuss only the MOD1 mock data-set and the \lcdm reconstruction. As similar considerations can be done for other fitting models, we discuss further cases in Appendix~\ref{ap:det_comparison}.
The uncertainty in the cosmological parameters are the 2$\sigma$ relative uncertainties obtained from the MCMC chains (\textit{e.g.},~$\Delta H_\mathrm{0}/H_\mathrm{0} = 2\sigma_{H_\mathrm{0}}/H_\mathrm{0}$)\footnote{We adopt a different convention with respect to Ref.~\cite{Belgacem:2018lbp}}. 
Following Eq.~\eqref{eq:epstilde}, we quantify the constraining power of each configuration through $\tilde{\epsilon}$, which we refer to as FoM for brevity. Since $\tilde{\epsilon}\propto \mathrm{FoM}^{-1/2}$, a smaller $\tilde{\epsilon}$ corresponds to a larger FoM, i.e.\ to tighter constraints. Unless otherwise stated, the improvement factors quoted in the following refer to ratios of $\tilde{\epsilon}$.

In the left panel of Figure~\ref{fig:LVK_ET_LL} we show the constraints extracted from the LVK (O5), LVKI A$^\#$ and ET$\Delta$ analyses. For LVK, we find a 5.5$\%$ uncertainty on \Hnot and a 4.4$\%$ for LVKI A$^\#$, while ET$\Delta$ stands at $1.3\%$ accuracy. Concerning $\Omega_\mathrm{m}$, one can clearly see that both in the LVK and LVKI A$^\#$ cases it is practically unconstrained, \textit{i.e.}, with a relative uncertainty $\gtrsim 100\%$. The FoMs summarize these considerations: $\tilde{\epsilon}_{\mathrm{LVK}}=14.1\%$ , $\tilde{\epsilon}_{\mathrm{LVKI A}^{\#}}=8.4\%$ and $\tilde{\epsilon}_{\mathrm{ET}\Delta}=2.1\%$.
These numbers and the relative uncertainties on the cosmological parameters are reported in Figure~\ref{fig:whisk_mod1_lcdm} (but see also Appendix~\ref{ap:tables}).
To overcome the issue of an unconstrained $\Omega_\mathrm{m}$, as it is due to the limited redshift coverage of 2G detectors, it is worth performing a joint analysis with a tracer of $H(z)$ at $z\gtrsim 1$, such as the uncalibrated SNIa data catalog from Pantheon$\Plus$. These explore the regime where the $\{d_L-z\}$ relation is influenced by the \Omegam and \Omegade values. We show the results of the joint GW+SNIa analyses in the right panel of Figure~\ref{fig:LVK_ET_LL}. In other words, as the $z_{\rm max}$ reached by 2G detectors like LVK is $\approx$ 0.25, they can only constrain the Taylor expansion at low redshift of Eq.~\eqref{eq:H2} (\textit{i.e.},~the Hubble's law). Therefore, for LVK, the addition of \panp data does improve the constraints on \Omegam down to 11$\%$ with just 8 GW events. Note that the resulting $\tilde{\epsilon}_{\mathrm{PP}+\mathrm{LVK}}=3.6\%$ is still larger than the one obtained with 3G detectors alone ($\tilde{\epsilon}_{\mathrm{ET}\Delta}=2.1\%$), \textit{i.e.}~the constraining power remains weaker, as also shown in Figure~\ref{fig:whisk_mod1_lcdm}.
From Figure~\ref{fig:LVK_ET_LL} it is clear that 3G detectors, being able to observe a larger volume, are themselves a good tracer of $H(z)$ up to $z_{\rm max} \approx 3$. However, one can appreciate the enhanced constraining capabilities provided by the addition of \panp even when combined with 3G GW data. The corresponding FoM is $\tilde{\epsilon}_{\mathrm{PP}+\mathrm{ET}\Delta}=1.4\%$, to be compared with $\tilde{\epsilon}_{\mathrm{ET}\Delta}=2.1\%$.
We finally notice that the closest MM events, for which 3G detectors would provide the lowest uncertainty in the $d_L$ estimation (see the orange curve in Figure \ref{fig:peculiar}), are the ones for which the peculiar velocity is the dominant component of the uncertainty. Such an ever-present source of uncertainty at low redshift could be mitigated by a better modeling of the host galaxy peculiar motion. In Appendix~\ref{ap:tables} we show a comparison with the ideal case of no peculiar velocities, namely without the correction that we introduced in Section~\ref{subsec:GWfish+}.

\subsection{Comparison among different 3G GW detectors}\label{subsec:3G}
In this Section we focus our discussion on comparing the performances of different configurations of 3G GW detector networks, with and without additional external SNIa data from Pantheon$\Plus$. To begin with, we present the results of the reconstruction of an underlying \lcdm cosmology (MOD1) with parameters ~$\theta_c^{\Lambda\mathrm{CDM}}=\{H_\mathrm{0}=67.81~\mathrm{km ~Mpc}^{-1}\mathrm{s}^{-1}, \, \Omega_\mathrm{m}=0.309\}$. We do not explicitly show the case of a PEDE fitting model over MOD1 because this is completely similar to carry out a \lcdm fit on MOD2, that is the second case that we discuss in detail in this Section.

In the left panel of Figure~\ref{fig:triangle_mod1_lcdm} we compare the ET$\Delta$, ET2L,  ET$\Delta$+CE and ET2L+CE configurations. A full comparison of all the configurations examined in this paper is available in Appendix~\ref{ap:tables} (Table~\ref{tab:mod1_lcdm} and Figure~\ref{fig:whisker_LL_complete}). From the left panel of Figure~\ref{fig:triangle_mod1_lcdm} we notice that the ET2L configuration performs better than ET$\Delta$, both alone and in a network with CE. For ET$\Delta$ (ET2L) we obtain an uncertainty on \Hnot of 1.3$\%$ (1.2$\%$), on \Omegam of 22$\%$ (19$\%$) and a FoM of 2.1$\%$ (1.9$\%$). Notice also that the addition of CE shrinks the contours, improving the constraining power with respect to ET alone in either of its configurations, reaching a sub-percent level precision on $H_\mathrm{0}$. This results in a highly similar FoM in the two cases: 1.4$\%$ for ET$\Delta$+CE and 1.3$\%$ for ET2L+CE. The improvement observed in the latter two configurations can be attributed to the factors briefly mentioned at the end of Subsection~\ref{subsec:datasets}. Specifically, these improvements result from the combined effect of the duty cycle and the increased sensitivity of the network with the introduction of the CE observatory.

In the right panel of Figure~\ref{fig:triangle_mod1_lcdm}, we present the results obtained by combining the \panp data set with ET$\Delta$.
These results are compared to the current measurements of $H_0$ and $\Omega_m$ from \panp data calibrated with both SH$_0$ES and BAO.
In the analysis with SH$_0$ES, both the mock GW and mock \panp data sets are generated with the same MOD1 underlying cosmology, and jointly analyzed to accurately reconstruct $\Omega_\mathrm{m}$. On the contrary, actual BAO data are combined with the actual \panp data set. This strategy, outlined in Subsection~\ref{subsec:mock}, enables a fair visual comparison between current and future experiments\footnote{Note that, even when combining real data with mock data, the FoM reflecting the constraining power would remain unchanged. This is because the precision is not influenced by whether the injected cosmological parameters are correctly recovered.}. As shown by the contours in the right panel of Figure~\ref{fig:triangle_mod1_lcdm}, combining current SNIa data with GWs yields constraining capabilities similar to those provided by BAO. We underline that the BAO constraints on \Omegam outperform the ones from GW data, while the contrary happens for the estimates on $H_\mathrm{0}$, resulting in comparable FoMs for the two cases, namely $\tilde{\epsilon}_{\mathrm{PP}+\mathrm{BAO}}=1.1\%$ and $\tilde{\epsilon}_{\mathrm{PP}+\mathrm{ET}\Delta}=1.4\%$.
As testified by the tightness of the contours that we report, GW observations from future interferometers are a promising way to finally unambiguously set the \Hnot (or cosmic calibrator) tension debate~\cite{Chen:2017rfc, Muttoni:2023prw, Borghi:2023opd, Mancarella:2022cnu, Mancarella:2024qle}.
From Figure~\ref{fig:whisk_mod1_lcdm} we notice that the extended catalog (containing 53 events, rather than 38, in the ET$\Delta$+CE configuration) consistently provides more precise limits, while using the very extended catalog (60 events) does not significantly improve the constraints. The improvement is in fact primarily driven by the inclusion of high-$z$ GRBs, rather than by the larger statistics. However, since such improvement is always of the same order, we stick to discussing only the results obtained with the fiducial catalog. We refer the reader to Appendix~\ref{ap:grbt_comparison} for a comprehensive comparison among all catalogs.

\begin{figure}[t]
    \centering
    \includegraphics[width=0.9\linewidth]{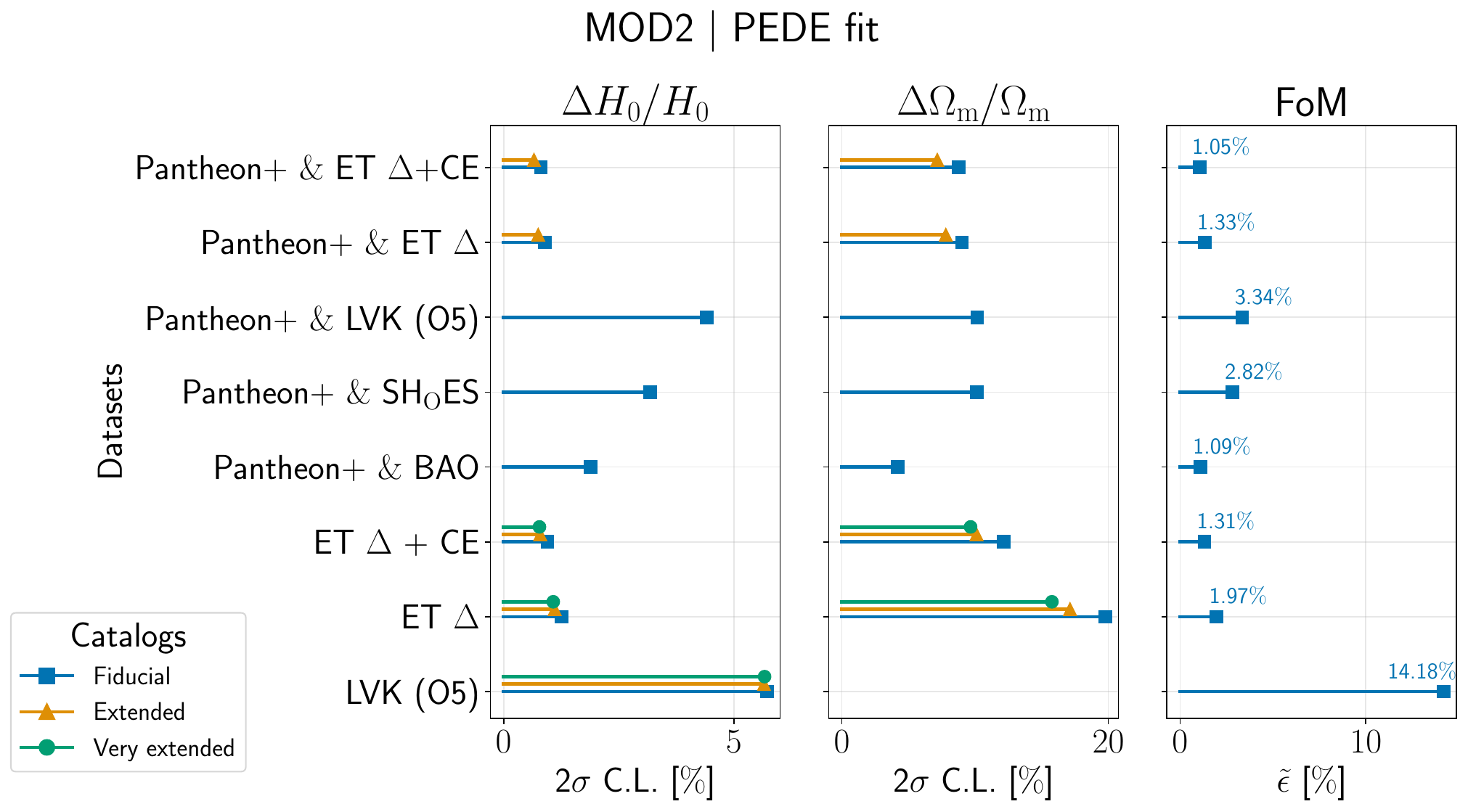}
    \caption{Whisker plot reporting the 2$\sigma$ relative uncertainties on \Hnot and $\Omega_\mathrm{m}$, as well as the corresponding FoMs, obtained by fitting an underlying MOD2 cosmology assuming PEDE. The absence of line is related to an uncertainty greater than $\gtrsim100\%$. The complete version of this plot is in Appendix~\ref{ap:tables}.}
    \label{fig:whisk_mod2_pede}
\end{figure}
\sloppy
We now discuss the MOD2 underlying cosmology, that we produced considering a conservative set of parameters for the PEDE scenario, namely $\theta_c^{\mathrm{PEDE}}=\{H_\mathrm{0}=70.3~\mathrm{km ~Mpc}^{-1}\mathrm{s}^{-1},\Omega_\mathrm{m}=0.295\}$. As explained in Section~\ref{subsec:mock}, these fiducial values correspond to the best-fit parameters from a joint analysis with Pantheon$\Plus$, BAO data from eBOSS, and \textit{Planck} 2018 CMB data.
In Figure~\ref{fig:triangle_mod2} we compare constraints obtained by fitting MOD2 with PEDE and with $\Lambda$CDM, for ET$\Delta$ (filled contours) and ET$\Delta$+CE (empty contours), with and without \panp (left and right panel, respectively). One can easily see that the size of the contours are similar. In fact, the ET$\Delta$ FoMs for PEDE and \lcdm are $\tilde{\epsilon}_{\mathrm{ET}\Delta}=1.9\%$ and $\tilde{\epsilon}_{\mathrm{ET}\Delta}=2.0\%$, respectively.
As stated before, we underline that the network of ET$\Delta$ and CE outperforms the ET$\Delta$ alone, for both the fitting cosmologies. The difference becomes even less evident in the joint analysis with \panp (right panel). The value of the FoMs for ET$\Delta$ are  $\tilde{\epsilon}_{\mathrm{PP}+\mathrm{ET}\Delta}=1.3\%$ and $\tilde{\epsilon}_{\mathrm{PP}+\mathrm{ET}\Delta}=1.4\%$, for PEDE and \lcdm respectively.
These two GW detector configurations give similar constraints on $H_\mathrm{0}$, whereas ET$\Delta$+CE is a factor 1.6 better in constraining \Omegam (see Table~\ref{tab:mod2_pede}). This can be attributed to the ability of the ET$\Delta$+CE configuration to detect GRBs at higher redshifts compared to the ET$\Delta$ configuration alone. Indeed, $\tilde{\epsilon}$ decreases from $\tilde{\epsilon}_{\mathrm{ET}\Delta}=2.0\%$ to $\tilde{\epsilon}_{\mathrm{ET}\Delta+\mathrm{CE}}=1.3\%$, corresponding to a FoM more than a factor 2 larger for ET$\Delta$+CE.
As expected, the addition of \panp data drives the \Hnot and \Omegam constraints to become very similar, so that the FoMs in this case read 
$\tilde{\epsilon}_{\mathrm{PP}+\mathrm{ET}\Delta}=1.3\%$ and $\tilde{\epsilon}_{\mathrm{PP}+\mathrm{ET}\Delta+\mathrm{CE}}=1.0\%$. We report the list of FoMs and uncertainties discussed in this Section in Figure~\ref{fig:whisk_mod2_pede}, while the full list is shown in Appendix~\ref{ap:tables} (Table~\ref{tab:mod2_pede} and Figure~\ref{fig:whisker_PP_complete}).

Notice that, even though MOD1 and MOD2 are characterized by relatively similar phenomenologies, when MOD2 is fitted within $\Lambda$CDM, we do not retrieve the injected values of the cosmological parameters. This is due to the fact that \lcdm is not able to reproduce the phenomenology of a dynamical DE described by the $w(z)$ evolution (Eq.~\eqref{eq:wPEDE}). Therefore, the net effect of wrongly assuming \lcdm as a fitting model is a shift in the inferred values of \Omegam and \Hnot towards values lower than the injected ones. Such a shift clearly points toward a crucial fact: assuming an arbitrary model to fit data under an unknown underlying cosmology might produce biased results.
From the left panel of Figure~\ref{fig:triangle_mod2} it is apparent that in the ET$\Delta$ configuration such biases are present but not statistically significant, whereas in the most constraining case of ET$\Delta$+CE the shift between \lcdm and PEDE contours is larger than 2$\sigma$. Analogous considerations apply to the joint GW+SNIa analysis (right panel). 
We explicitly checked that the same issue occurs when PEDE is fitted on the MOD1 mock data set. In the next Subsections~\ref{subsec:cpl_gp_mod1} and \ref{subsec:cpl_gp_mod2} we will show how to overcome this problem by means of model-independent reconstruction techniques.

\subsection{Model-independent reconstruction of a \tpdf{$\Lambda$CDM} Universe (MOD1)}\label{subsec:cpl_gp_mod1}

\begin{figure}
    \centering
    \includegraphics[width=0.5\textwidth]{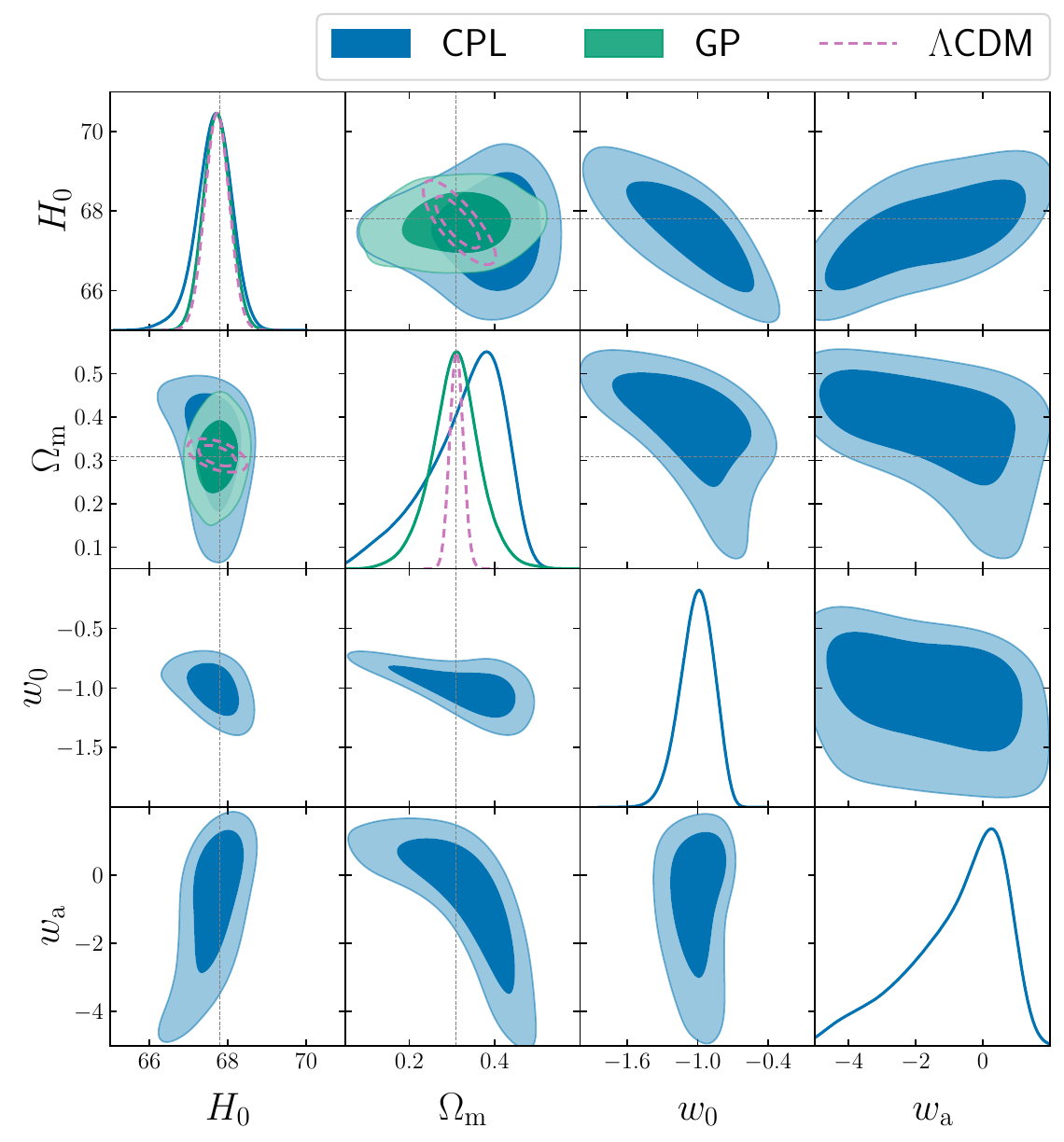}
    \caption{CPL (blue), GP (green) and \lcdm (empty pink) constraints on $H_\mathrm{0}$, $\Omega_\mathrm{m}$, \wnot and \wa obtained by fitting a $\Lambda$CDM cosmology (MOD1) over mock MM and Pantheon$\Plus$ data. (\textit{Upper corner}) Contour comparison for ET$\Delta$ alone. (\textit{Lower corner}) Contour comparison for ET$\Delta$ joint analysis with uncalibrated SNIa data. The 1-dimensional posterior refers to the lower corner plot.}
    \label{fig:ET_delta_MOD1_all_cosmo}
\end{figure}

\begin{figure}
    \centering
    \includegraphics[width=0.9\textwidth]{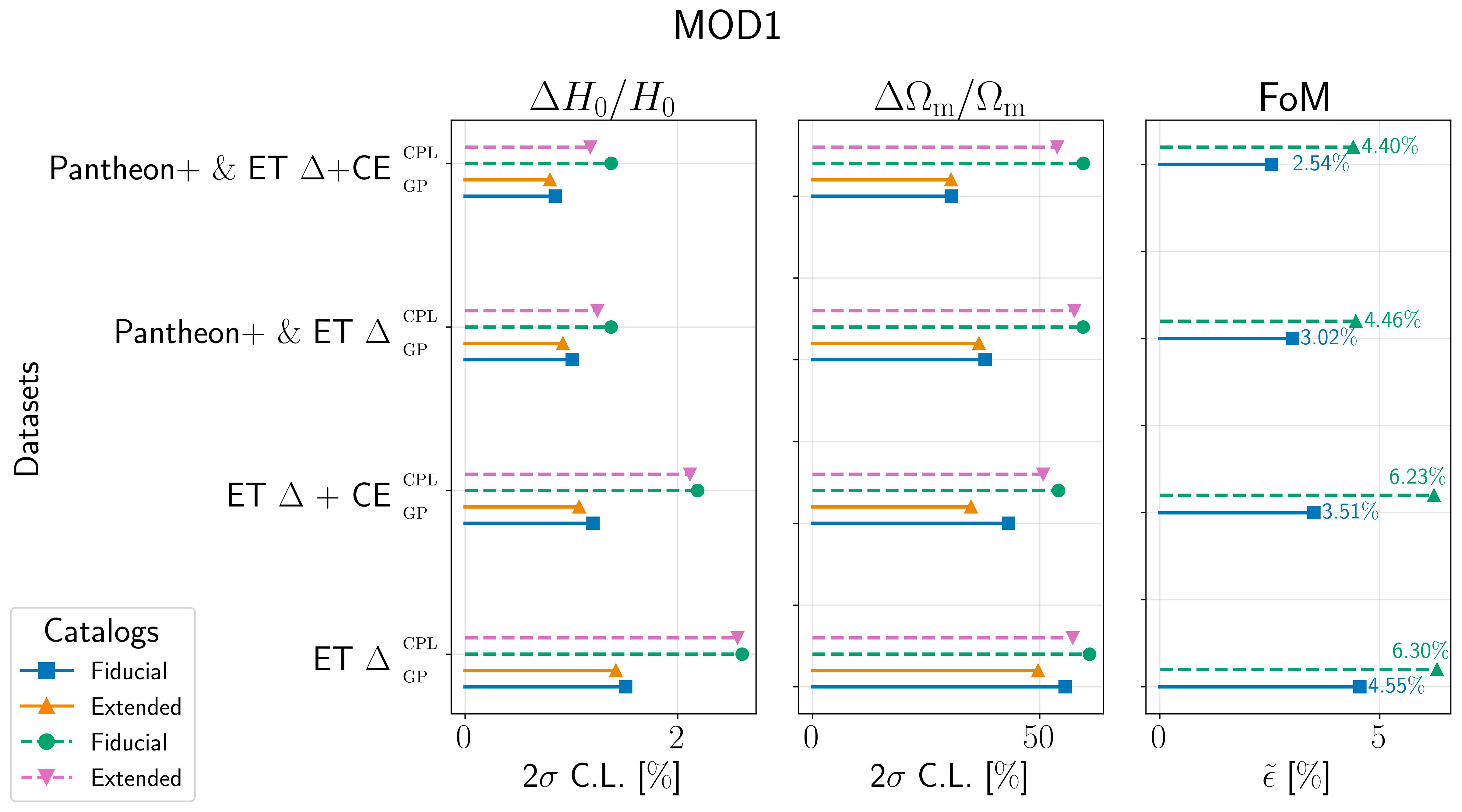}
    \caption{Whisker plot reporting the 2$\sigma$ relative uncertainties on \Hnot and $\Omega_\mathrm{m}$, as well as the corresponding FoMs, obtained by fitting an underlying MOD1 cosmology. We compare the CPL (dashed lines) and GP (solid lines) estimates for fiducial and extended catalogs. The FoMs are reported only for the fiducial catalog.}
    \label{fig:whisker_33}
\end{figure}

In this Section we focus on the comparison of parametric (CPL) and non-parametric (GP) fitting methods over MOD1. When comparing GP with CPL, besides $H_0$ and $\Omega_\mathrm{m}$, we show the constraints on the DE EoS parameters, \wnot and $w_\mathrm{a}$. Only when discussing the GP results alone, we show the constraints on the GP hyperparameters $\sigma_f$ and $\ell_f$ (in logarithmic scale for visualization purposes). 
In the case of MOD1 we focus on the results for ET$\Delta$ and its combination with Pantheon$\Plus$, while a fully comprehensive set of results can be found in Appendix~\ref{ap:tables} (Tables~\ref{tab:mod1_cpl}, \ref{tab:mod1_gp}, and Figures~\ref{fig:whisker_LC_complete}, \ref{fig:whisker_LGP_complete}). 
From Figure~\ref{fig:ET_delta_MOD1_all_cosmo}, as expected, one can see that the constraining power on \Hnot and \Omegam from both GP and CPL is lower with respect to the cases presented in Subsection~\ref{subsec:3G}, due to the larger parameter space explored. In particular, the upper corner plot displays the constraints for ET$\Delta$, while the lower corner one shows the combination of \panp with ET$\Delta$. The one-dimensional marginalized posteriors refer to the lower corner plot,~\emph{i.e.},~to the most constraining configuration. We compare the CPL, GP and \lcdm fitting models. As just stated, having more freedom, both CPL and GP perform worse in terms of constraining power with respect to the $\Lambda$CDM fit. Nonetheless, both of them are able to recover the injected cosmology. In the case of ET$\Delta$, the CPL constraint on \Hnot
is $2.6\%$, while the one given by GP is $1.5\%$.
For what concerns $\Omega_\mathrm{m}$, we get a relative uncertainty of $61\%$ in CPL, and of $58\%$ for GP. The FoMs are $\tilde{\epsilon}_{\mathrm{ET}\Delta}=6.3\%$ and $\tilde{\epsilon}_{\mathrm{ET}\Delta}=4.6\%$, for CPL and GP respectively.
Thus, we can conclude that the agnostic reconstruction of GP outperforms the standard parametric CPL approach. The difference between the two is a factor $\sim 1.3$.

The constraining capability and eventually the FoM further improve when we include Pantheon$\Plus$, as the additional constraint on \Omegam allows to better constrain the DE evolution. The tightening of the contours translates into more precise estimates of both \Hnot and $\Omega_\mathrm{m}$. The \Hnot ($\Omega_{\mathrm{m}}$) uncertainties are $1.4\%$ ($60\%$) and $1.0\%$ ($40\%$) for CPL and GP respectively.
The FoMs are $\tilde{\epsilon}_{\mathrm{PP}+\mathrm{ET}\Delta}=4.5\%$ for CPL and $\tilde{\epsilon}_{\mathrm{PP}+\mathrm{ET}\Delta}=3.0\%$ for GP. The precision of the GP constraints in this case improves over CPL by a factor $\sim 1.5$. In Figure \ref{fig:whisker_33} we report a comprehensive comparison of the CPL and GP relative uncertainties and FoMs.

\begin{figure}
    \centering
    \includegraphics[width=\textwidth]{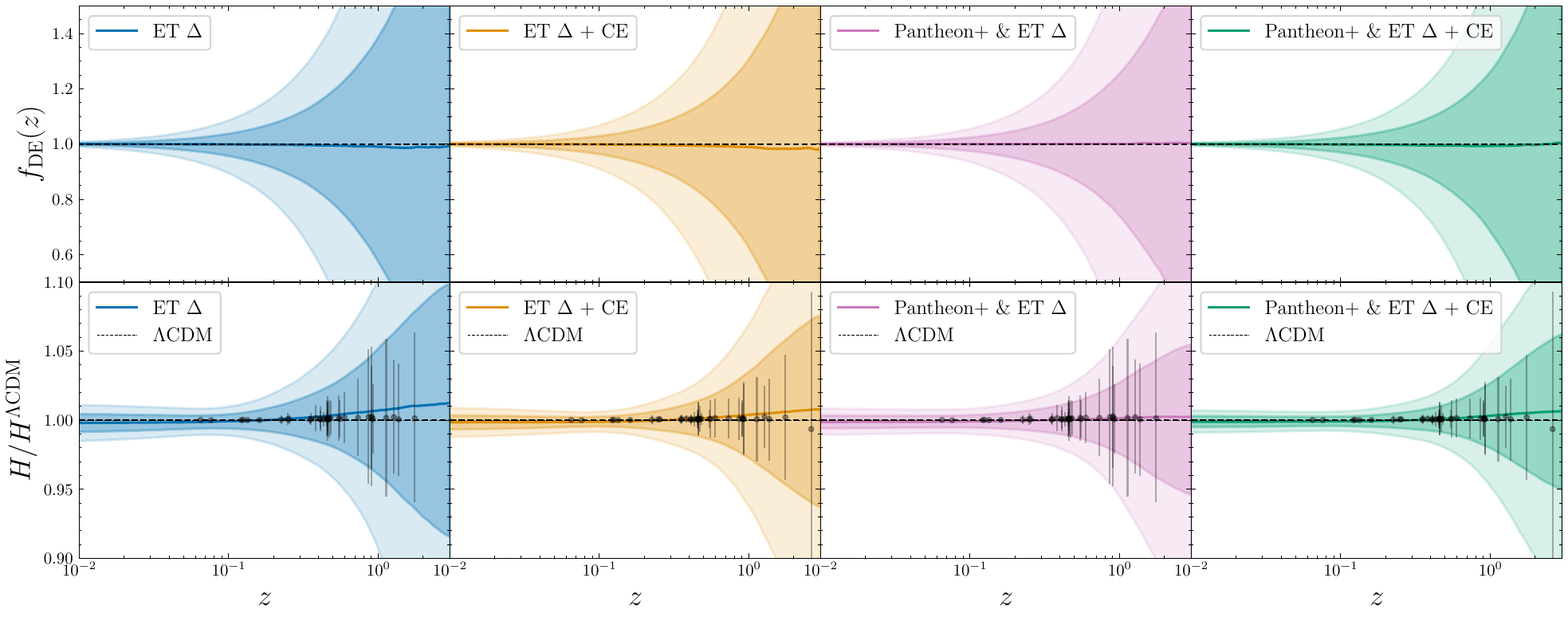}
    \caption{Marginalized posterior distributions on $\fde(z)$ and $H(z)/H^{\Lambda\textrm{CDM}}$ obtained with GP for an underlying MOD1 Universe.
    The dashed lines show the \lcdm behaviour of $\fde(z)$ and $H(z)/H^{\Lambda\textrm{CDM}}$. The colored solid lines show the mean GP reconstruction of those functions, while the shaded regions show the $1\sigma$ and $2\sigma$ uncertainty bands. 
    From left to right we display the ET$\Delta$ (+CE) configuration and the same but including the \panp mock data set (last two plots). These results are relative to the fiducial GRB catalog.}
    \label{fig:LCDM-GP-recos}
\end{figure}

We now discuss the GP reconstruction of the function $f_\mathrm{DE}(z)$, as shown in the upper panels of Figure \ref{fig:LCDM-GP-recos}, obtained by marginalizing over the two cosmological parameters (\Hnot and $\Omega_{\mathrm{m}}$) and the two hyperparameters ($\sigma_f$ and $\ell_f$). 
The \fde reconstruction is perfectly in agreement with the MOD1 underlying model (reported as a dashed black line in the plots), for the entire redshift range spanned by our data.\footnote{This was expected, as the GP mean function corresponds to the true underlying cosmology used to generate the data. We will show that the method is able to recover the true cosmology even when the latter differs from the GP mean function in the next Subsection.} The colored line represents the mean reconstruction, while the shaded areas around it are the $1\sigma$ and $2\sigma$ bands. We show the reconstruction performed for ET$\Delta$, ET$\Delta$+CE, both with and without Pantheon$\Plus$. 
As explained in Subsection~\ref{subsec:GPth} we have adopted an untrained GP procedure~\cite{Calderon2022,Calderon2023}. For this reason, the \fde is well-inferred even where no data points are present. As reflected by the FoM quoted above, the more the cosmological probes are constraining, the more the \fde uncertainty decreases. However, all of them show a worsening of the uncertainties at $z>1$, reflecting the uncertainty associated with the GRB-GW mock catalogs. The main limitation preventing a more precise inference of the DE phenomenology is given by the poor estimates of the $d_L$ parameter at high-$z$. A larger number of GRBs at higher $z$ with respect to the ones currently available would significantly reduce the uncertainties on the reconstructed $f_{\mathrm{DE}}$. This is clearly visible by looking at how $\tilde{\epsilon}$ decreases (\textit{i.e.}~the FoM increases) for all detector configurations when the extended GRB catalog is used instead of the fiducial one (see Table~\ref{tab:mod1_gp}).
A similar discussion can be done regarding the reconstructed Hubble parameter $H(z)$, normalized to the MOD1 \lcdm fiducial expansion rate, displayed in the lower panels of Figure~\ref{fig:LCDM-GP-recos}. 
The results reflect the ones seen for \fde: the most constraining configuration is given by \panp joint with ET$\Delta$+CE. The data points and error bars are derived by propagating the uncertainties from the $d_L$ posterior, which was estimated using our GW detector simulator, assuming the same underlying cosmology injected to generate the mock data.

\subsection{Model-independent reconstruction of a PEDE Universe (MOD2)}\label{subsec:cpl_gp_mod2}

\begin{figure}
    \centering
    \includegraphics[width=0.49\textwidth]{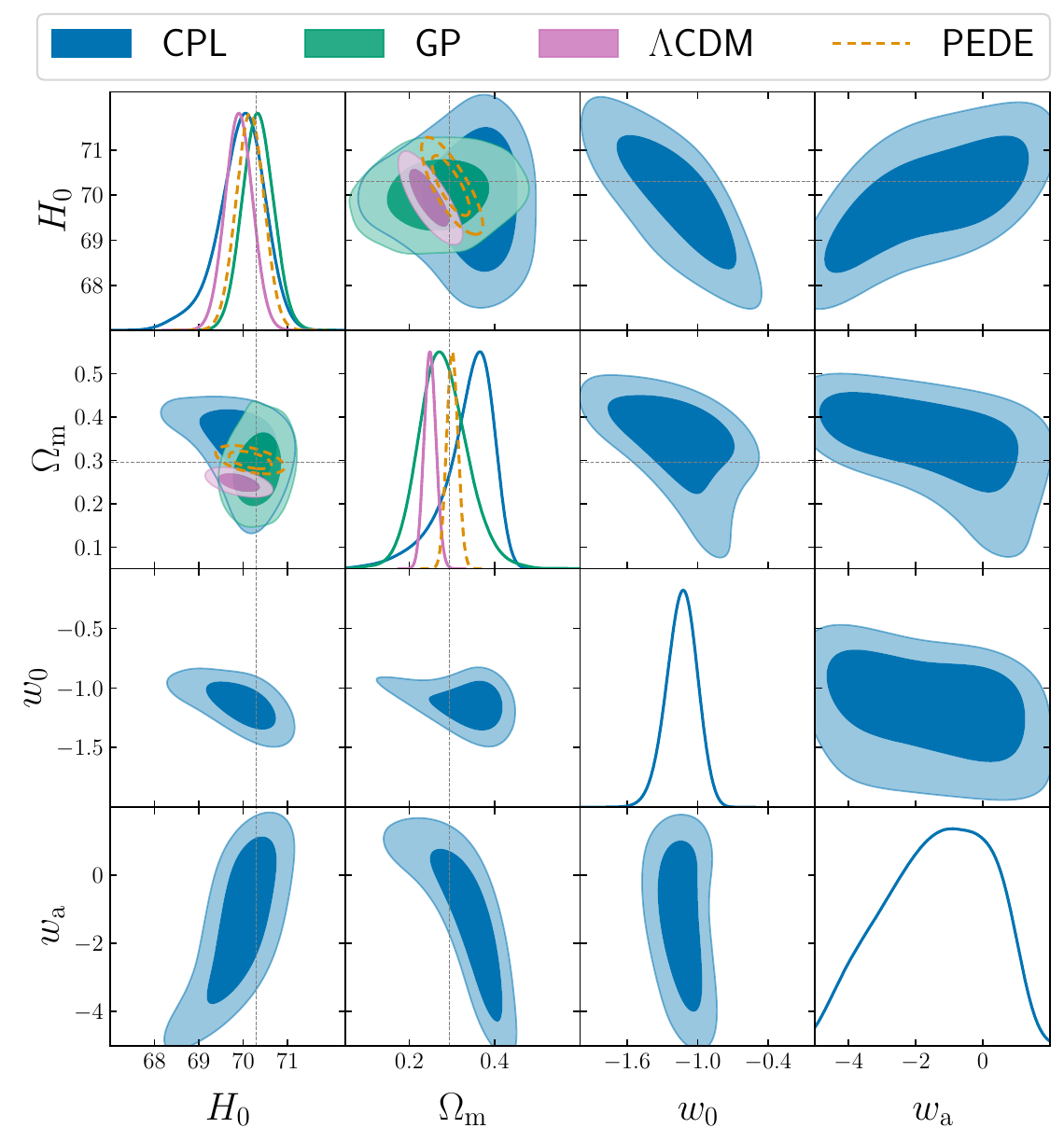}
    \includegraphics[width=0.49\textwidth]{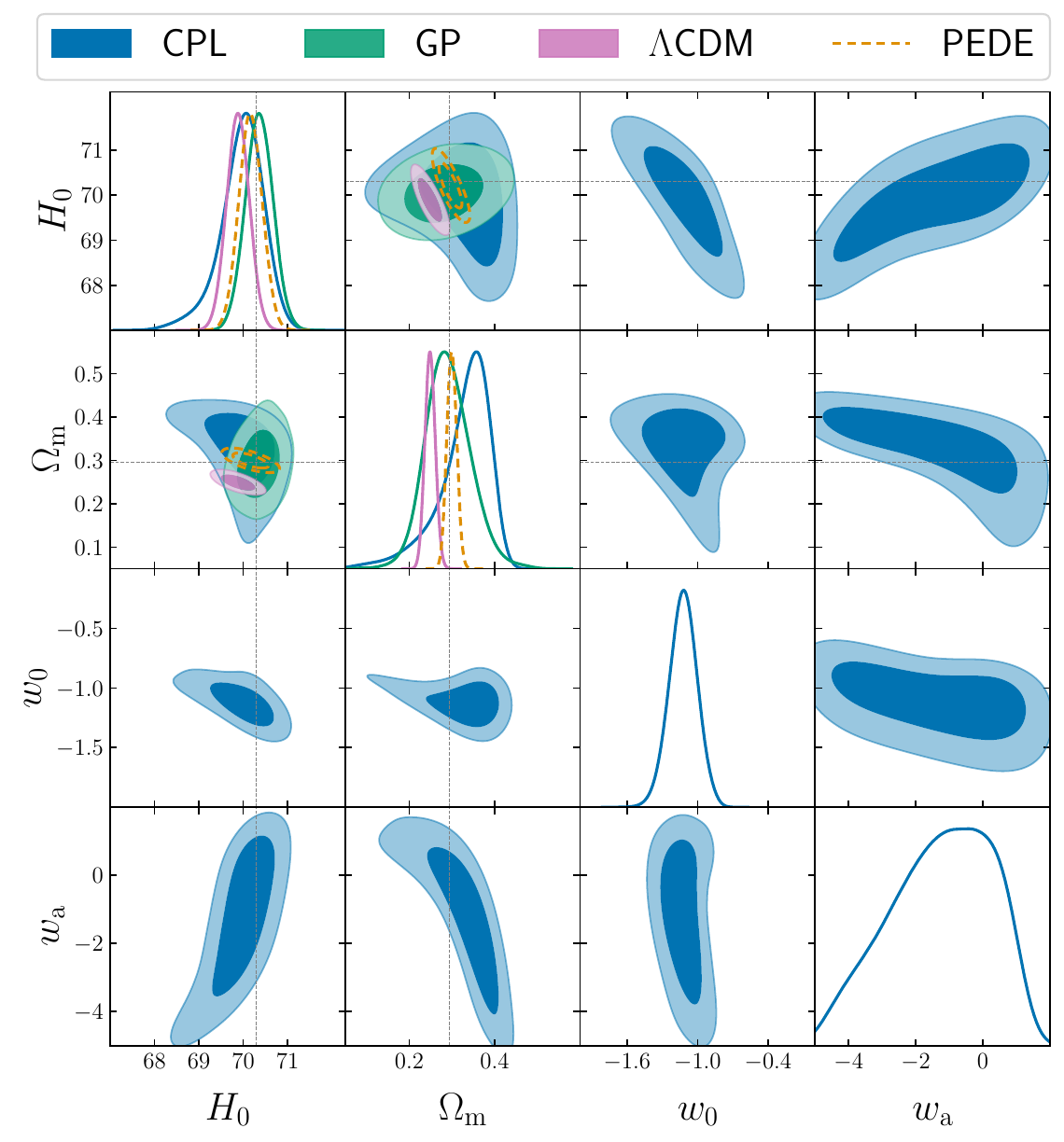}
    \caption{CPL (blue), GP (green), \lcdm (pink) and PEDE (empty yellow) constraints on $H_\mathrm{0}$, $\Omega_\mathrm{m}$, \wnot and \wa obtained by fitting a PEDE cosmology (MOD2) over mock MM and Pantheon$\Plus$ data. (\textit{Left})
    (\textit{Upper corner}) Contour comparison for ET$\Delta$ alone. (\textit{Lower corner}) Contour comparison for ET$\Delta$ joint analysis with uncalibrated SNIa data. (\textit{Right}) Same as the left panel but for the ET$\Delta$+CE network. The 1-dimensional posterior refers to the lower corner plot.}
    \label{fig:ET_delta/CE_MOD2_all_cosmo}
\end{figure}

\begin{figure}
    \centering
    \includegraphics[width=0.9\textwidth]{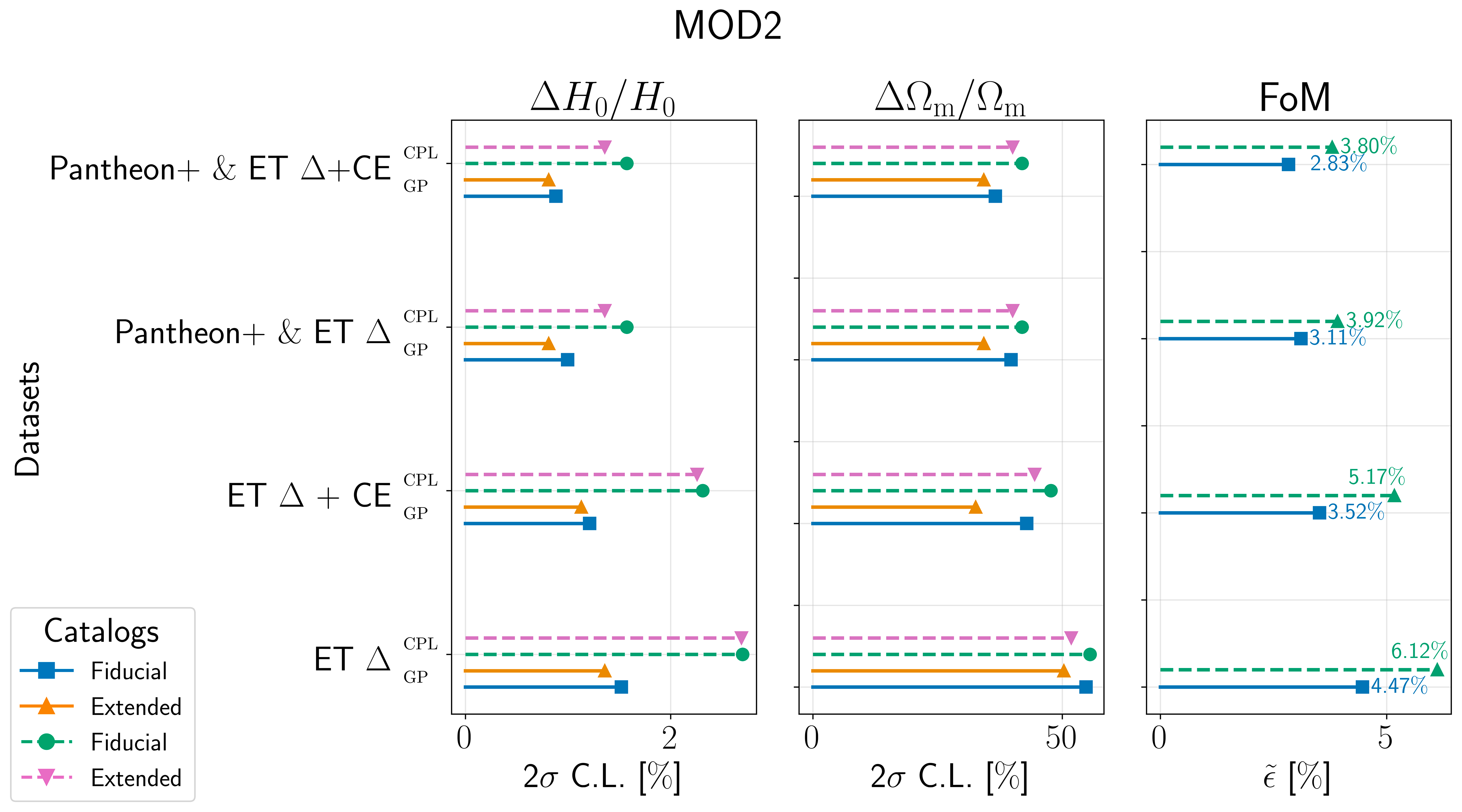}
    \caption{Whisker plot reporting the 2$\sigma$ relative uncertainties on \Hnot and $\Omega_\mathrm{m}$, as well as the corresponding FoMs, obtained by fitting an underlying MOD2 cosmology. We compare the CPL (dashed lines) and GP (solid lines) estimates for fiducial and extended catalogs. The FoMs are reported only for the fiducial catalog.}
    \label{fig:whisker_34}
\end{figure}

\begin{figure}
    \centering
    \includegraphics[width=0.6\textwidth]{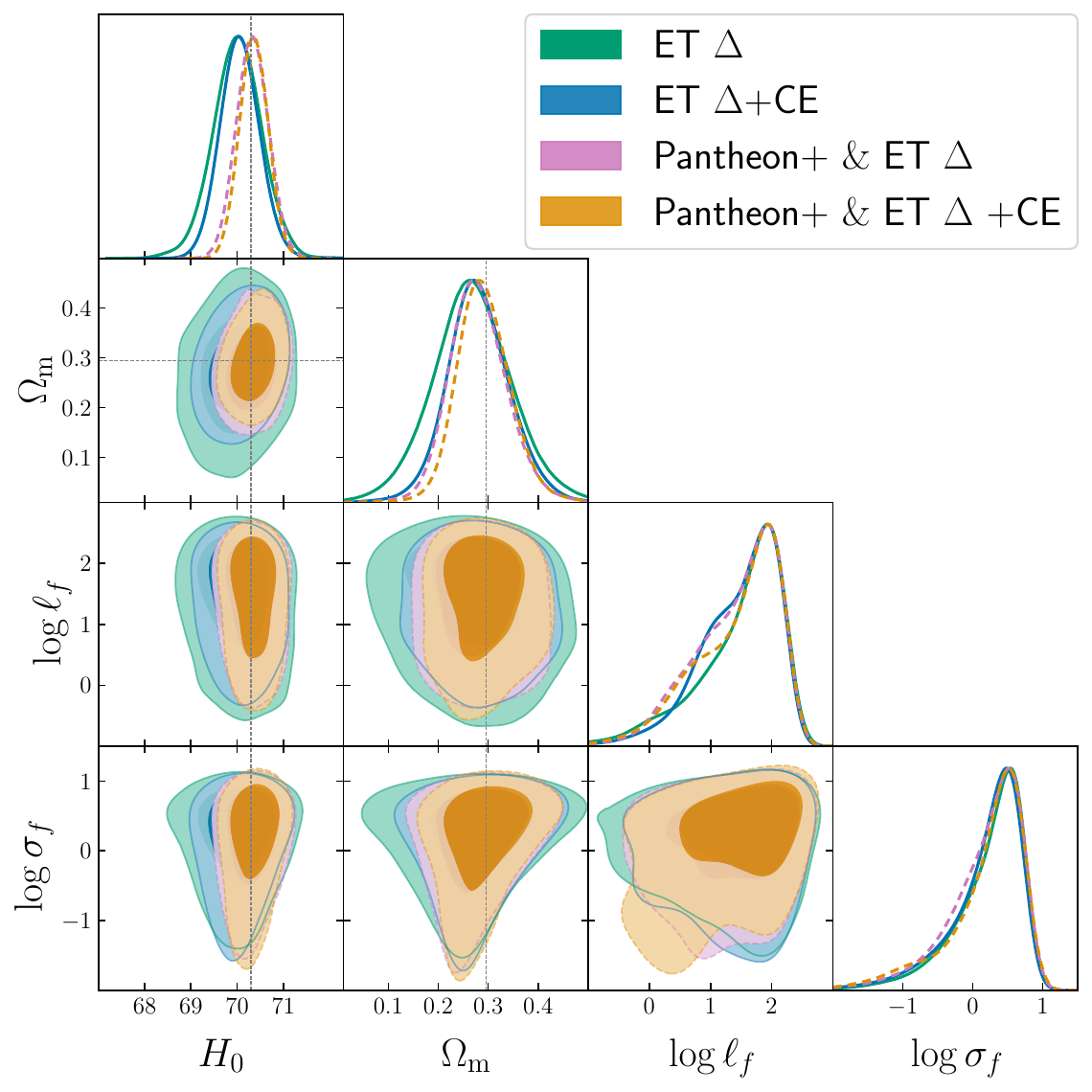}
    \caption{GP constraints (68\% and 95\% C.L.) on the physical parameters \Hnot and $\Omega_\mathrm{m}$, and the hyperparameters log$\sigma_f$ and log$\ell_f$ obtained by fitting MOD2 (dashed lines) over mock MM and Pantheon$\Plus$ data. We report the contours for ET$\Delta$ and ET$\Delta$+CE with and without Pantheon$\Plus$.}
    \label{fig:GP_confront_MOD2}
\end{figure}

In this Subsection we discuss the results obtained with CPL and GP on the GRB-GW mock data set generated assuming PEDE (MOD2). We report the results of the ET$\Delta$ and ET$\Delta$+CE configurations in the left and right panels of Figure~\ref{fig:ET_delta/CE_MOD2_all_cosmo}, respectively. We show the contours for all the fitting models considered in this work, \textit{i.e.} CPL, GP, \lcdm and PEDE. 
In Figure~\ref{fig:ET_delta/CE_MOD2_all_cosmo}, the results are displayed as in Subsection~\ref{subsec:cpl_gp_mod1}: the corner plot in the upper part contains the results with GW data alone, while the lower part is dedicated to the joint analysis with Pantheon$\Plus$. As already stated in Subsection~\ref{subsec:3G}, the \lcdm and PEDE fits are comparable in terms of precision, but the \lcdm fit clearly presents a bias in the inferred cosmological parameters. In fact, the injected values of \Hnot and \Omegam lie outside the 95\% C.L. contours of $\Lambda$CDM. In the case of the ET$\Delta$+\panp joint analysis (left panel of Figure~\ref{fig:ET_delta/CE_MOD2_all_cosmo}) we report a relative uncertainty on \Hnot of $1.0\%$ ($1.6\%$) for GP (CPL). For \Omegam the uncertainty of the parametric and non-parametric fits are comparable: $40\%$ ($42\%$) for GP (CPL) (see Figure~\ref{fig:whisker_34}). The FoM is $\tilde{\epsilon}_{\mathrm{PP}+\mathrm{ET}\Delta}=3.9\%$ in the CPL case, while it is $\tilde{\epsilon}_{\mathrm{PP}+\mathrm{ET}\Delta}=3.1\%$ in the case of GP. These results correspond to a GP reconstruction that is a factor 1.3 better than CPL. The same trend is present also for the ET$\Delta$ configuration alone (left panel of Figure~\ref{fig:ET_delta/CE_MOD2_all_cosmo}, upper corner). In the right panel of 
Figure~\ref{fig:ET_delta/CE_MOD2_all_cosmo} we instead show the result of the ET$\Delta$+CE configuration. We again confirm the enhanced constraining power of this configuration with respect to the ET$\Delta$ one. The relative uncertainty on \Hnot is $0.9\%$ ($1.4\%$) in the GP (CPL) case. The reconstructed values of \Omegam are again comparable for the two reconstructions, giving an uncertainty of $37\%$ ($43\%$) for GP (CPL). This gives us a FoM of $\tilde{\epsilon}_{\mathrm{PP}+\mathrm{ET}\Delta \mathrm{+ CE}}=3.8\%$ for CPL and of $\tilde{\epsilon}_{\mathrm{PP}+\mathrm{ET}\Delta \mathrm{+ CE}}=2.8\%$. In other words, GP again outperforms CPL by a factor 1.3. The full set of results is available in Appendix~\ref{ap:tables} (Tables~\ref{tab:mod2_cpl}, ~\ref{tab:mod2_gp}, and Figures~\ref{fig:whisker_PC_complete},~\ref{fig:whisker_PGP_complete}). For completeness, in Figure~\ref{fig:GP_confront_MOD2} we present constraints obtained with the GP fit over MOD2 for ET$\Delta$ and ET$\Delta$+CE, both with and without the \panp data set.
Note that we require the GP samples to satisfy $\fde(z=0) \equiv 1$ by definition. This condition, combined with the relatively large correlation lengths $\log \ell_f \sim 2$ favored by the data, leads to a narrowing of uncertainty around $z = 0$. At the same time, it allows for relatively large deviations from the mean function, with $\log \sigma_f \sim 1$, as long as these deviations occur outside the range probed by the data. This behavior is evident in the posteriors shown in Figure~\ref{fig:GP_confront_MOD2}.

\begin{figure}
    \centering
    \includegraphics[width=\textwidth]{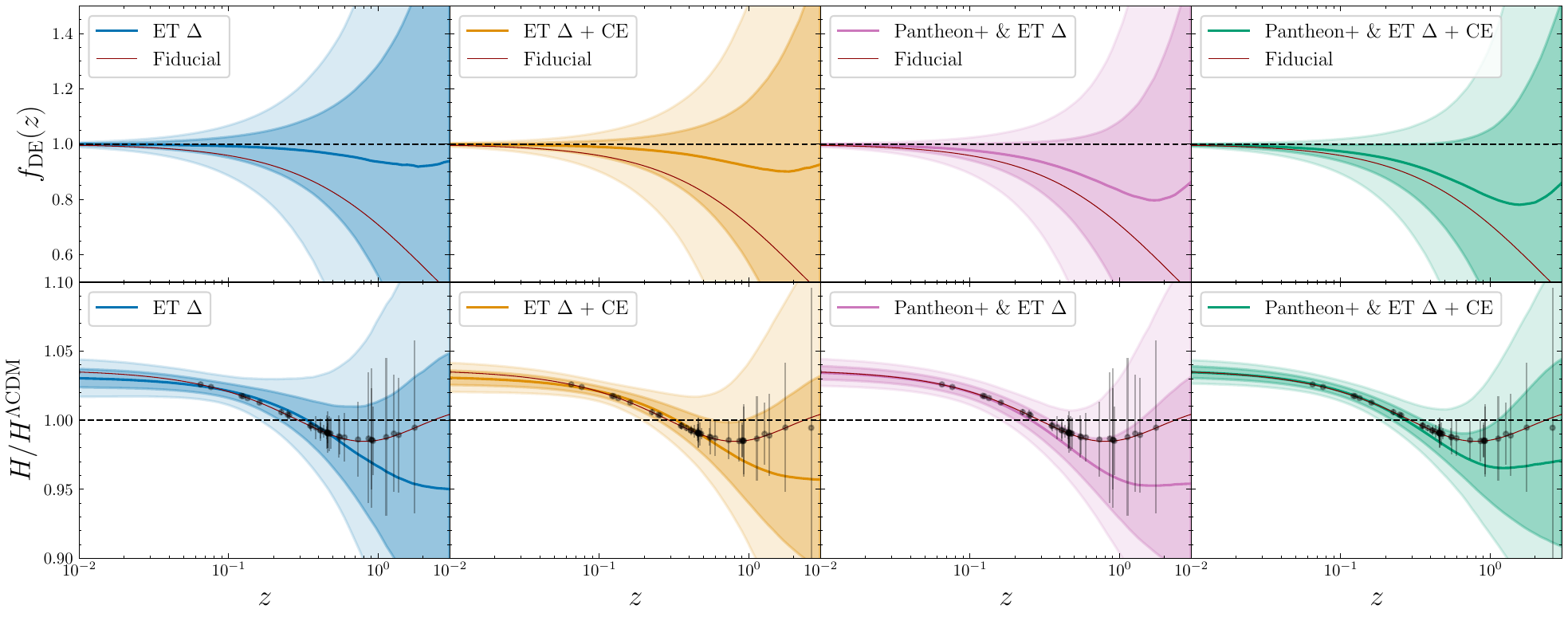}

    \caption{Marginalized posterior distributions on $\fde(z)$ and $H(z)/H^{\Lambda\textrm{CDM}}$ obtained with GP for an underlying MOD2 Universe.
    The dashed lines show the \lcdm behaviour of $\fde(z)$ and $H(z)/H^{\Lambda\textrm{CDM}}$. The colored solid lines show the mean GP reconstruction of those functions, while the shaded regions show the $1\sigma$ and $2\sigma$ uncertainty bands. The red solid line reports the fiducial cosmology with which the data are generated. The data themselves are reported in black in the $H(z)/H^{\Lambda\textrm{CDM}}$ panel. 
    From left to right we display the ET$\Delta$ (+CE) configuration and the same but including the \panp mock data set (last two plots). These results are relative to the fiducial GRB catalog.}
    \label{fig:PEDE-reconstructions-delta}
\end{figure}

In Figure~\ref{fig:PEDE-reconstructions-delta}, we present the reconstructed DE density $\fde(z)$ and corresponding expansion history $H(z)$ normalized to the \lcdm one. The dashed line refers to the \lcdm model evolution (MOD1, \fde = 1), while the red solid line reflects the underlying cosmology of the mock data (MOD2). 
The configurations shown are increasingly more constraining from left to right. In the case of \panp and ET$\Delta$+CE joint analysis, we see that the reconstructed cosmology is in perfect agreement with the one used for generating the mock data. In this case, the \fde = 1 line lies on the edge of the 1-$\sigma$ band. In the lower panel, we note a deviation from the reference $H(z)$ behaviour set by MOD1 in the $0.01<z<0.2$ region. However, this is expected because it is driven by the difference in the injected parameters (\Hnot and $\Omega_\mathrm{m}$) between MOD1 and MOD2.

The limitations to a more refined reconstruction are primarily given by having only 38 GRB-GW data points, with a sparse representation at high redshifts, as well as by MOD2 being very similar to \lcdm in terms of free parameters and phenomenology. In other words -- although not statistically significant -- these constraints are quite promising, as demonstrated by their tightening when the fiducial GRB catalog is replaced by the extended one (see Appendix~\ref{ap:tables}). It is important to note that the \fde reconstruction typically aligns with the \lcdm value at $z\gtrsim 2.5$. This behavior is intrinsic to the pipeline, which assumes that our Universe should resemble an Einstein-de Sitter model during the matter-dominated era. Indeed, not only do the error bars increase in size, but also the DE contribution becomes negligible at those redshifts where the Universe is matter dominated.

As a final remark, we note that the observational set-up considered in this context is conservative, as we are using a much smaller sample of EM counterparts compared to what will be available in the era of 3G detectors. A large number of KNe, from ten to several hundreds, are expected to be detected in association with GW signals~\cite{Loffredo:2024}. 
Cosmological measurements from KN detection can be considered complementary to those of GRBs. Although the number of detections per year is higher, the redshifts achieved are limited to the relatively close Universe ($z \lesssim 0.3$).
Furthermore, unlike the present work on GRBs, the number of KN detections has large uncertainties as it is based on population synthesis data sets and poor constraints on local BNS merger rates. 
Another aspect to take into account is that our realistic scenario is based on the number of GRB detections by current detectors. Innovative projects are expected to be operative at the time of next generation GW detectors, for example cube-satellite constellations such as HERMES which will detect hundreds short GRBs per year~\cite{Ghirlanda:2024ayb}, and observatories such as THESEUS. The latter will detect fewer events compared to HERMES (order of few tens short GRBs per year) but with an unprecedented spectral coverage, from NIR to MeV, and sky-localization accuracy, from the arcmin to arcsec level, allowing to drive the ground-based follow-up to obtain the GRB redshifts~\cite{THESEUS:2021uox,Ciolfi:2021gzg,Ronchini:2022gwk}. Wide field-of-view (FoV) X-ray telescopes such as Einstein Probe~\citep{EP2018} and THESEUS-SXI~\cite{THESEUS:2021uox}, are expected to detect a non-negligible fraction of BNS mergers observed off-axis, whose relativistic jets are not pointing to us to have a detectable $\gamma$-ray emission \citep{Ronchini:2022gwk}. 
Instruments such as the wide FoV Vera Rubin Observatory~\cite{2022arXiv220802781B} will be fully operational allowing efficient follow-up campaigns even for GW-GRB events that are not perfectly localised. Operating together with telescopes such as the Extremely Large Telescope~\cite{Liske2014tmt, Marconi2022SPIE} or project such as the Wide Field Spectroscopic Telescope~\cite{WST:2024zvm} will enhance host galaxy identification and redshift measurements.

In addition, we are combining and confronting GW data with currently operating SNIa and BAO surveys, whose performances will soon be surpassed by a large extent from new generation surveys such as DESI, Euclid and Nancy Grace Roman Space Telescope. We will focus on this topic in the following Subsection.

\subsection{Future prospects}\label{subsec:future_prospects}

We now discuss cosmological constraints obtained by jointly analyzing the MM data sets presented in this work with mock data from next-generation BAO surveys such as DESI(Y5). As discussed in Subsection \ref{subsec:cpl_gp_mod2}, integrating next-generation cosmological surveys with data from 3G detectors will yield even more precise insights into the nature of DE, possibly identifying potential deviations from the standard \lcdm model. It is important to note that even the results presented in the present Subsection remain somewhat conservative, as we have not yet accounted for expected improvements in detecting GRBs and other EM counterparts providing the redshift of the sources. Such advancements will significantly enhance the precision of our constraints, meaning that our current findings may still be underestimating the true potential of future observations.

\begin{figure}[h]
    \centering
    \includegraphics[width=\textwidth]{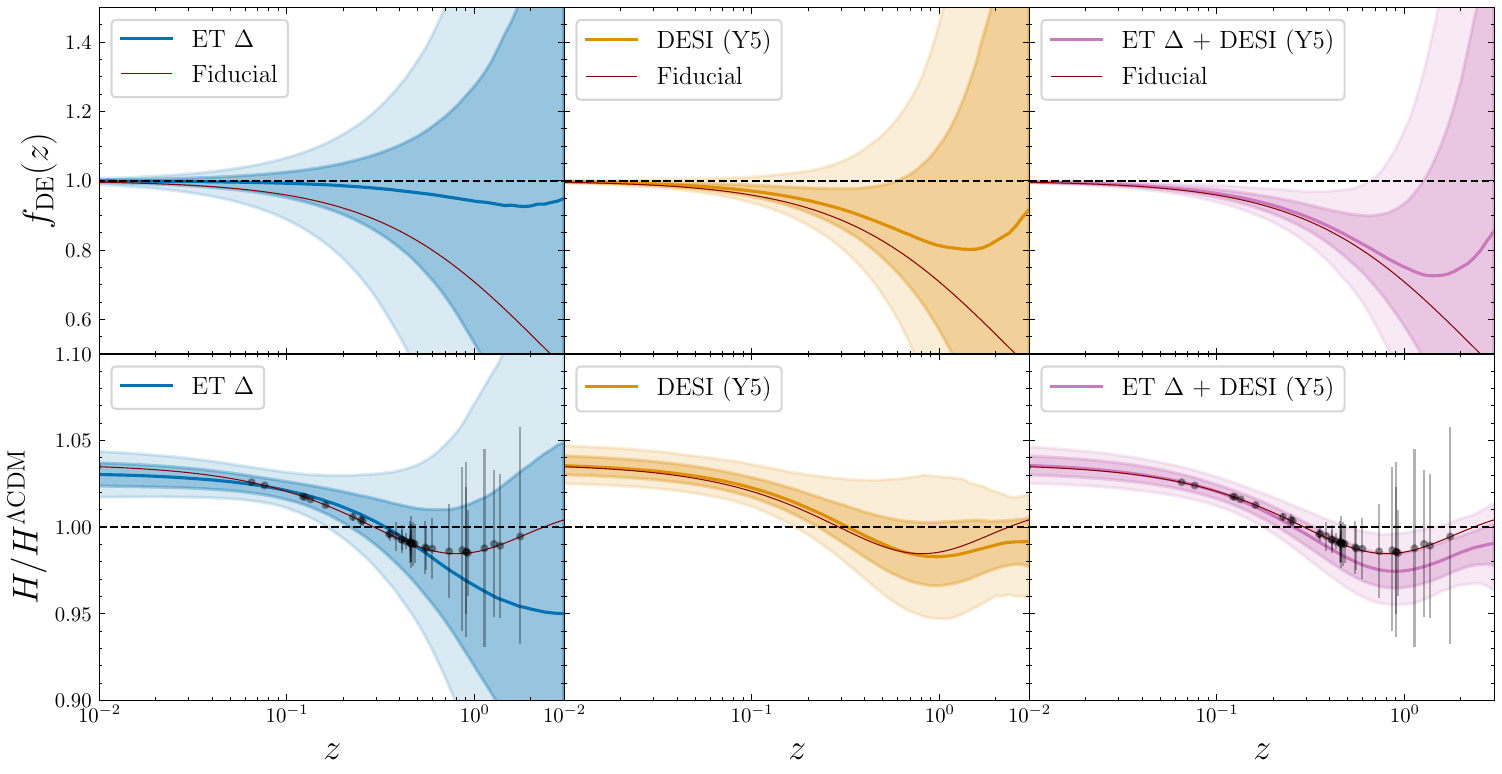}
    \caption{Marginalized posterior distributions on $\fde(z)$ and $H(z)/H^{\Lambda\textrm{CDM}}$ obtained with GP for an underlying MOD2 Universe.
    The dashed lines show the \lcdm behaviour of $\fde(z)$ and $H(z)/H^{\Lambda\textrm{CDM}}$. The colored solid lines show the mean GP reconstruction of those functions, while the shaded regions show the $1\sigma$ and $2\sigma$ uncertainty bands. The GW data themselves are reported in black in the $H(z)/H^{\Lambda\textrm{CDM}}$ panel.
    From left to right we display the ET$\Delta$, DESI (Y5) and the combination of the two. These results are relative to the fiducial GRB catalog.}
    \label{fig:StageIV-GP-recos}
\end{figure}

In Figure~\ref{fig:StageIV-GP-recos} we compare the reconstruction capabilities of $f_\mathrm{DE}(z)$ and $H(z)$ for ET$\Delta$ and DESI(Y5), both alone and in combination (details on the data generation are provided in Section~\ref{sec:methods}). As in the previous Subsections, we show in the upper panels the reconstructed \fde, while the corresponding $H(z)/H^{\Lambda\textrm{CDM}}$ is displayed in the lower panels, normalized to the MOD1 expansion history. The underlying PEDE cosmology injected to generate the data (MOD2) is shown as a red solid line. In the left panel we show the constraints provided by ET$\Delta$ alone. In the middle panel we highlight the constraints provided by DESI(Y5), focusing solely on BAO measurements. In the right panel we show the results of the joint analysis of the two. The latter analysis explicitly shows that the \fde GP reconstruction can distinguish deviations from \lcdm cosmology with precision greater than $2\sigma$, within the $0.06 \lesssim z \lesssim 0.7$ range. As already stated, the deviation from \lcdm at low redshift is driven by the difference in the injected parameters (\Hnot and $\Omega_\mathrm{m}$) between MOD1 and MOD2. It is interesting to notice that neither ET$\Delta$ nor DESI(Y5) alone can distinguish the dynamical DE evolution in the intermediate region ($0.06 \lesssim z \lesssim 0.7$ range). On the other hand, focusing on the $H(z)$ estimates, one can clearly notice that the combination of ET$\Delta$+DESI(Y5) is capable of distinguishing the \lcdm departure at more than $2\sigma$ level in the range $0.6 \lesssim z \lesssim 1$. As a result, the \fde reconstruction is also very precise, clearly demonstrating the crucial role that will be played in the future by multi-probe model-independent approaches to disentangle different DE phenomenologies, even when almost fully degenerate one to another.

\section{Summary and conclusions}\label{sec:summary}

In this work we presented cosmological constraints obtained by employing GRB-GW MM data. We constructed a catalog of GRB data observed in the past two decades by \textit{Fermi}-GBM and \textit{Swift}-BAT/XRT/UVOT. The use of real EM data prevents uncertainties associated with BNS populations generated by population synthesis codes. We assumed that all GRB in our catalogs are merger-driven events, thus associated to a GW counterpart. 
The latter is analyzed in the context of the next 2G and 3G GW detectors. We carried out an extensive analysis both alone, and in combination with currently available BAO and uncalibrated SNIa data, as well as next-generation simulated ones.

To accomplish this, we devised a new numerical tool that is able to output the posterior distribution of the MM events luminosity distance taking into account the correction for the host galaxies' peculiar velocity. The GW parameter estimation is made possible by the use of \texttt{GWFish}, a GW parameter estimation tool based on a recently introduced prior-informed Fisher matrix approach~\cite{Dupletsa:2024gfl}. The latest version of \texttt{GWFish} guarantees a more precise analysis with respect to standard Fisher matrix analyses, as it includes the physical priors of the GW parameters. The posteriors obtained through this procedure are then fitted using a KDE interpolation. 
As a final step, we developed a likelihood function to analyze the full KDE posterior generated by our pipeline, in order to extract cosmological constraints by means of a comprehensive set of MCMC analyses.

In the following, we outline our main results:

\begin{itemize}
    \item we showed that even the more advanced configuration of 2G GW detectors (LVK(O5) and LVKI A$^{\#}$) will not be able to provide competitive cosmological constraints with BNS data alone. However, the joint analysis with the \panp data set provides comparable precision compared to that of current BAO data and calibrated SNIa (SH$_0$ES). A good metric for this comparison is the \lcdm fit over MOD1, which has a FoM approximately $\tilde{\epsilon} \sim 3\%$. 3G GW detectors provide instead outstanding constraints compared to 2G detectors. We recap the uncertainties on the cosmological parameters and the relative FoM for the \lcdm fit over MOD1 of LVK+\panp (8 GW events), ET$\Delta$ and  ET$\Delta$+\panp (38 GW events), in this order: for $\Delta H_{\mathrm{0}}/ H_{\mathrm{0}}$ we have $4.5\%$, $1.3\%$ and  $0.9\%$; for $\Delta \Omega_{\mathrm{m}}/ \Omega_{\mathrm{m}}$ we get $11\%$, $22\%$ and  $10\%$; the relative FoM are $3.6\%$, $2.1\%$ and  $1.4\%$;
    
    \item we compared different 3G detector configurations and identified three key findings: $(i)$ the ET non-cryogenic configuration, dubbed HF, provides cosmological constraints very similar to the full cryogenic configuration; $(ii)$ the constraining capabilities of the ET2L are better than those coming from ET$\Delta$; $(iii)$ adding CE to the network further improves the constraining power mostly due to the better overall sensitivity;

    \item we explicitly showed that fitting the data with an incorrect cosmological model could lead to misinterpretations of the underlying physics.  This biased reconstruction can occur even when choosing a conservative alternative cosmology, as demonstrated in our analysis of the PEDE scenario (MOD2). The latter model has a very similar phenomenology with respect to the \lcdm best fit to \emph{Planck} 2018 data. Despite this, the fit of \lcdm over MOD2 leads to an incorrect reconstruction. When the analysis is carried out with ET$\Delta$+CE, it reveals a statistically significant bias ($> 2\sigma$) even in the absence of \panp data;

    \item we assessed the constraining capabilities of non-parametric methods (GP) against standard parametric ones (CPL). Both CPL and GP always recover the injected cosmology. However, GP systematically provides better constraints with respect to CPL. In the case of MOD2, the constraints of GP are from 1.2 to 1.5 times tighter than CPL. For the ET$\Delta$+CE+\panp configuration, the GP (CPL) provides the following constraints: for $\Delta H_{\mathrm{0}}/ H_{\mathrm{0}}$ we have $0.88\%$ ($1.4\%$); for $\Delta \Omega_{\mathrm{m}}/ \Omega_{\mathrm{m}}$ we get $37\%$ ($43\%$); the relative FoM is $2.8\%$ ($3.8\%$). Our model-independent approach is significantly more robust and less prone to biases in reconstructing the underlying cosmology;
    
   \item we explored the potential for joint analysis of next-generation GW data sets and future cosmological probes. By the time ET becomes operational, more precise data on \Omegam will be available from sources such as SNIa observations from the Vera Rubin Observatory, Nancy Grace Roman Space Telescope, and BAO measurements from DESI (Y5)~\cite{Mitra_2021, Mitra:2022ykq, Wang:2021oec, Mitra:2024ahj}. In particular, our analysis with DESI(Y5) indicates that the \fde GP reconstruction can identify deviations from \lcdm cosmology with a precision greater than $2\sigma$ in the redshift range $0.06 \lesssim z \lesssim 0.7$;

    \item we emphasized that obtaining significant insights into cosmology using bright sirens necessitates future generations of GW interferometers to be complemented by high-energy satellites capable of precise sky localization, similar to the combined efforts of \textit{Fermi} and \textit{Swift}. Additionally, a robust follow-up effort to ensure the redshift determination of events is crucial.
    As demonstrated by the improvements seen with the extended GRB catalog, even a few high-redshift data points can significantly enhance the constraining power. In the case of GP, the use of the extended catalog results in a 7.7$\%$ increase in precision for the ET$\Delta$+CE + \panp case.

\end{itemize}
All our results are obtained under the most conservative assumptions. With just 38 GRBs with known redshift -- that could be achieved in 4-5 years by an instrument such as THESEUS -- our method of analysis will enable to pinpoint at the percent level deviations from the \lcdm expansion history. 
Furthermore, the methods and tools presented in this paper are highly flexible, so that they can easily be adapted to incorporate data catalogs from many other tracers of the distance-redshift relation.
The devised software is publicly available on GitHub~\gitlink{https://github.com/AndreaCozzumbo/bright_cosmology}, together with the data catalogs used in this paper. The repository will be regularly updated with new catalogs from different $H(z)$ tracers, such as BBHs cross-correlated with Large Scale Structure (LSS) data\footnote{Note that the likelihood in Eq.~\eqref{eq:lkl_mpt}, while perfectly suitable to fit mock GW data as in this work, might lead to biased results when applied to real GW data (due \textit{e.g.} to selection biases)~\cite{Essick:2023upv}. Designing a likelihood encompassing all these effects is beyond the scope of this paper.}.

\section*{Acknowledgements and data availability}
We are very thankful to Biswajit Banerjee, Jan Harms, Samanta Macera, Simone Mastrogiovanni, Alessio Mei, Costantino Pacilio for useful discussions.
We would also like to acknowledge all the participants to the \href{https://indico.gssi.it/event/606/}{ModIC 2024 workshop} for the fruitful interdisciplinary exchange of ideas. In particular, we thank Eric V. Linder for helpful input on the treatment of peculiar velocities. We are thankful to the INFN Section, the University, and the Observatory of Cagliari (Sardinia) for their hospitality during the
development of this project. M.B. acknowledges financial support from the Italian Ministry of University and Research (MUR) for the PRIN grant METE under contract no. 2020KB33TP. R.C. is funded by the Czech Ministry of Education, Youth and Sports (MEYS) and European Structural and Investment Funds (ESIF) under project number CZ.02.01.01/00/22\_008/0004632. We also acknowledge the Director and the Computing and Network Service of Laboratori Nazionali del Gran Sasso (LNGS-INFN), and the high-performance computing cluster Seondeok at the Korea Astronomy and Space Science Institute.
We acknowledge the use of {\tt Matplotlib}~\cite{Hunter:2007}, {\tt NumPy}~\cite{harris2020array}, {\tt SciPy}~\cite{Virtanen2020}, {\tt{AstroPy}}~\cite{Astropy2022}. The results of our analyses are fully reproducible by making use of publicly available catalogs of GRB archival data from \href{https://bright.ciera.northwestern.edu/welcome/}{BRIGHT}, \href{https://www.mpe.mpg.de/~jcg/grbgen.html}{MPE GRB catalog}, \href{https://user-web.icecube.wisc.edu/~grbweb_public/Summary_table.html}{GRBweb}. The code needed to reproduce our results is available on GitHub~\gitlink{https://github.com/AndreaCozzumbo/bright_cosmology}.

\appendix

\section{Full catalog of merger-driven GRBs with known redshift (circa 2024)}\label{ap:grbt_comparison} 

This Appendix is dedicated to a comparison of the results obtained considering different GRB catalogs classified according to their $P_{cc}$. The full catalog of GRBs considered in this work is presented in Table~\ref{tab:GRBall}.
As illustrated in the main text, we follow Ref.~\cite{Fong:2022mkv} and we provide three catalogs: the \textit{fiducial} catalog (containing only those events with $P_{cc} \leq 0.02$), the \textit{extended} catalog (events with $P_{cc} \leq 0.10$), and the \textit{very extended} catalog ($P_{cc} \leq 0.20$). 

\begin{figure}[h]
    \centering
    \includegraphics[width=0.7\linewidth]{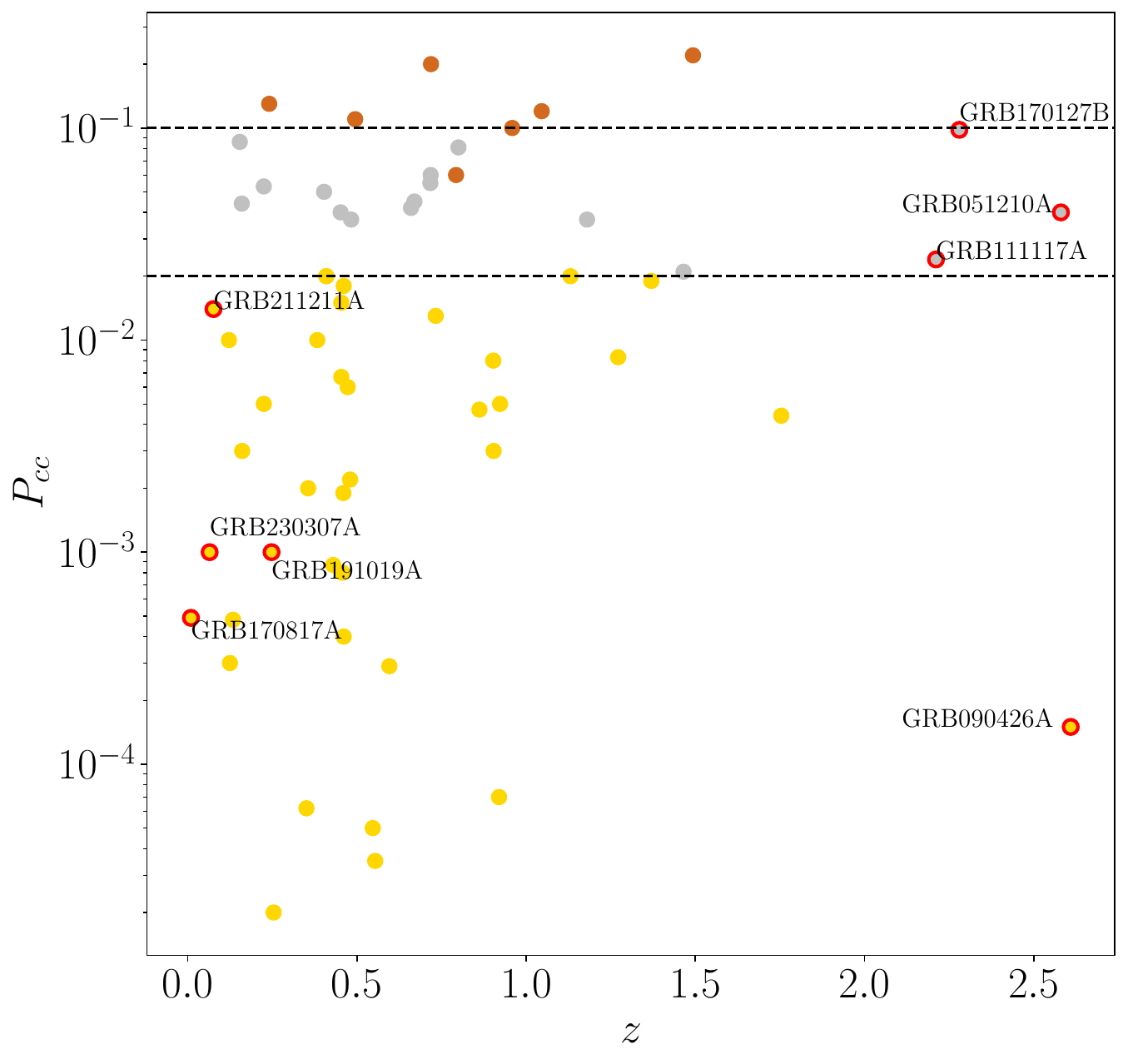}

    \caption{The probability of chance coincidence distribution (see main text) as a function of redshift. GRB datapoints are organized by color following \cite{Fong:2022mkv} categorization. The points under the lower dashed line represent our fiducial 38 events catalog. The ones residing under the upper dashed line compose the extended 54 events catalog. All together they compose the very extended 61 event catalog. In the left part we have circled some of the most important GRBs like the three associated to a KN event and the SNless long GRB. In the right part we can find the furthest GRB event of our catalog at $z=2.609$ and three GRB that, even if associated with a small $P_{cc}$ value, can be crucial in the high-$z$ region to constrain the expansion history.
    }
    \label{fig:olympics_pic}
\end{figure}

Concerning the GRB emission, its prompt $\gamma$/hard X -ray signal is emitted on-axis, beamed with an angle $| \theta | \lesssim 10^\circ$~\cite{Ronchini:2022gwk, Banerjee:2022gkv}, and followed by an optical/soft X-ray afterglow off-axis emission. The observational strategy to attribute a redshift to a GRB source is the following: the GRB event is firstly detected by a wide-field high-energy telescope, such as \textit{Fermi}-GBM or \textit{Swift}-BAT. The reconstructed sky localization ($\approx$ arcmin) is then passed to a focusing instrument like \textit{Swift}-XRT/UVOT, that constrain the GRB position with $\approx$ arcsec precision, following the GRB afterglow emission~\cite{Abbott2017Multi}. The resulting sky position is finally passed to ground-based deep-field optical surveys, in order to identify the GRB host galaxy and measure its redshift, with a given uncertainty. The redshift of GRBs can also be determined using optical or NIR
spectroscopy of their afterglow emission.
In the local Universe ($z \lesssim 0.6$) the off-axis GRB prompt emission~\cite{Ascenzi:2020kxz} will provide the main contribution to the observed events, as clearly shown in \cite{Branchesi:2023mws}. On the contrary, the off-axis emission of GRBs at higher redshift is too faint to be detected. In general, not knowing a priori the true value of the GRB inclination angle is a crucial source of uncertainty, due to its intrinsic degeneracy with the luminosity distance $d_L$~\cite{Chassande-Mottin:2019nnz, Usman:2018imj}, and consequently with \Hnot. Therefore, the prior choice on the inclination angle can in principle bias the \Hnot estimate.
This problem can be addressed properly as explained in Refs.~\cite{Chen:2023dgw, Mancarella:2024qle, Muller:2024wzl}. Our results, however, are completely unaffected: our pipeline effectively removes this degeneracy, ensuring that any shift from the true \Hnot value is not due to this issue. We always have explicitly checked that the inferred \Hnot is consistent with the injected one. 
An additional source of uncertainty is related to the GRB-host galaxy association, and typically parameterized via the so-called probability of chance coincidence ($P_{cc}$), that encodes the probability of a wrong association~\cite{Bloom:2000pq}. In Figure~\ref{fig:olympics_pic} we display this information as a function of $z$ and categorizing the GRBs in our data set. 

\begin{figure}
    \centering
    \includegraphics[width=0.42\textwidth]{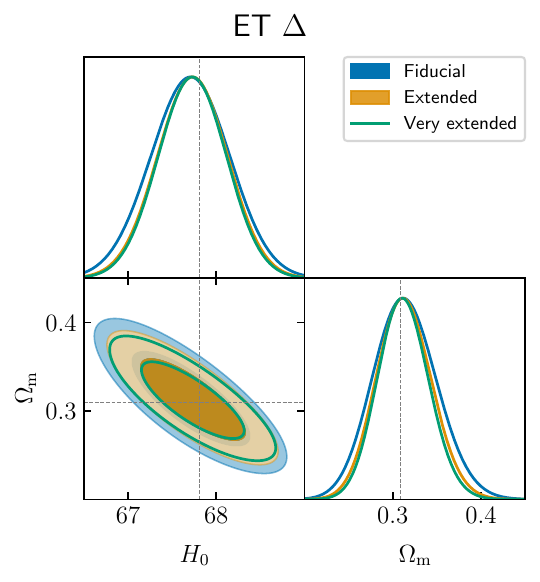}
    \includegraphics[width=0.42\textwidth]{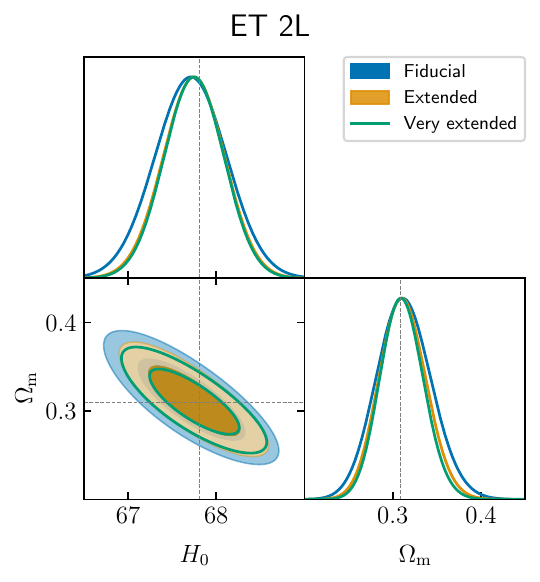}
    
    \caption{\lcdm constraints (68\% and 95\% C.L.) on the relevant cosmological parameters, obtained by fitting the underlying MOD1 cosmology. Comparison of the constraints obtained by the three GRB catalogs with ET$\Delta$ (\textit{left}) and ET2L (\textit{right}).}
    \label{fig:catalog_comp_lcdm}
\end{figure}

\begin{figure}
    \centering
    \includegraphics[width=0.42\textwidth]{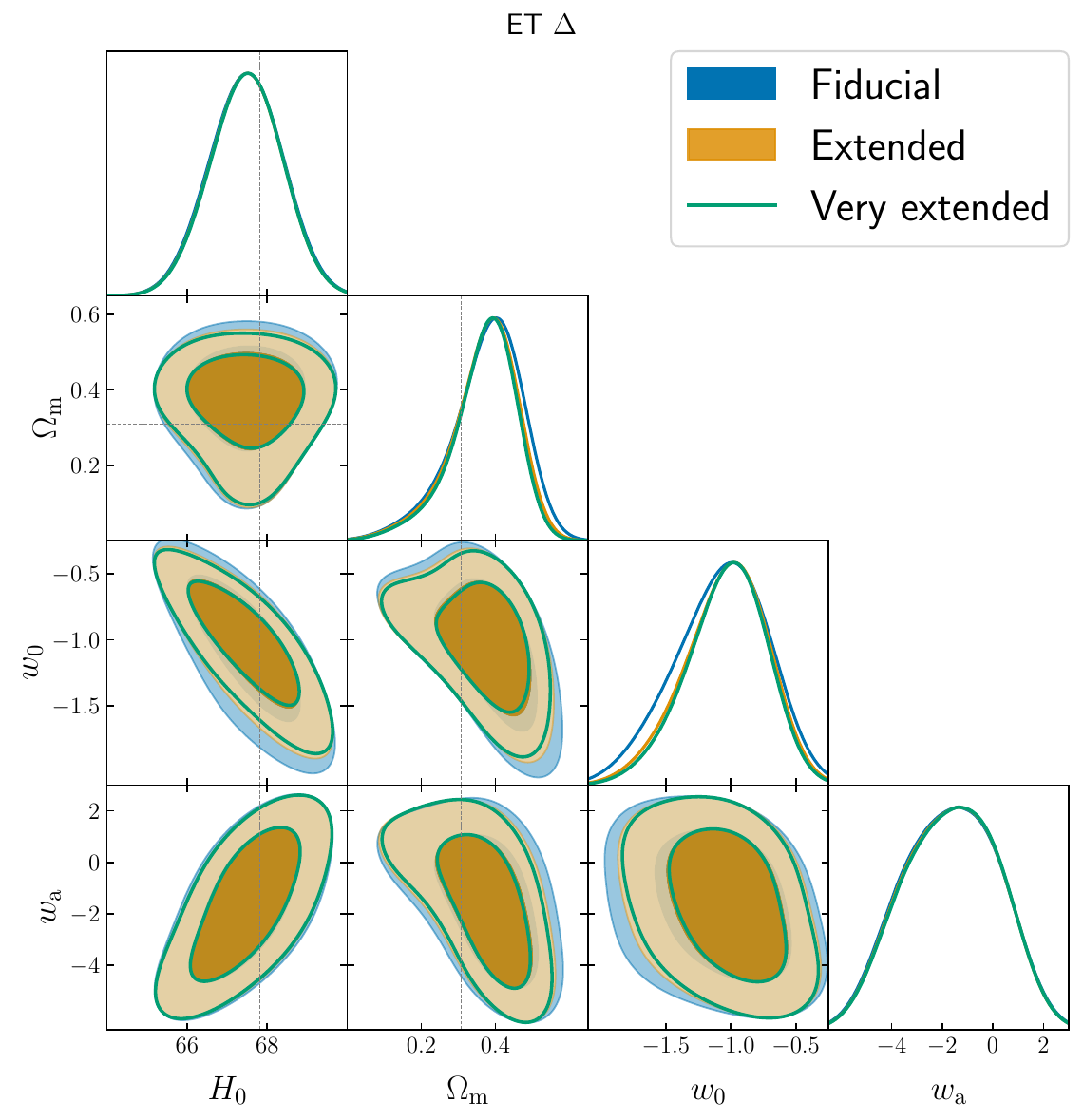}
    \includegraphics[width=0.42\textwidth]{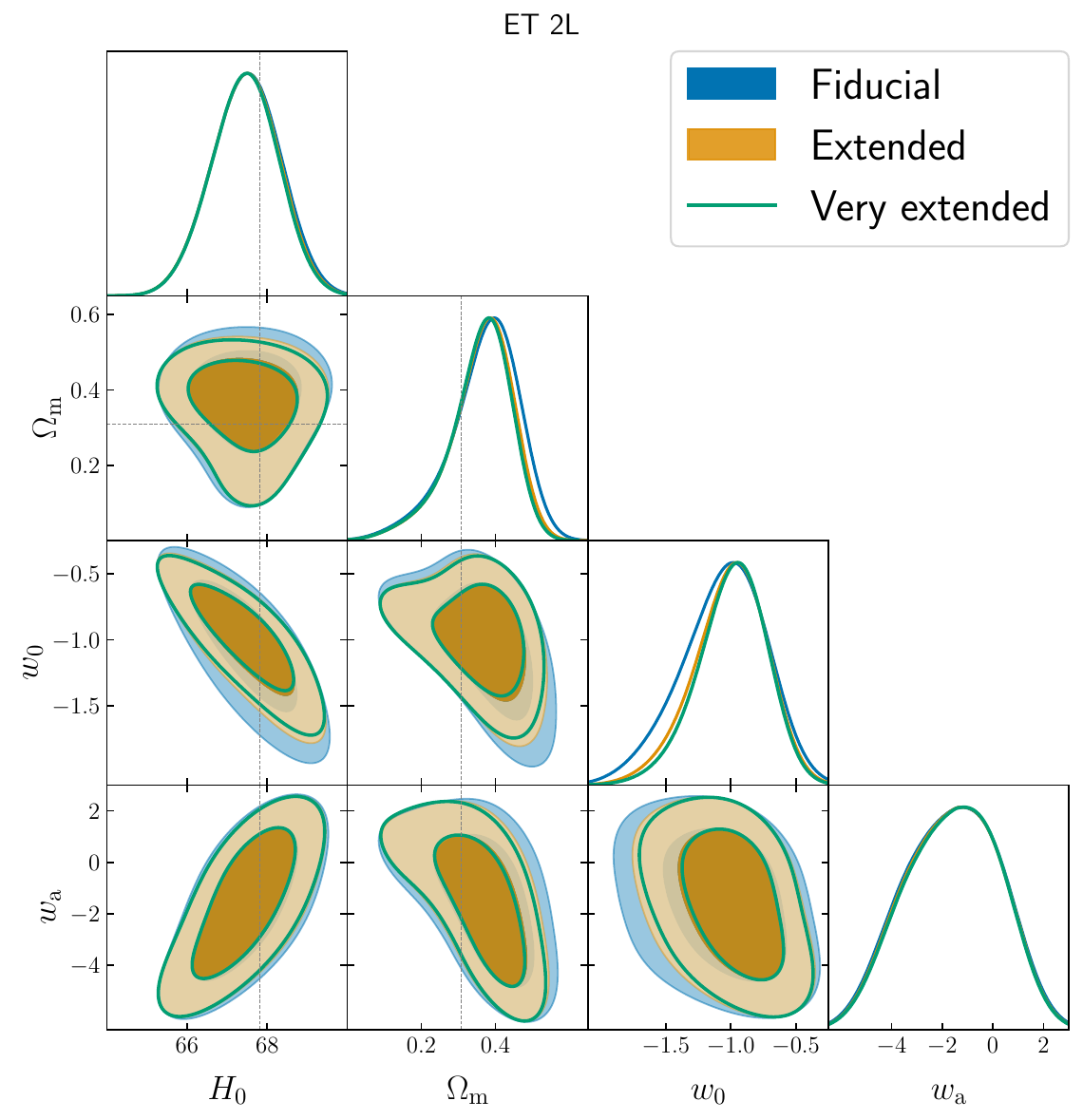}
    
    \caption{CPL constraints (68\% and 95\% C.L.) on the relevant cosmological parameters, obtained by fitting the underlying MOD1 cosmology. Comparison of the constraints obtained by the three GRB catalogs with ET$\Delta$ (\textit{left}) and ET2L (\textit{right}).}
    \label{fig:catalog_comp_cpl}
\end{figure}

\begin{figure}
    \centering
    \includegraphics[width=0.42\textwidth]{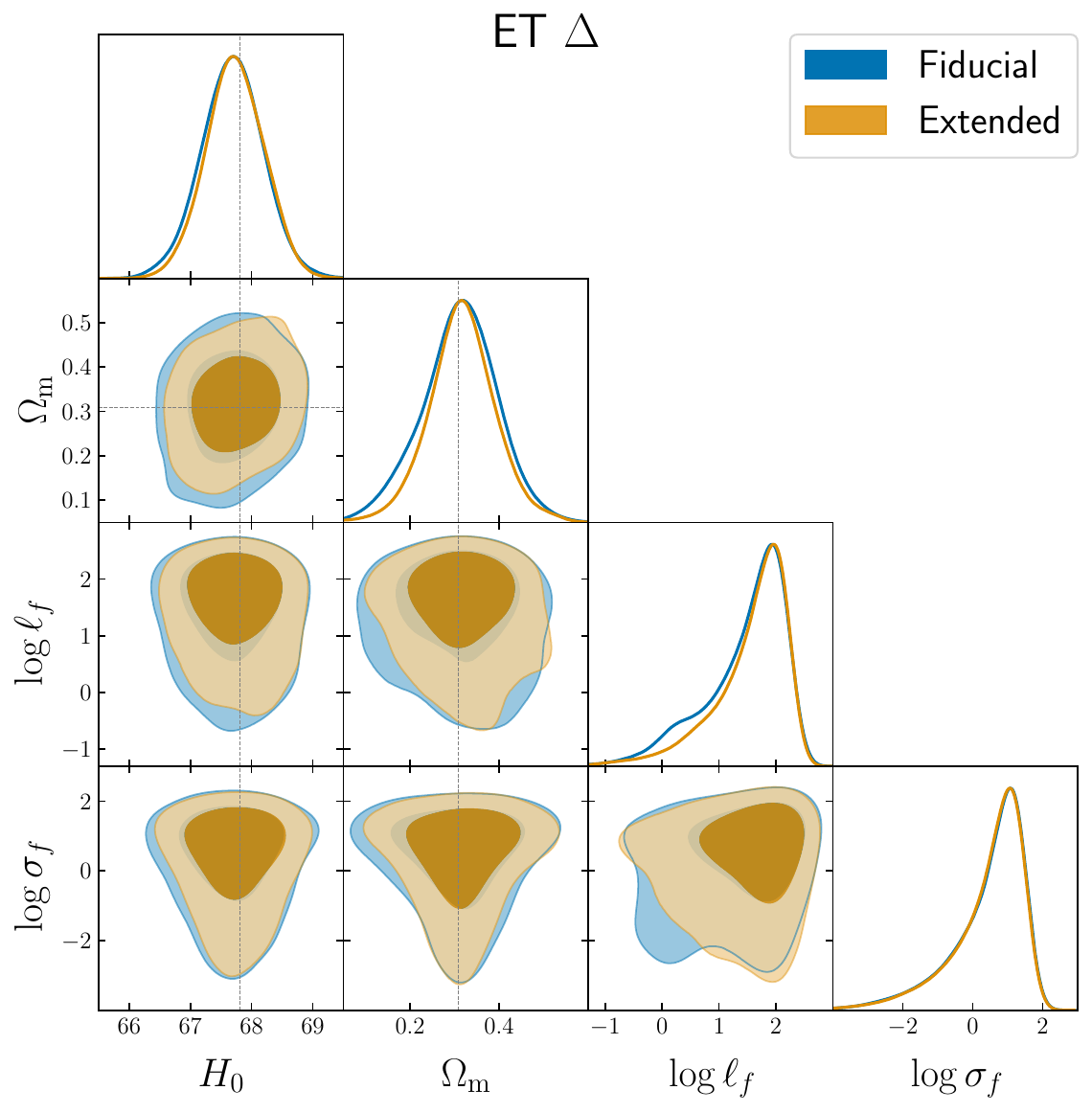}
    \includegraphics[width=0.42\textwidth]{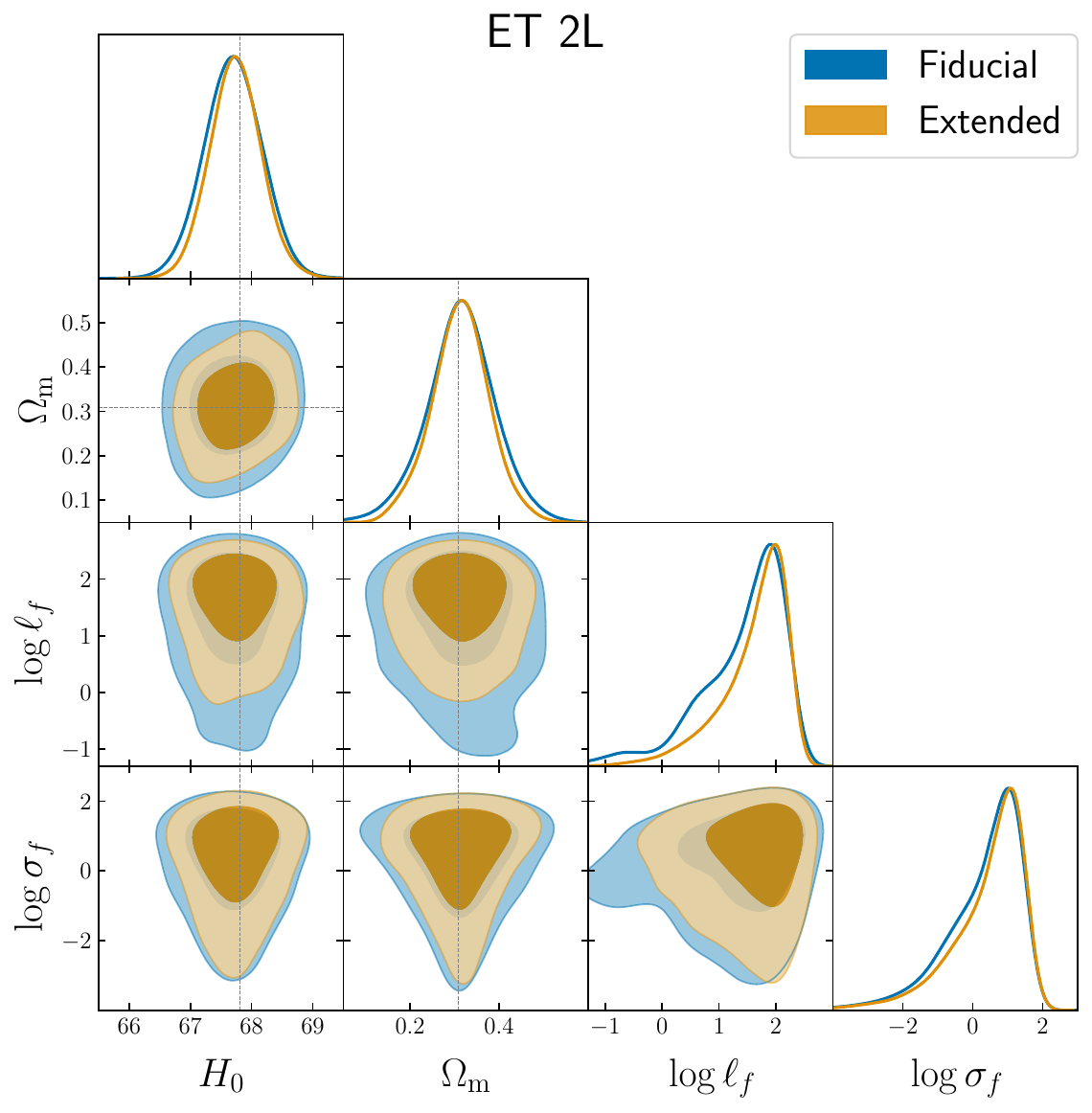}
    
    \caption{GP constraints (68\% and 95\% C.L.) on the relevant cosmological parameters, obtained by fitting the underlying MOD1 cosmology. Comparison of the constraints obtained by the GRB catalogs with ET$\Delta$ (\textit{left}) and ET2L (\textit{right}).}
    \label{fig:catalog_comp_gp}
\end{figure}

In Figure~\ref{fig:catalog_comp_lcdm} we display the comparison of the constraining capabilities of the three different catalogs of mock data for a fixed detector configuration. We show the ET$\Delta$ and the ET2L configurations in the case of a \lcdm fit over MOD1 mock data. The difference in the constraining power is mainly present between the fiducial and the extended catalogs and must be sought in the addition of three GRB events at $z>2$ (Figure \ref{fig:olympics_pic}).
We quote the following FoM for the ET$\Delta$ configuration: $\tilde{\epsilon}_{\mathrm f} =2.1\%$ $\tilde{\epsilon}_{\mathrm e}=1.8\%$ $\tilde{\epsilon}_{\mathrm{ve}}=1.7\%$ for fiducial, extended and very extended, respectively.
More details can be found in Appendix~\ref{ap:tables}, Figure~\ref{fig:whisker_LL_complete} and Table~\ref{tab:mod1_lcdm}.
We can see that the additional information given by more extended catalogs will not drastically change the results. This is a clear sign on how important are the high $z$ GRBs for constraining the expansion history of the Universe. 
We report the same type of analysis in the case of a CPL fit over MOD1 mock data. This is shown in Figure~\ref{fig:catalog_comp_cpl}. We quote the following FoM for the ET$\Delta$ configuration: $\tilde{\epsilon}_{\mathrm f} =6.3\%$ $\tilde{\epsilon}_{\mathrm e}=6.1\%$ $\tilde{\epsilon}_{\mathrm{ve}}=5.9\%$. The details are reported in Appendix~\ref{ap:tables}, Figure~\ref{fig:whisker_LC_complete} and Table~\ref{tab:mod1_cpl}. We can observe that the CPL constraints do not significantly improve from one catalog to another. Finally, Figure~\ref{fig:catalog_comp_gp} shows the GP fit over MOD1 mock data for the fiducial and extended catalogs. We quote the following FoM for the ET$\Delta$ configuration: $\tilde{\epsilon}_{\mathrm f} =4.6\%$ $\tilde{\epsilon}_{\mathrm e}=4.2\%$. We can appreciate a good improvement from the fiducial to the extended catalog and a general shrink of the contours with respect to CPL. The details are reported in Appendix~\ref{ap:tables}, Figure~\ref{fig:whisker_LGP_complete} and Table~\ref{tab:mod1_gp}. 
We suggest the reader to refer to the main text for a more detailed illustration of the analysis implications.

\begin{table}[h]
    \centering \tiny

\caption{Full catalog of 61 GRBs from BNS mergers measured by \textit{Fermi}-GBM and \textit{Swift}-BAT/XRT with an uncertainty $< 7\%$. The event marked with a * was discovered by HETE-2~\cite{2003AIPC..662....3R}.  The GRBs are sorted in chronological order. The first column contains the GRB name. The second column contains the GRB type. $T_0$ represents the UTC time of the event trigger. R.A. and DEC. are the two angles specifying the sky position, namely right ascension and declination, respectively. The $T_{\mathrm{90}}$ column refers to the \textit{Swift}-BAT/\textit{Fermi}-GBM time interval over which 90$\%$ of the gamma-ray emission is measured. The $z$ column contains the redshift of the objects. The catalog association column refers to the three different catalogs used in this paper and it is directly connected to the $P_{cc}$ column, representing the probability of chance coincidence from \cite{Fong:2022mkv}. Events marked with an apex $a$, although having $T_{90} \gtrsim 2 \,\mathrm{s}$, are also classified as sGRB, accordingly to Refs.~\cite{Fong:2022mkv,Nugent:2022paq}. Redshifts marked with an apex $p$ are determined through photometry.\\}

\begin{tabular}{|c|c|c|c|c|c|c|c|c|}
\hline
\bf{GRB name}  & \bf{GRB class}  &  $T_0~[\mathrm{UTC}]$	&   R.A.~$[\mathrm{deg}]$ &     DEC.~$[\mathrm{deg}]$	&	    $T_{\mathrm{90}}~[\mathrm{s}]$    &   $z$   &   Catalog association  & $P_{\mathrm{cc}}$      \\
\hline
GRB050509B  &  sGRB       &     4:00:19    &   189.06  &  28.98     &      0.073   &    0.23    &    fiducial         &      5e-3  \\
GRB050709A  &  sGRB$^*$       &     22:36:37   &   23.03   &  -38.99    &      0.070    &    0.16    &    fiducial         &      3e-3  \\
GRB050724A  &  sGRB+E.E.  &     12:34:09   &   246.18  &  -27.54    &      98   &    0.25    &    fiducial         &      2e-5  \\
GRB050813A  &  sGRB       &     6:45:10    &   241.99  &  11.25     &      0.45    &    0.72    &    very extended    &      2e-1  \\
GRB051210A  &  sGRB$^p$       &     5:46:21    &   330.17  &  -57.61    &      1.3     &    2.6     &    extended         &      4e-2  \\
GRB051221A  &  sGRB       &     1:51:16    &   328.70  &  16.89     &      1.4     &    0.55    &    fiducial         &      5e-5  \\
GRB060614A  &  sGRB+E.E.  &     12:43:49   &   320.88  &  -53.03    &      110   &    0.13    &    fiducial         &      3e-4  \\
GRB060801A  &  sGRB       &     12:16:15   &   213.01  &  16.98     &      0.49    &    1.1    &    fiducial         &      2e-2  \\
GRB061006A  &  sGRB+E.E.  &     16:45:51   &   111.03  &  -79.20    &      130   &    0.46    &    fiducial         &      4e-4  \\
GRB061210A  &  sGRB+E.E.  &     12:20:39   &   144.52   &  15.61      &      85    &    0.41   &    fiducial         &      2e-2  \\
GRB070429B  &  sGRB       &     3:09:04    &   328.02  &  -38.83    &      0.47    &    0.90    &    fiducial         &      3e-3  \\
GRB070714B  &  sGRB+E.E.  &     4:59:29    &   57.84   &  28.30     &      64.0    &    0.92    &    fiducial         &      5e-3  \\
GRB070724A  &  sGRB       &     10:53:50   &   27.81   &  -18.59    &      0.40     &    0.46    &    fiducial         &      8e-4  \\
GRB070809A  &  sGRB       &     19:22:17   &   203.77   &  -22.12     &      1.3     &    0.47    &    fiducial         &      6e-3  \\
GRB071227A  &  sGRB       &     20:13:47   &   58.13   &  -55.98    &      1.8     &    0.38    &    fiducial         &      1e-2  \\
GRB080123A  &  sGRB+E.E.  &     4:21:57    &   338.94  &  -64.90    &      110   &    0.50    &    very extended    &      1.1e-1 \\
GRB080905A  &  sGRB       &     11:58:54   &   287.67  &  -18.88    &      0.96    &    0.12    &    fiducial         &      1e-2  \\
GRB090426A  &  sGRB       &     12:48:47   &   189.08  &  32.99      &      1.2     &    2.6    &    fiducial         &      1.5e-4    \\
GRB090510A  &  sGRB       &     0:22:59    &   333.55  &  -26.58    &      0.96    &    0.90    &    fiducial         &      8e-3  \\
GRB090515A  &  sGRB       &     4:45:09    &   164.15  &  14.44     &      0.036   &    0.40    &    extended         &      5e-2  \\
GRB100117A  &  sGRB       &     21:06:19   &   11.27    &  -1.59     &      0.26   &    0.92     &    fiducial         &      7e-5  \\
GRB100206A  &  sGRB       &     13:30:05   &   47.16   &  13.16      &      0.18   &    0.41     &    fiducial         &      2e-2  \\
GRB100625A  &  sGRB       &     18:32:27   &   15.80   &  -39.09    &      0.24    &    0.45    &    extended         &      4e-2  \\
GRB101219A  &  sGRB       &     2:31:29    &   74.59   &  -2.54     &      0.60     &    0.72    &    extended         &      6e-2  \\
GRB101224A  &  sGRB       &     5:27:13    &   285.92  &  45.71     &      1.7   &    0.45    &    fiducial         &      1.5e-2    \\
GRB111117A  &  sGRB       &     12:13:41   &   12.70   &  23.02     &      0.43   &    2.2    &    extended         &      2.4e-2    \\
GRB120305A  &  sGRB       &     19:37:30   &   47.54   &  28.49     &      0.10     &    0.23    &    extended         &      5.3e-2    \\
GRB121226A  &  sGRB$^p$       &     19:09:43   &   168.64  &  -30.41    &      1.0     &    1.4     &    fiducial         &      1.9e-2    \\
GRB130515A  &  sGRB$^p$       &     1:21:17    &   283.44  &  -54.28    &      0.26   &    0.80      &    extended         &      8.1e-2    \\
GRB130603B  &  sGRB       &     15:49:14   &   172.20  &  17.07     &      0.18    &    0.36    &    fiducial         &      2e-3  \\
GRB130822A  &  sGRB       &     15:54:17   &   27.92   &  -3.21     &      0.04    &    0.15    &    extended         &      8.6e-2    \\
GRB131004A  &  sGRB       &     21:41:03   &   296.11  &  -2.96     &      1.2   &    0.72    &    extended         &      5.5e-2    \\
GRB140129B  &  sGRB       &     12:51:09   &   326.76  &  26.21     &      1.4    &    0.43     &    fiducial         &      8.7e-4    \\
GRB140622A  &  sGRB       &     9:36:04    &   317.17  &  -14.42    &      0.13    &    0.96    &    very extended    &      1e-1   \\
GRB140903A  &  sGRB       &     15:00:30   &   238.01  &  27.60     &      0.30     &    0.35    &    fiducial         &      6.2e-5    \\
GRB140930B  &  sGRB       &     19:41:42   &   6.35    &  24.29     &      0.84    &    1.5    &    extended         &      2.1e-2    \\
GRB141212A  &  sGRB       &     12:14:01   &   39.12   &  18.15      &      0.3     &    0.60    &    fiducial         &      2.9e-4    \\
GRB150101B  &  sGRB       &     15:23:00   &   188.02  &  -10.93     &      0.08    &    0.13     &    fiducial         &      4.8e-4    \\
GRB150120A  &  sGRB$^a$       &     2:57:46    &   10.32   &  33.99     &      3.3     &    0.46     &    fiducial         &      1.9e-3    \\
GRB150728A  &  sGRB       &     12:51:11   &   292.23  &  33.92     &      0.83    &    0.46     &    fiducial         &      1.8e-2    \\
GRB150831A  &  sGRB       &     10:34:09   &   221.02  &  -25.63     &      1.2    &    1.2     &    extended         &      3.7e-2    \\
GRB160624A  &  sGRB       &     11:27:01   &   30.19   &  29.64     &      0.38   &    0.48     &    extended         &      3.7e-2    \\
GRB160821B  &  sGRB       &     22:29:13   &   279.98  &  62.39     &      1.1   &    0.16     &    extended         &      4.4e-2    \\
GRB161001A  &  sGRB$^a$       &     1:05:16    &   71.92     &  -57.26    &      2.6     &    0.67     &    extended         &      4.5e-2    \\
GRB161104A  &  sGRB       &     9:42:26    &   77.89   &  -51.46    &      0.10     &    0.79    &    very extended    &      6e-2  \\
GRB170127B  &  sGRB       &     15:13:28   &   19.98   &  -30.36    &      1.7   &    2.3     &    extended         &      9.8e-2    \\
GRB170428A  &  sGRB       &     9:13:38    &   330.08  &  26.92     &      0.20     &    0.45     &    fiducial         &      6.7e-3    \\
GRB170728A  &  sGRB       &     6:53:28    &   58.89   &  12.18     &      1.3    &    1.5     &    very extended    &      2e-1 \\
GRB170728B  &  sGRB+E.E.  &     23:03:18   &   237.98  &  70.12     &      47.7    &    1.27     &    fiducial         &      8.3e-3    \\
GRB170817A  &  GRB+KN       &     12:41:06   &   13.16   &  -23.38    &      2.6    &    0.010  &    fiducial         &      4.9e-4    \\
GRB180805B  &  sGRB       &     13:02:36   &   25.78   &  -17.49    &      0.96    &    0.66     &    extended         &      4.2e-2    \\
GRB181123B  &  sGRB       &     5:33:03    &   184.37  &  14.60      &      0.26    &    1.75    &    fiducial         &      4.4e-3    \\
GRB191019A  &  SNless lGRB       &     15:12:33   &   340.02  &  -17.33     &      64   &    0.25    &    fiducial         &      --  \\
GRB200219A  &  sGRB$^p$       &     7:36:45    &   342.64  &  -59.12    &      1.2   &    0.48     &    fiducial         &      2.2e-3    \\
GRB200522A  &  sGRB       &     11:41:34   &   5.68    &  -0.28     &      0.62    &    0.55    &    fiducial         &      3.5e-5    \\
GRB201221D  &  sGRB       &     23:06:34   &   171.06   &  42.14     &      0.14   &    1.05    &    very extended    &      1.2e-1 \\
GRB210323A  &  sGRB       &     22:02:14   &   317.95  &  25.37     &      0.96    &    0.73    &    fiducial         &      1.3e-2    \\
GRB210919A  &  sGRB       &     0:28:33    &   80.25   &  1.31      &      0.16    &    0.24    &    very extended    &      1.3e-1 \\
GRB211023B  &  sGRB       &     21:05:52   &   170.31  &  39.14     &      1.3     &    0.86    &    fiducial         &      4.7e-3    \\
GRB211211A  &  GRB+KN       &     13:09:59   &   212.29   &  27.89     &      34   &    0.076   &    fiducial         &      1.4e-2    \\
GRB230307A  &  GRB+KN      &     15:44:05   &   60.82   &  -75.38    &      35   &    0.065    &    fiducial         &      -- \\

\hline
    
\end{tabular}

\label{tab:GRBall}
\end{table}

\newpage

\section{Full comparison among GW detector configurations}\label{ap:det_comparison} 

This Appendix is dedicated to comparing the results obtained from an extensive analysis of all GW detector configurations. In these cases, we focus on the fiducial GRB catalog. The Appendix is structured as follows: we present the results for the \lcdm, CPL and GP fitting models over a MOD1 Universe. For each scenario, we provide a comparison of contours between 2G and 3G detectors (if not already shown in the main text), along with an analysis involving combinations of \panp and GW detectors, as well as various configurations of 3G detectors.

\begin{figure}
    \centering
    \includegraphics[width=0.42\textwidth]{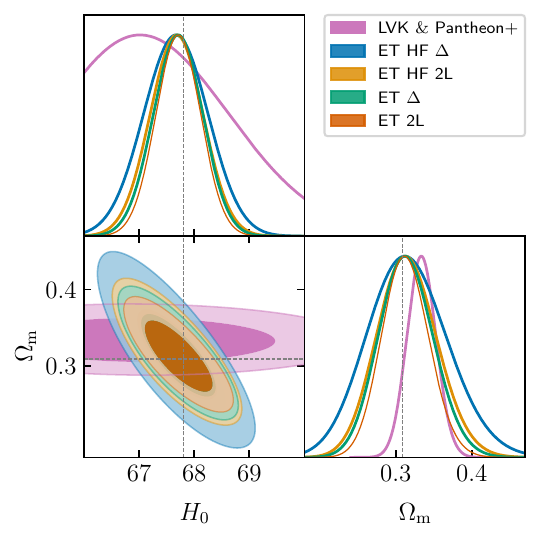}
    \includegraphics[width=0.42\textwidth]{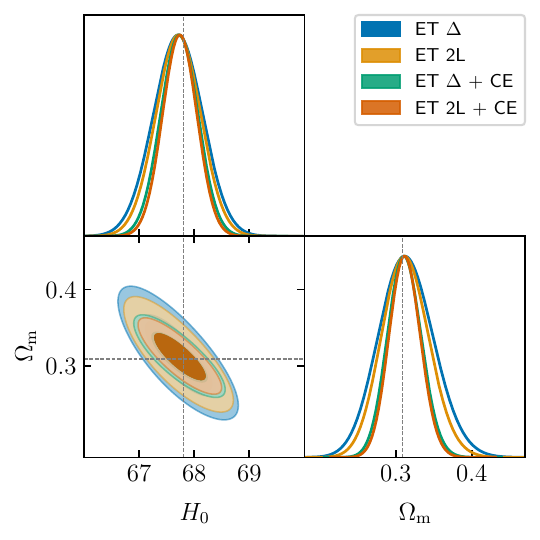}
    \caption{\lcdm constraints (68\% and 95\% C.L.) on the relevant cosmological parameters, obtained by fitting the underlying MOD1 cosmology. (\textit{Left}) Comparison of the cosmological constraints obtained by LVK+Pantheon$\Plus$, the ET HF $\Delta$/2L and ET$\Delta$/2L configurations. (\textit{Right}) The comparison between ET$\Delta$/2L and ET$\Delta$/2L + CE. }
    \label{fig:config_comp_lcdm}
\end{figure}

In Figure \ref{fig:config_comp_lcdm} we compare different configurations for a \lcdm fit over MOD1. A clear improvement with respect to LVK+\panp joint analysis can be perceived. We can observe some general features: the ET2L configuration performs better than the ET$\Delta$ ($\tilde{\epsilon}_{\mathrm{ET}\Delta}=2.1\%$ $\tilde{\epsilon}_{\mathrm{ET2L}}=1.9\%$). The HF configurations present comparable results with respect to their full cryogenic configurations (left panel of Figure \ref{fig:config_comp_lcdm}). In the case of ET$\Delta$(2L) the uncertainty on \Hnot is 1.3$\%$(1.2$\%$) while the ETHF$\Delta$(2L) configuration obtain 1.7$\%$(1.4$\%$). The difference increases in the case of \Omegam constraints: ET$\Delta$(2L) the uncertainty on \Omegam is 22$\%$(19$\%$) while the ETHF$\Delta$(2L) configuration obtain 33$\%$(25$\%$). The FoM are $\tilde{\epsilon}_{\mathrm{ETHF}\Delta}=2.9\%$ $\tilde{\epsilon}_{\mathrm{ETHF 2L}}=2.3\%$. In the right panel of Figure \ref{fig:config_comp_lcdm} we report the comparison with and without the CE detector. The FoM improves as follows $\tilde{\epsilon}_{\mathrm{ET}\Delta \mathrm{+CE} }=1.4\%$ $\tilde{\epsilon}_{\mathrm{ET2L +CE}}=1.3\%$. The full set of results is presented in Appendix~\ref{ap:tables}, Figure~\ref{fig:whisker_LL_complete} and Table~\ref{tab:mod1_lcdm}.

\begin{figure}
    \centering
    \includegraphics[width=0.45\textwidth]{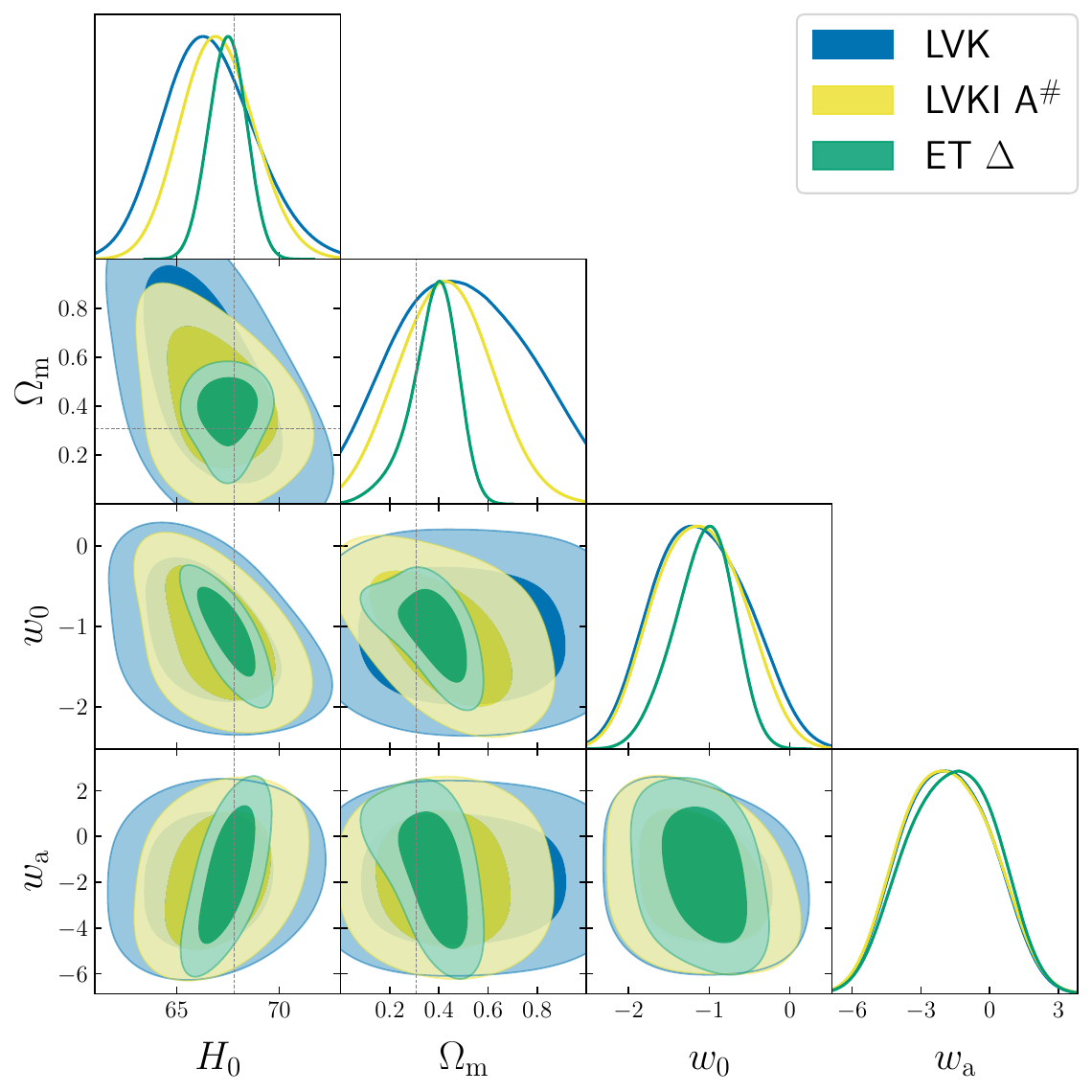}
    \includegraphics[width=0.45\textwidth]{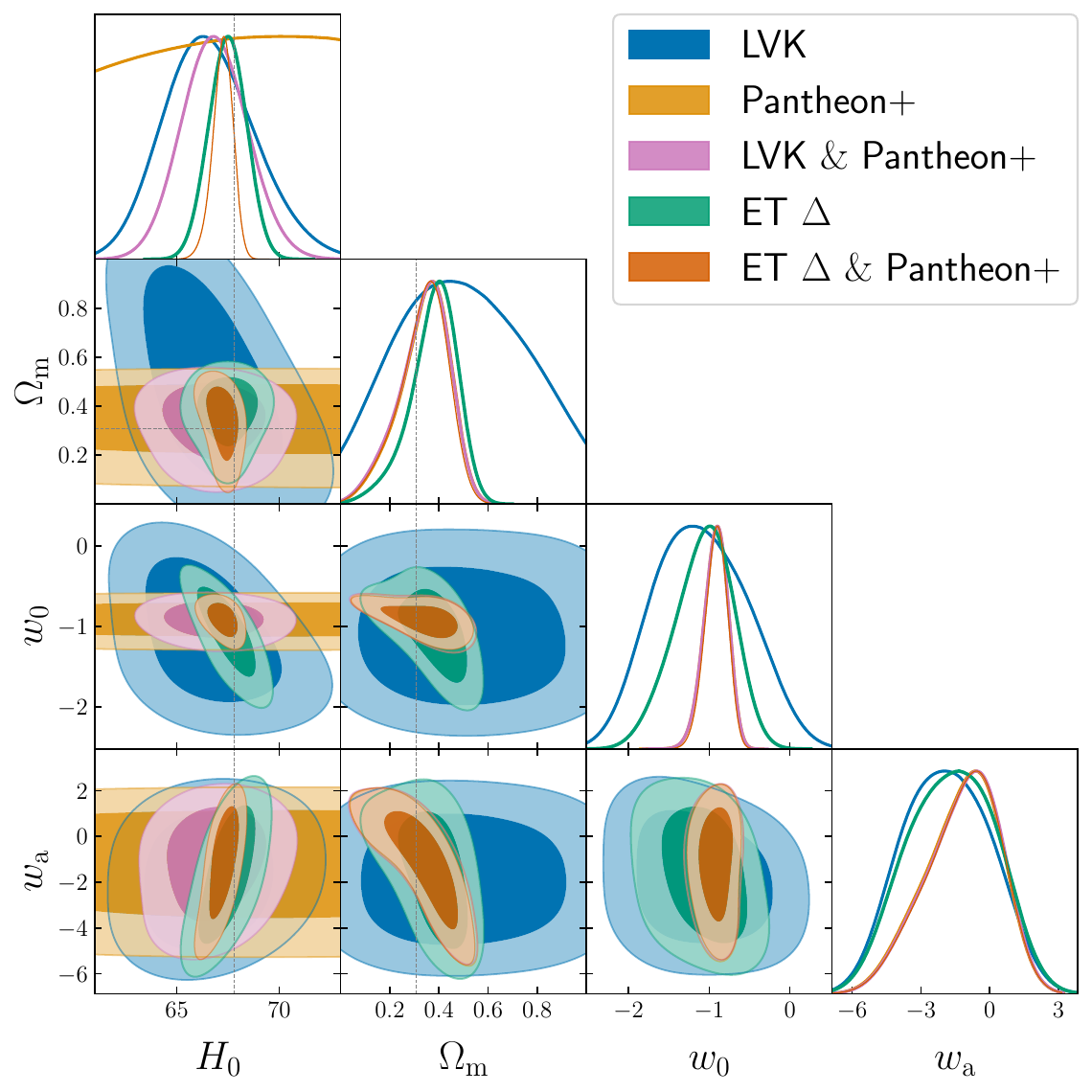}
    \caption{CPL constraints (68\% and 95\% C.L.) on relevant cosmological parameters, obtained by fitting the underlying MOD1 cosmology. (\textit{Left}) Comparison between LVK (O5) (blue) and ET$\Delta$ (green). (\textit{Right}) As in the left panel but over-plotting the constraints obtained with the \panp data set alone (orange) and in a joint analysis with LVK (O5) (pink).}
    \label{fig:LVK_ET_LC}
\end{figure}

In Figure~\ref{fig:LVK_ET_LC} we show the constraints obtained by fitting CPL over the MOD1 data set. In the left panel we can see again how 3G detectors outperform 2G detectors. We report the FoM of the three shown configurations, namely $\tilde{\epsilon}_{\mathrm{LVK}}=16\%$
$\tilde{\epsilon}_{\mathrm{LVKI} \mathrm{A}^\#}=12\%$
$\tilde{\epsilon}_{\mathrm{ET}\Delta}=6.3\%$. In the right panel instead, we show the results obtained with the \panp combination. 
We report an improvement quantified by the FoM as follows: 
$\tilde{\epsilon}_{\mathrm{PP+LVK}}=8.3\%$, $\tilde{\epsilon}_{\mathrm{PP+ET}\Delta}=4.5\%$. 
Note the following interesting fact, in the right panel of Figure~\ref{fig:LVK_ET_LC}: in the $w_\mathrm{0}$-\Omegam space the contours of ET$\Delta$+\panp and LVK+\panp are equal. This means that the use of the \panp SNIa data set is driving the choice of \wnot as a consequence of the better constraint on $\Omega_\mathrm{m}$.
The full 2$\sigma$ uncertainty on $H_{\mathrm{0}}$, \Omegam and \wnot are shown in Figure~\ref{fig:whisker_LC_complete} and Table~\ref{tab:mod1_cpl} of Appendix \ref{ap:tables}. We do not report the uncertainty on \wa because they derive unconstrained values for all the detector configurations and GRB catalogs.

\begin{figure}
    \centering
    \includegraphics[width=0.45\textwidth]{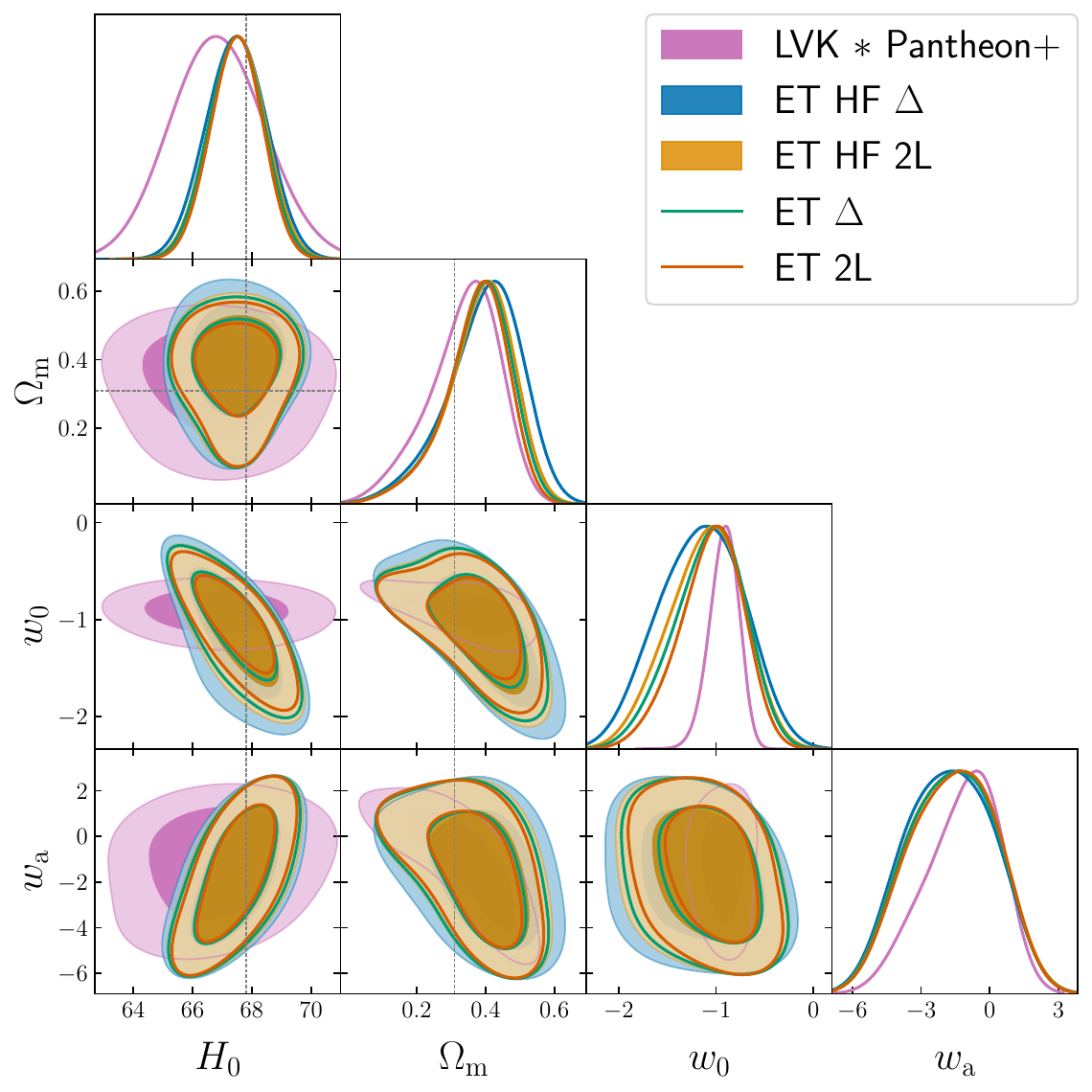}
    \includegraphics[width=0.45\textwidth]{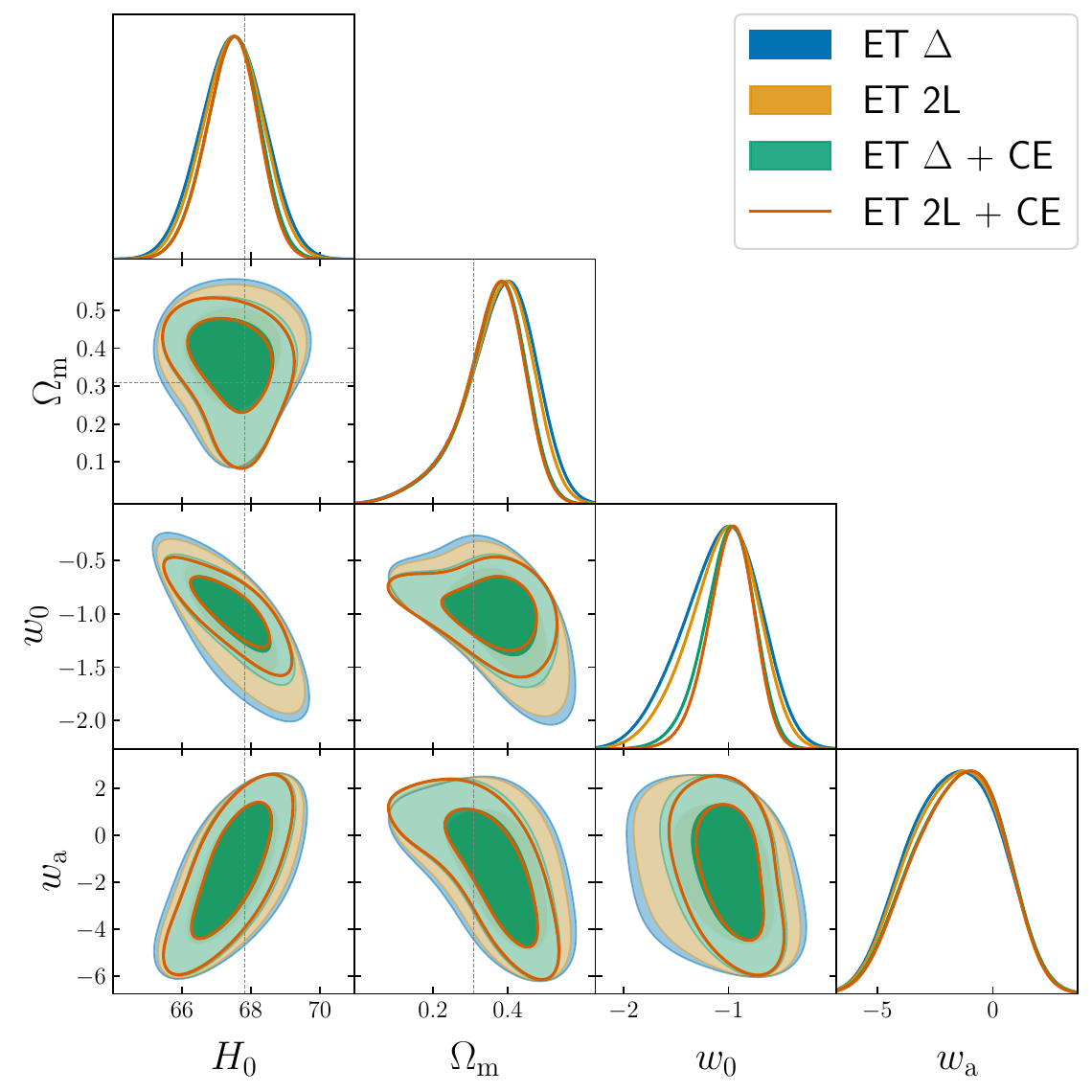}
    \caption{CPL constraints (68\% and 95\% C.L.) on the relevant cosmological parameters, obtained by fitting the underlying MOD1 cosmology. (\textit{Left}) Comparison of the cosmological constraints obtained by LVK+Pantheon$\Plus$, the ET HF $\Delta$/2L and ET$\Delta$/2L configurations. (\textit{Right}) The comparison between ET$\Delta$/2L and ET$\Delta$/2L + CE.}
    \label{fig:config_comp_cpl}
\end{figure}

In Figure~\ref{fig:config_comp_cpl}, we delve into a comparison of various GW detector configurations. It is worth emphasizing once again the slight advantage of ET2L over ET$\Delta$. However, with respect to the \lcdm fit on MOD1, transitioning to the CPL scenario reveals a noticeable decline in constraining power, attributed to the wider parameter space explored. The relative FoMs read: $\tilde{\epsilon}_{\mathrm{ETHF}\Delta}=7.1\%$, $\tilde{\epsilon}_{\mathrm{ET}\Delta}=6.3\%$,
$\tilde{\epsilon}_{\mathrm{ET}\Delta+\mathrm{CE}}=6.2\%$ and 
$\tilde{\epsilon}_{\mathrm{ETHF 2L}}=6.5\%$, $\tilde{\epsilon}_{\mathrm{ET2L}}=6.0\%$,
$\tilde{\epsilon}_{\mathrm{ET2L+CE}}=5.3\%$ 
 
\begin{figure}
    \centering
    \includegraphics[width=0.45\textwidth]{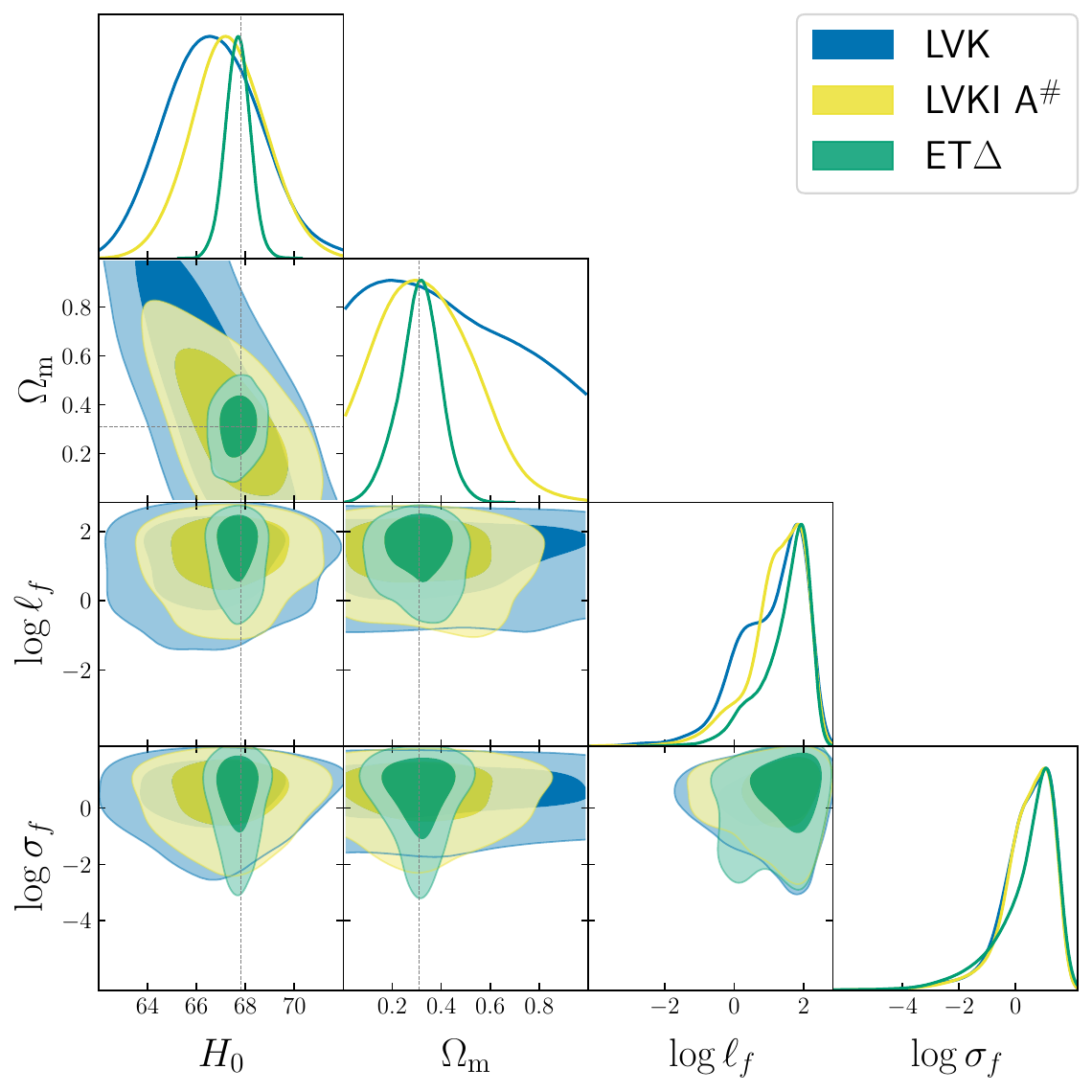}
    \includegraphics[width=0.45\textwidth]{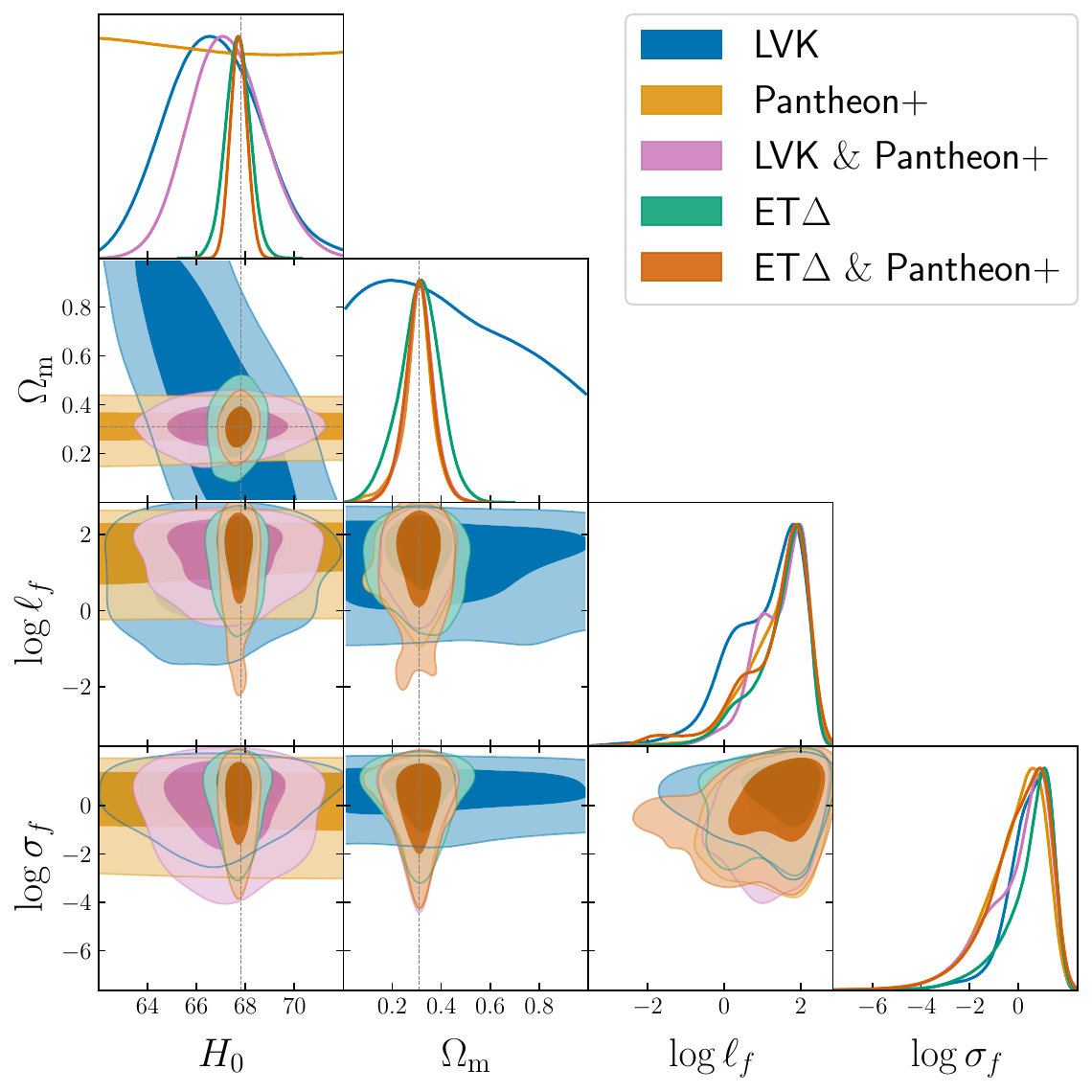}
    \caption{GP constraints (68\% and 95\% C.L.) on relevant cosmological parameters, obtained by fitting the underlying MOD1 cosmology. (\textit{Left}) Comparison between LVK (O5) (blue) and ET$\Delta$ (green). (\textit{Right}) As in the left panel but over-plotting the constraints obtained with the \panp data set alone (orange) and in a joint analysis with LVK (O5) (pink).}
    \label{fig:LVK_ET_LGP}
\end{figure}

\begin{figure}
    \centering
    \includegraphics[width=0.6\textwidth]{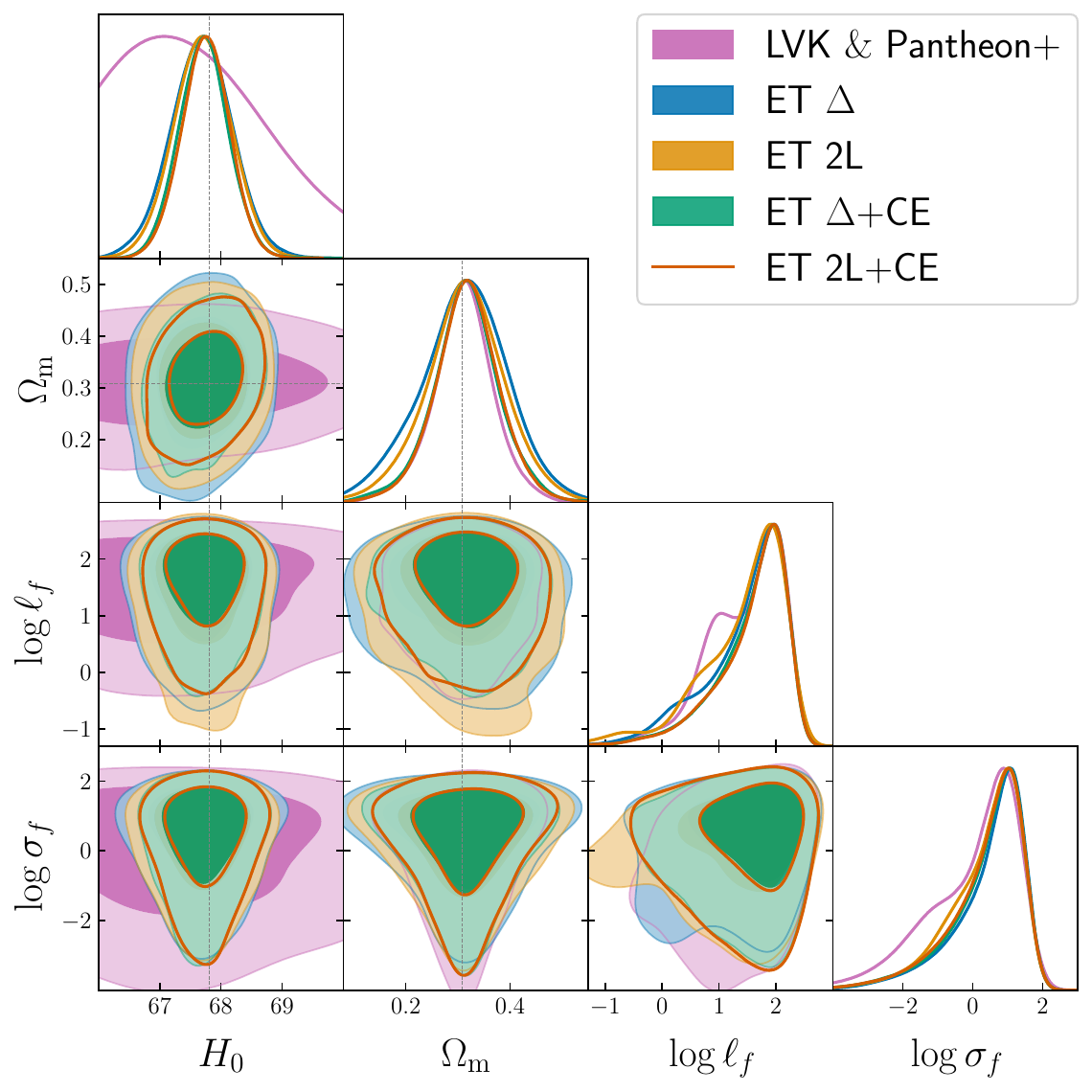}
    \caption{GP constraints (68\% and 95\% C.L.) on the relevant cosmological parameters, obtained by fitting the underlying MOD1 cosmology. Comparison of the cosmological constraints obtained by LVK+Pantheon$\Plus$, the ET$\Delta$/2L and ET$\Delta$/2L  + CE configurations.}
    \label{fig:config_comp_gp}
\end{figure}

Finally, we report the GP comparison among GW detectors with a  MOD1 underlying cosmology. In the left panel of Figure~\ref{fig:LVK_ET_LGP} we report 2G and 3G detector comparison. The FoMs for the configurations reported in the Figure read $\tilde{\epsilon}_{\mathrm{LVK}}=14.5\%$
$\tilde{\epsilon}_{\mathrm{LVKI} \mathrm{A}^\#}=10.6\%$
$\tilde{\epsilon}_{\mathrm{ET}\Delta}=4.6\%$. In the right panel of Figure~\ref{fig:LVK_ET_LGP} we report the \panp calibration with GW. The resulting FoMs are $\tilde{\epsilon}_{\mathrm{PP+LVK}}=6.7\%$, $\tilde{\epsilon}_{\mathrm{PP+ET}\Delta}=3.0\%$. 
The comparison among 3G detectors different configurations is instead reported in Figure~\ref{fig:config_comp_gp}. The associated FoMs read $\tilde{\epsilon}_{\mathrm{ET}\Delta}=4.6\%$,
$\tilde{\epsilon}_{\mathrm{ET}\Delta + \mathrm{CE}}=3.5\%$ and 
$\tilde{\epsilon}_{\mathrm{ET2L}}=4.2\%$,
$\tilde{\epsilon}_{\mathrm{ET2L+CE}}=3.4\%$. We disclose a consistent improvement with respect to CPL, as already underlined in the main text. The complete results are reported in Appendix~\ref{ap:tables}, Figure~\ref{fig:whisker_LGP_complete} and Table~\ref{tab:mod1_gp}.

\section{Further tables and plots}\label{ap:tables}

\begin{figure}[h]
    \centering
    \includegraphics[width=\linewidth]{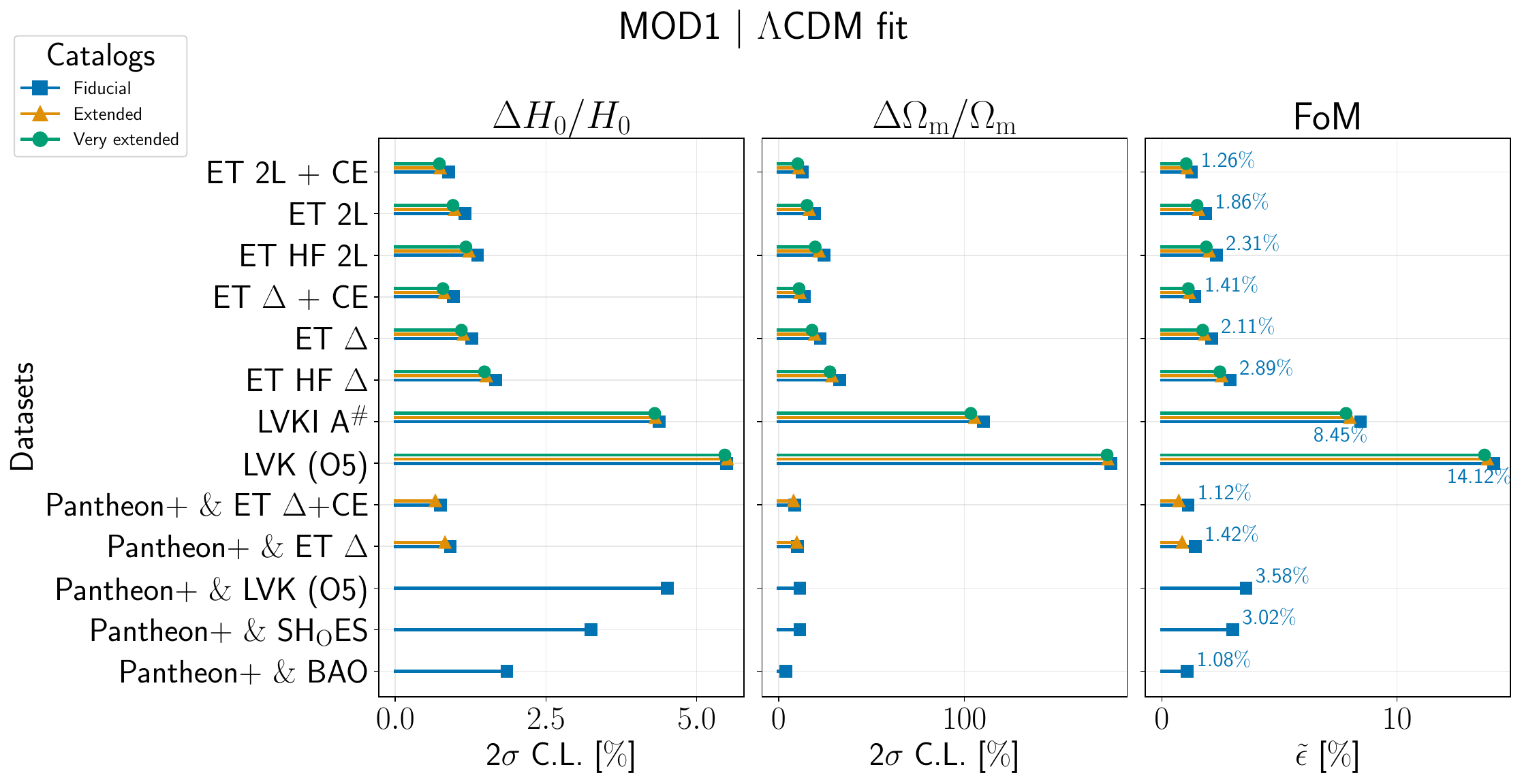}
    \caption{\lcdm 2$\sigma$ relative error on $H_\mathrm{0}$ and $\Omega_\mathrm{m}$ for an underlying MOD1 cosmology. The last panel presents the relative FoM. The associated values can be found in Table~\ref{tab:mod1_lcdm}.}
    \label{fig:whisker_LL_complete}
\end{figure}

\begin{figure}[h]
    \centering
    \includegraphics[width=1\linewidth]{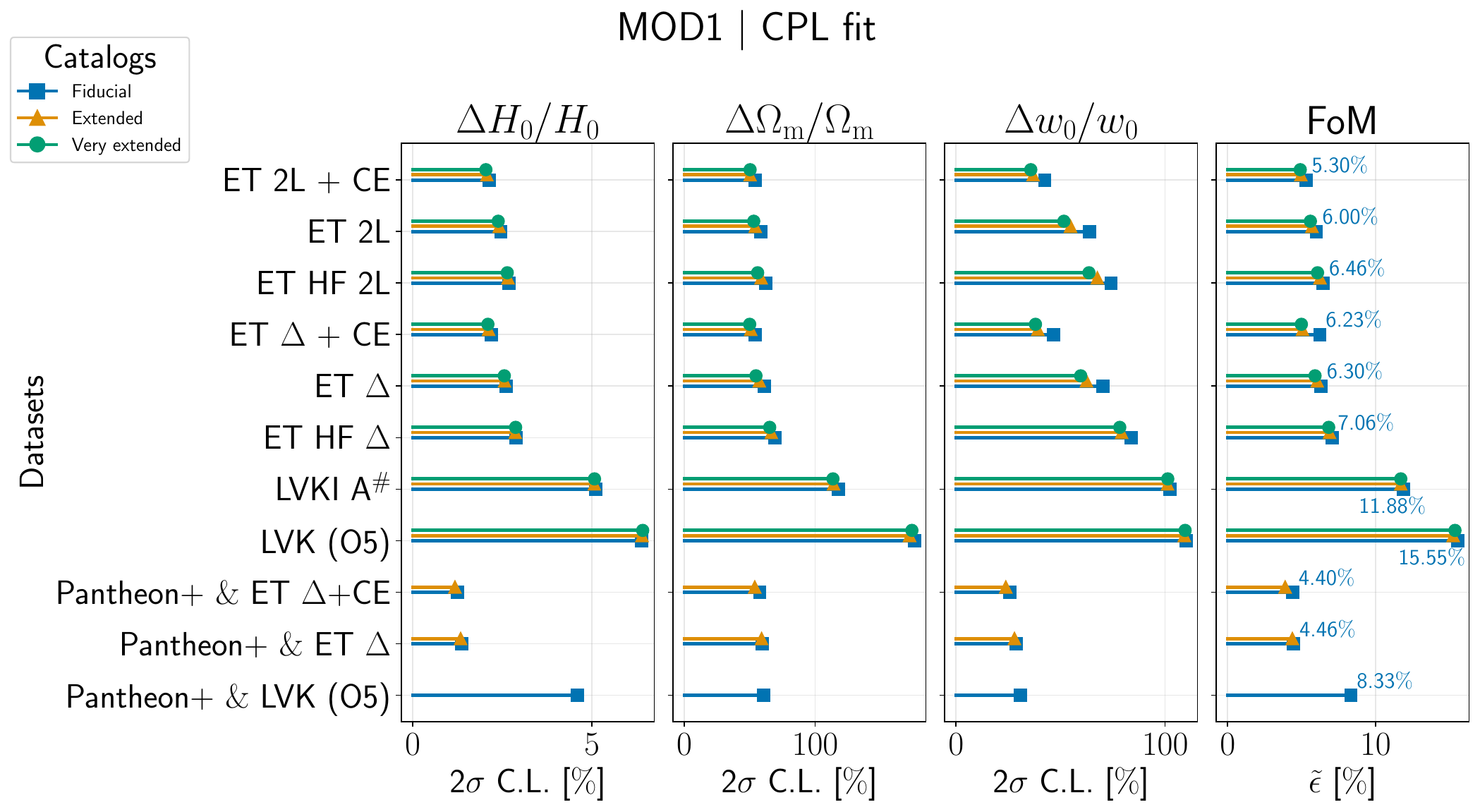}
    \caption{CPL 2$\sigma$ relative error on $H_\mathrm{0}$, $\Omega_\mathrm{m}$ and \wnot for an underlying MOD1 cosmology. The last panel presents the relative FoM. The associated values can be found in Table~\ref{tab:mod1_cpl}.}
    
    \label{fig:whisker_LC_complete}
\end{figure}

\begin{figure}[h]
    \centering
    \includegraphics[width=1\linewidth]{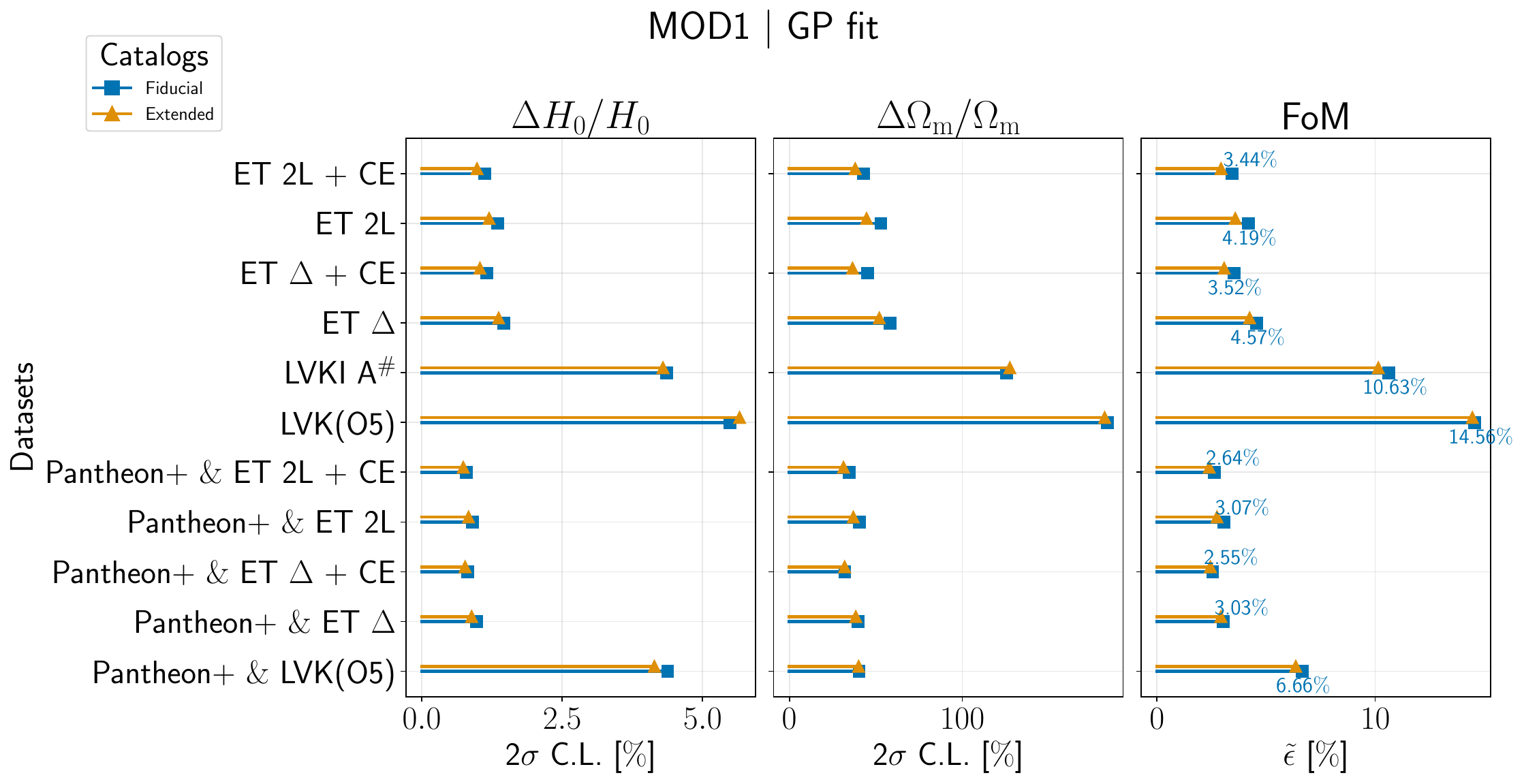}
    \caption{GP 2$\sigma$ relative error on $H_\mathrm{0}$ and $\Omega_\mathrm{m}$ for an underlying MOD1 cosmology. The last panel presents the relative FoM. The associated values can be found in Table~\ref{tab:mod1_gp}.}
    
    \label{fig:whisker_LGP_complete}
\end{figure}

\begin{figure}[h]
    \centering
    \includegraphics[width=1\linewidth]{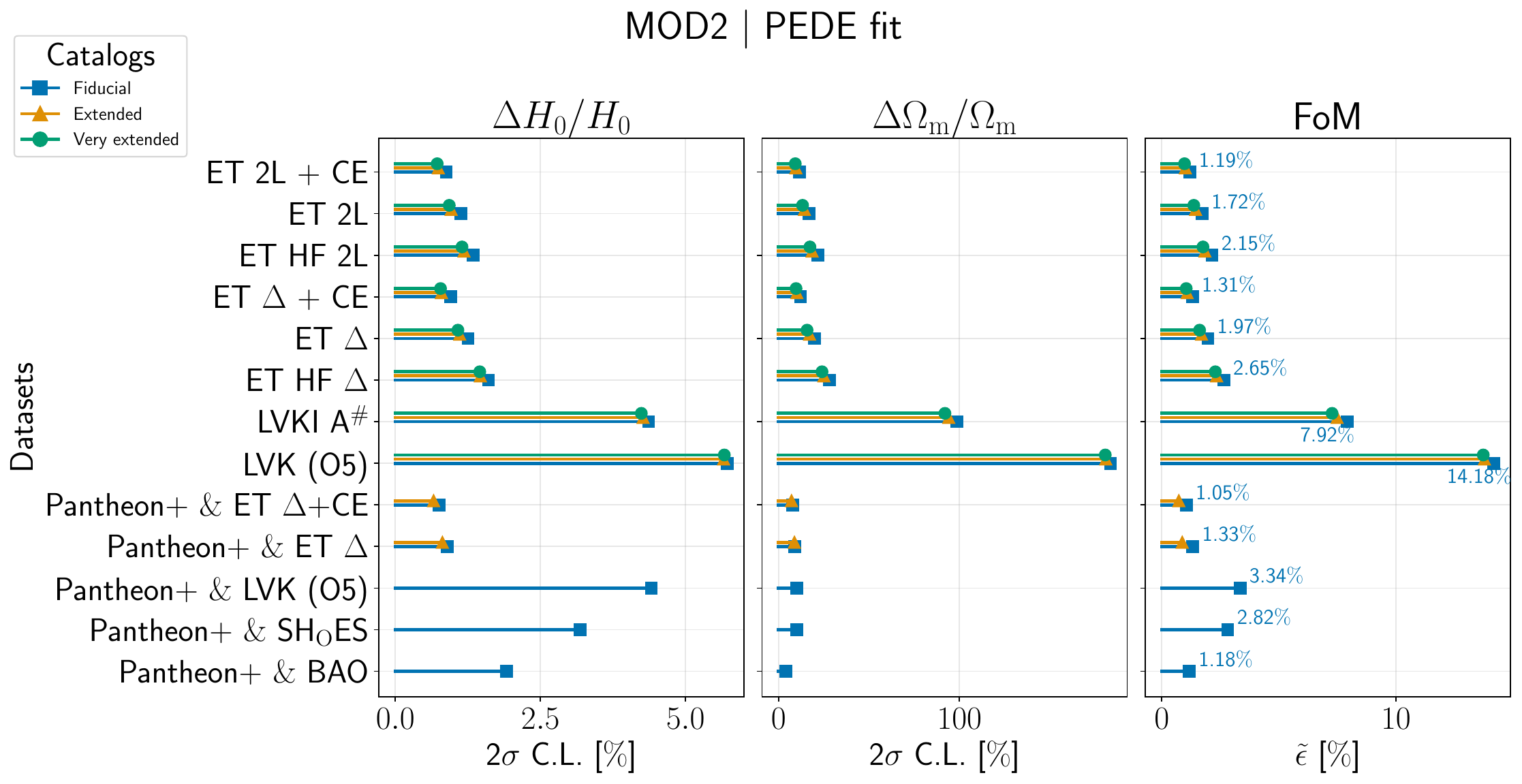}
    \caption{PEDE 2$\sigma$ relative error on $H_\mathrm{0}$ and $\Omega_\mathrm{m}$ for an underlying MOD2 cosmology. The last panel presents the relative FoM. The associated values can be found in Table~\ref{tab:mod2_pede}.}
    \label{fig:whisker_PP_complete}
\end{figure}

\begin{figure}[h]
    \centering
    \includegraphics[width=1\linewidth]{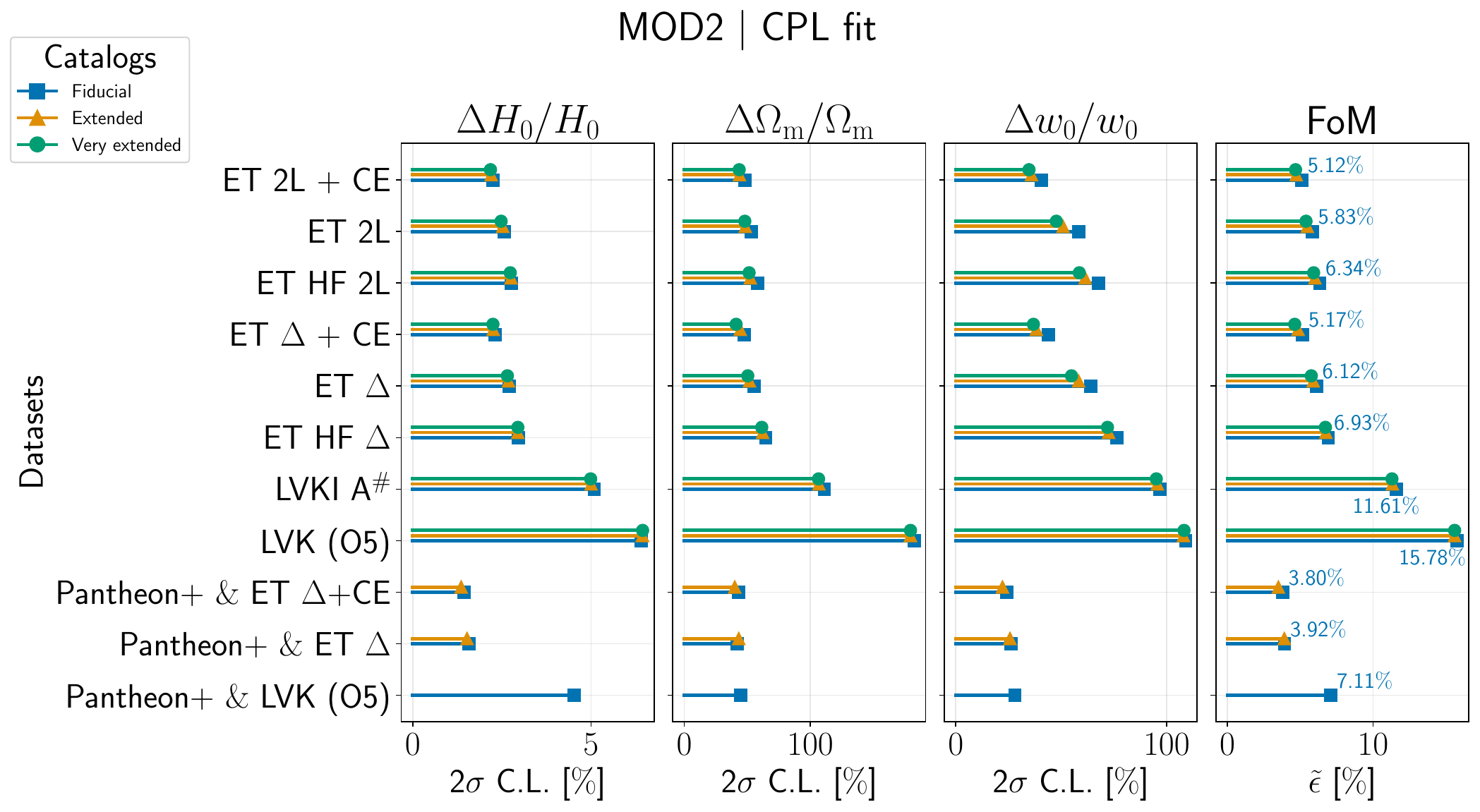}
    \caption{CPL 2$\sigma$ relative error on $H_\mathrm{0}$, $\Omega_\mathrm{m}$ and \wnot for an underlying MOD2 cosmology. The last panel presents the relative FoM. The associated values can be found in Table~\ref{tab:mod2_cpl}. }
    \label{fig:whisker_PC_complete}
\end{figure}

\begin{figure}[h]
    \centering
    \includegraphics[width=1\linewidth]{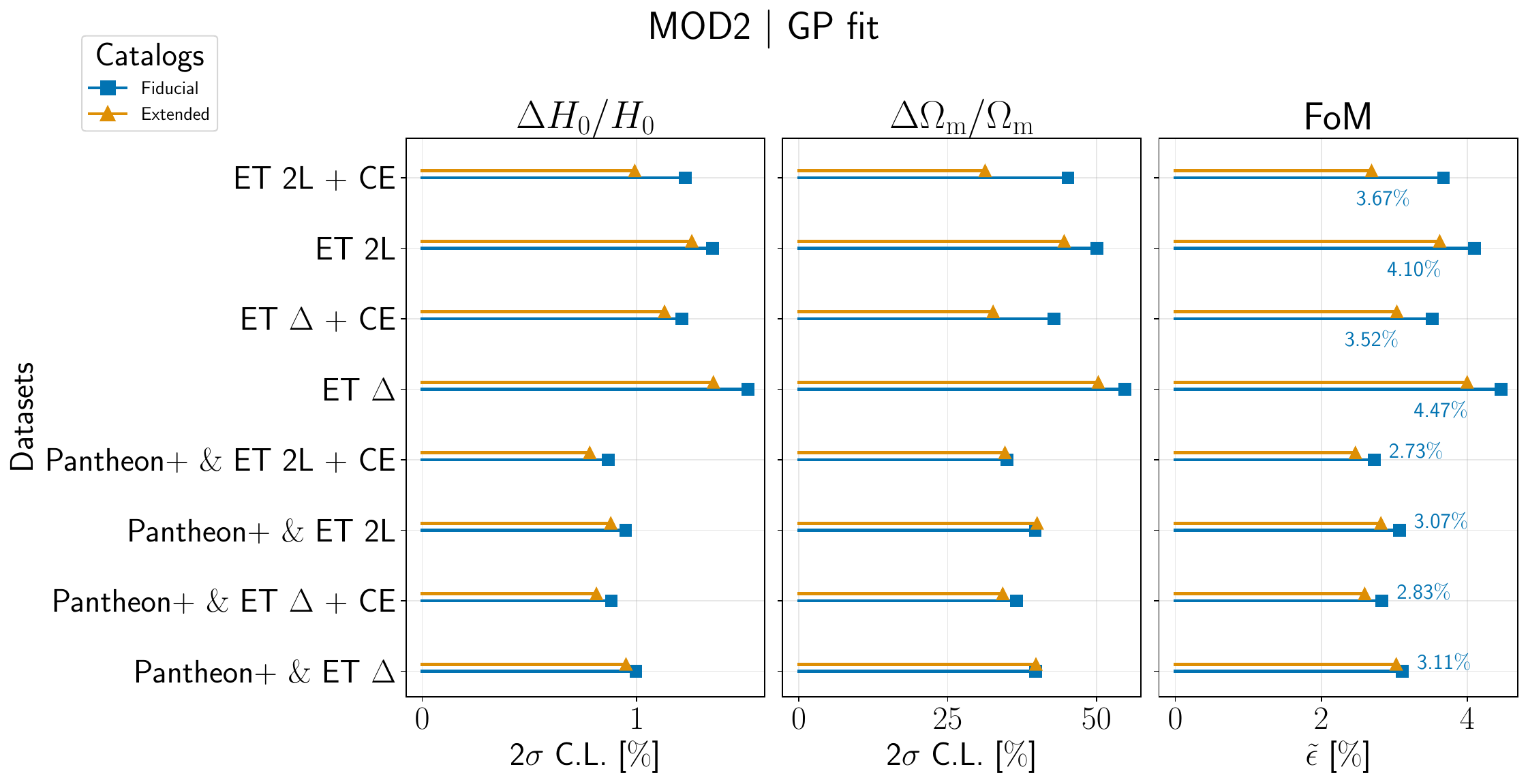}
    \caption{GP 2$\sigma$ relative error on $H_\mathrm{0}$ and $\Omega_\mathrm{m}$ for an underlying MOD2 cosmology. The last panel presents the relative FoM. The associated values can be found in Table~\ref{tab:mod2_gp}.}
    \label{fig:whisker_PGP_complete}
\end{figure}

This Appendix is dedicated to display plots and tables accompanying or completing the results found in the main text and the previous appendices. 
We report the results for the $\Lambda$CDM, CPL and GP fit over MOD1 in Figures~\ref{fig:whisker_LL_complete}, \ref{fig:whisker_LC_complete} and \ref{fig:whisker_LGP_complete}, respectively. The condensed numbers associated to these can be found in Tables~\ref{tab:mod1_lcdm}, \ref{tab:mod1_cpl} and \ref{tab:mod1_gp}. The same is done for MOD2 in Figures~\ref{fig:whisker_PP_complete}, \ref{fig:whisker_PC_complete}, \ref{fig:whisker_PGP_complete} and Tables~\ref{tab:mod2_pede}, \ref{tab:mod2_cpl} and \ref{tab:mod2_gp}.

\begin{figure}[h]
    \centering
    \includegraphics[width=0.5\textwidth]{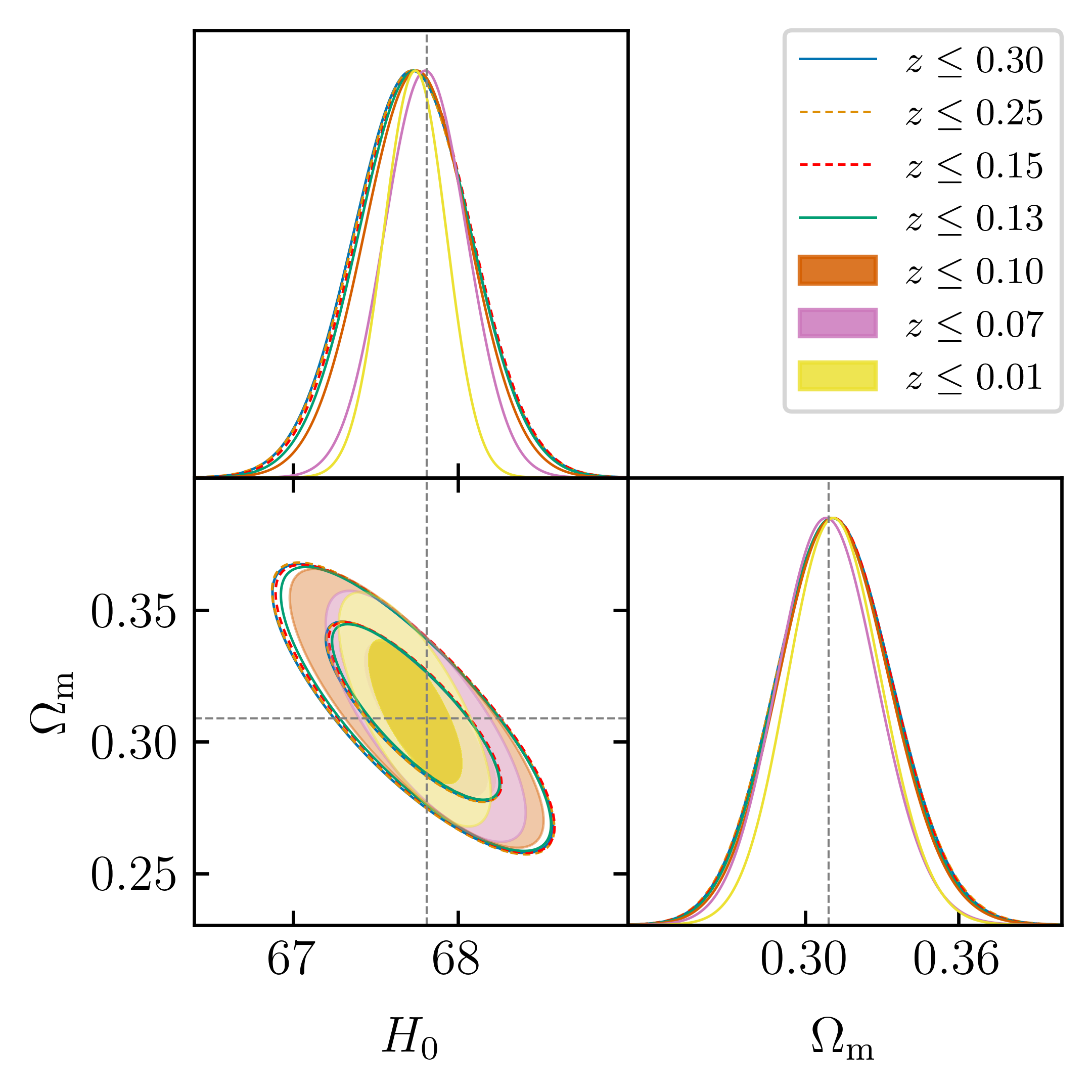}
    \caption{Comparison among different thresholds for the peculiar velocity analysis. These contours are all derived from fitting \lcdm over the MOD1 fiducial data set using the ET$\Delta$+CE configuration. The filled contours are obtained as follows: the first contour incorporates the correction for only the closest GRB, the second smallest contour includes corrections for the two closest GRBs, and the third contour accounts for the three closest GRBs. Observe that the constraints become stable when $z>0.15$.}
    \label{fig:z_vp_comaparison}
\end{figure}

\begin{figure}[b]
    \centering
    \includegraphics[width=0.5\textwidth]{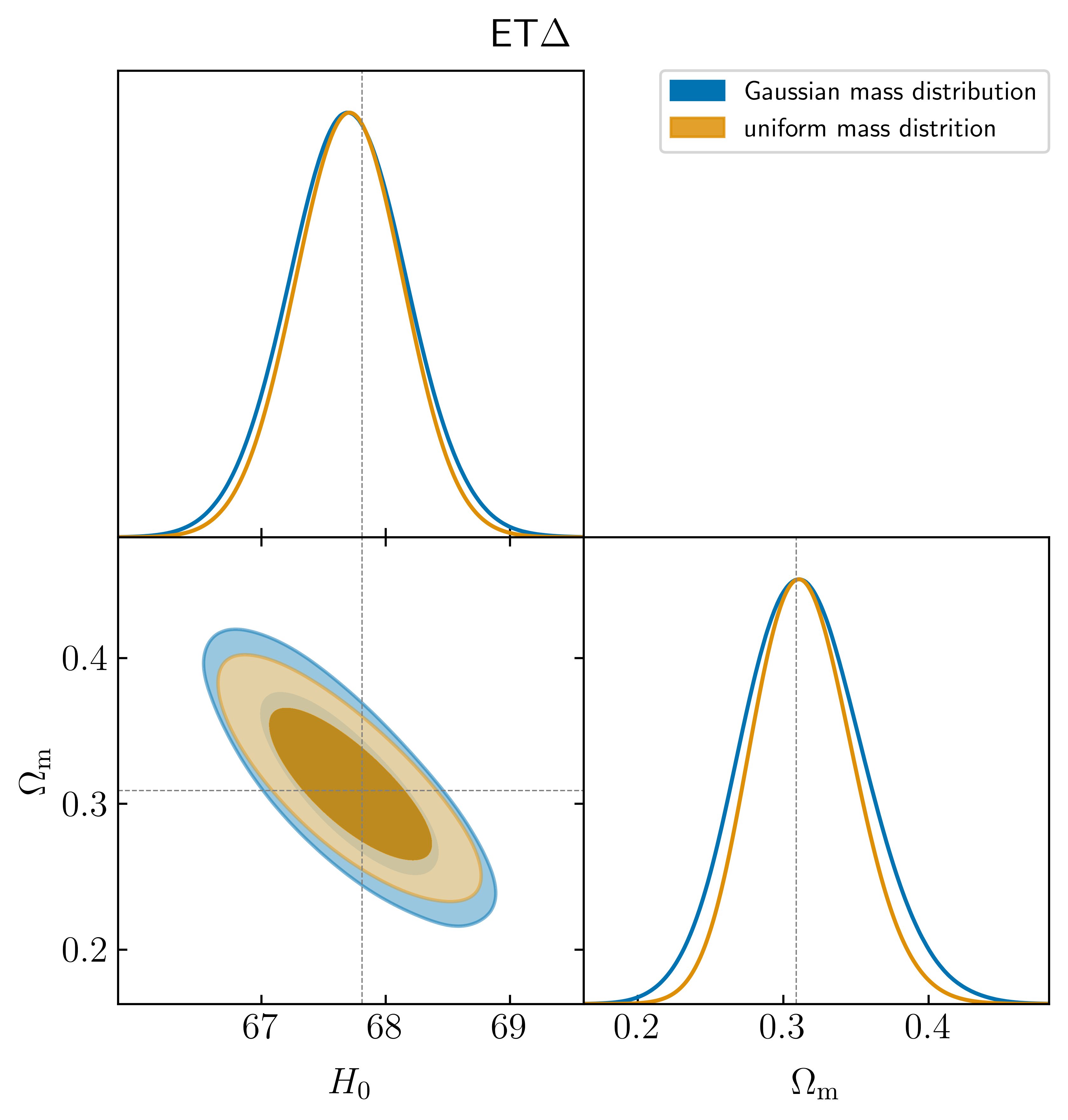}

    \caption{Comparison among different neutron star mass distributions. We show the case for ET$\Delta$ and the fiducial catalog. The uniform distribution refers to a catalog generated with $\mathcal{U}(1.1,2.1) \mathrm{M}_\odot$. The Gaussian distribution refers instead to a catalog generated drawing masses from a normal distribution $\mathcal{N}(1.33, 0.09)\mathrm{M}_\odot$.}
    \label{fig:gaussmass_ETD}
\end{figure}

In Figure~\ref{fig:z_vp_comaparison} we present the $H_{\mathrm 0}$-\Omegam contours from a \lcdm fit on MOD1, using the ET$\Delta$+CE configuration. This figure illustrates how the constraints vary by adjusting the redshift threshold applied for the peculiar velocity correction, as discussed in Subsection~\ref{subsec:GWfish+}. It is evident that the contours cease to expand when the threshold is set at $z\sim0.15$, which is why we selected this threshold for use throughout the paper. Notably, neglecting the proper peculiar velocity correction would result in significantly smaller contours compared to those that include the correction. In Figure~\ref{fig:gaussmass_ETD} we compare different BNS mass distributions, specifically examining the ET$\Delta$ configuration for the fiducial catalog. The uniform distribution refers to a data set generated by uniformly sampling BNS masses in the range of 1.1 to 2.1 $\mathrm{M}_\odot$. In contrast, the Gaussian distribution represents a data set generated by drawing masses from a normal distribution with a mean of 1.33 $\mathrm{M}_\odot$ and a standard deviation of 0.09 $\mathrm{M}_\odot$~\cite{Loffredo:2024}. The differences in the constraints are minimal and primarily driven by different number of detected events. In the case of the Gaussian mass distribution, the number of detected events is reduced to 33, compared to 36 events for the uniform distribution.

\begin{table}[t]
\centering
\renewcommand{\arraystretch}{1.4} 

\caption{MOD1 | \lcdm fit} 

\scriptsize{
\begin{tabular}{cc|c|c|c|c|c|c|c|c|}
\hline
\multicolumn{1}{|c|}{\multirow{2}{*}{\textbf{Data set}}}    
& \multicolumn{3}{c|}{\textbf{Fiducial}}      
& \multicolumn{3}{c|}{\textbf{Extended}}  
& \multicolumn{3}{c|}{\textbf{Very extended}} \\ \cline{2-10}

\multicolumn{1}{|c|}{}    
& $\boldsymbol{\frac{\Delta H_\mathrm{0}}{H_\mathrm{0}}}$ & $\boldsymbol{\frac{\Delta \Omega_\mathrm{m}}{\Omega_\mathrm{m}}}$ &$\boldsymbol{\tilde{\epsilon}}$ 
& $\boldsymbol{\frac{\Delta H_\mathrm{0}}{H_\mathrm{0}}}$ & $\boldsymbol{\frac{\Delta \Omega_\mathrm{m}}{\Omega_\mathrm{m}}}$&$\boldsymbol{\tilde{\epsilon}}$ 
& $\boldsymbol{\frac{\Delta H_\mathrm{0}}{H_\mathrm{0}}}$ & $\boldsymbol{\frac{\Delta \Omega_\mathrm{m}}{\Omega_\mathrm{m}}}$ &$\boldsymbol{\tilde{\epsilon}}$                        \\ \hline
\multicolumn{1}{|c|}{\panp $\&$ BAO} & 0.019 & 0.04 & 0.011 & -- & -- & -- & -- & -- & -- \\ 
\multicolumn{1}{|c|}{\panp $\&$ SH$_0$ES} & 0.032 & 0.11 & 0.03 & -- & -- & -- & -- & -- & -- \\ 
\multicolumn{1}{|c|}{\panp $\&$ LVK (O5)} & 0.045 & 0.11 & 0.036 & -- & -- & -- & -- & -- & -- \\ 
\multicolumn{1}{|c|}{\panp $\&$ ET$\Delta$} & 0.0091 & 0.1 & 0.014 & 0.0082 & 0.098 & 0.0086 & -- & -- & -- \\ 
\multicolumn{1}{|c|}{\panp $\&$ ET$\Delta$+CE} & 0.0075 & 0.088 & 0.011 & 0.0066 & 0.081 & 0.0072 & -- & -- & -- \\ 
\multicolumn{1}{|c|}{LVK (O5)} & 0.055 & 1.8 & 0.14 & 0.055 & 1.8 & 0.14 & 0.055 & 1.8 & 0.14 \\ 
\multicolumn{1}{|c|}{LVKI A$^\#$} & 0.044 & 1.1 & 0.084 & 0.043 & 1.1 & 0.08 & 0.043 & 1.0 & 0.078 \\ 
\multicolumn{1}{|c|}{ET HF $\Delta$} & 0.017 & 0.33 & 0.029 & 0.015 & 0.29 & 0.025 & 0.015 & 0.28 & 0.025 \\ 
\multicolumn{1}{|c|}{ET$\Delta$} & 0.013 & 0.22 & 0.021 & 0.011 & 0.19 & 0.018 & 0.011 & 0.18 & 0.017 \\ 
\multicolumn{1}{|c|}{ET$\Delta$ + CE} & 0.0096 & 0.14 & 0.014 & 0.0081 & 0.11 & 0.012 & 0.0079 & 0.11 & 0.011 \\ 
\multicolumn{1}{|c|}{ET HF 2L} & 0.014 & 0.25 & 0.023 & 0.012 & 0.22 & 0.02 & 0.012 & 0.2 & 0.019 \\ 
\multicolumn{1}{|c|}{ET2L} & 0.012 & 0.19 & 0.019 & 0.0098 & 0.17 & 0.016 & 0.0096 & 0.15 & 0.015 \\ 
\multicolumn{1}{|c|}{ET2L + CE} & 0.0088 & 0.13 & 0.013 & 0.0075 & 0.11 & 0.011 & 0.0073 & 0.1 & 0.01 \\ 
\hline
\end{tabular}
}

\label{tab:mod1_lcdm}

\end{table}

\begin{table}[b]
\centering
\renewcommand{\arraystretch}{1.4} 

\caption{MOD1 | CPL fit}

\scriptsize{
\begin{tabular}{cc|c|c|c|c|c|c|c|c|c|c|c|}
\hline
\multicolumn{1}{|c|}{\multirow{2}{*}{\textbf{Data set}}}    
& \multicolumn{4}{c|}{\textbf{Fiducial}}      
& \multicolumn{4}{c|}{\textbf{Extended}}  
& \multicolumn{4}{c|}{\textbf{Very extended}} \\ \cline{2-13}

\multicolumn{1}{|c|}{}    
& $\boldsymbol{\frac{\Delta H_\mathrm{0}}{H_\mathrm{0}}}$ & $\boldsymbol{\frac{\Delta \Omega_\mathrm{m}}{\Omega_\mathrm{m}}}$& $\boldsymbol{\frac{\Delta w_\mathrm{0}}{w_\mathrm{0}}}$ &$\boldsymbol{\tilde{\epsilon}}$ 
& $\boldsymbol{\frac{\Delta H_\mathrm{0}}{H_\mathrm{0}}}$ & $\boldsymbol{\frac{\Delta \Omega_\mathrm{m}}{\Omega_\mathrm{m}}}$& $\boldsymbol{\frac{\Delta w_\mathrm{0}}{w_\mathrm{0}}}$ &$\boldsymbol{\tilde{\epsilon}}$ 
& $\boldsymbol{\frac{\Delta H_\mathrm{0}}{H_\mathrm{0}}}$ & $\boldsymbol{\frac{\Delta \Omega_\mathrm{m}}{\Omega_\mathrm{m}}}$& $\boldsymbol{\frac{\Delta w_\mathrm{0}}{w_\mathrm{0}}}$ &$\boldsymbol{\tilde{\epsilon}}$ 
                                                \\ \hline
\multicolumn{1}{|c|}{\panp $\&$ LVK (O5)} & 0.046 & 0.61 & 0.31 & 0.083 & -- & -- & -- & -- & -- & -- & -- & -- \\ 
\multicolumn{1}{|c|}{\panp $\&$ ET$\Delta$} & 0.014 & 0.6 & 0.29 & 0.045 & 0.013 & 0.59 & 0.28 & 0.044 & -- & -- & -- & -- \\ 
\multicolumn{1}{|c|}{\panp $\&$ ET$\Delta$+CE} & 0.012 & 0.58 & 0.26 & 0.044 & 0.012 & 0.54 & 0.24 & 0.039 & -- & -- & -- & -- \\ 
\multicolumn{1}{|c|}{LVK (O5)} & 0.064 & 1.8 & 1.1 & 0.16 & 0.064 & 1.7 & 1.1 & 0.15 & 0.064 & 1.7 & 1.1 & 0.15 \\ 
\multicolumn{1}{|c|}{LVKI A$^\#$} & 0.051 & 1.2 & 1.0 & 0.12 & 0.051 & 1.1 & 1.0 & 0.12 & 0.051 & 1.1 & 1.0 & 0.12 \\ 
\multicolumn{1}{|c|}{ET HF $\Delta$} & 0.029 & 0.69 & 0.84 & 0.071 & 0.029 & 0.66 & 0.79 & 0.069 & 0.029 & 0.65 & 0.78 & 0.068 \\ 
\multicolumn{1}{|c|}{ET$\Delta$} & 0.026 & 0.61 & 0.7 & 0.063 & 0.026 & 0.57 & 0.62 & 0.061 & 0.026 & 0.55 & 0.6 & 0.059 \\ 
\multicolumn{1}{|c|}{ET$\Delta$ + CE} & 0.022 & 0.54 & 0.47 & 0.062 & 0.021 & 0.51 & 0.39 & 0.051 & 0.021 & 0.5 & 0.38 & 0.05 \\ 
\multicolumn{1}{|c|}{ET HF 2L} & 0.027 & 0.62 & 0.74 & 0.065 & 0.027 & 0.59 & 0.68 & 0.062 & 0.026 & 0.56 & 0.64 & 0.061 \\ 
\multicolumn{1}{|c|}{ET2L} & 0.025 & 0.59 & 0.64 & 0.06 & 0.024 & 0.54 & 0.55 & 0.057 & 0.024 & 0.53 & 0.52 & 0.056 \\ 
\multicolumn{1}{|c|}{ET2L + CE} & 0.021 & 0.54 & 0.42 & 0.053 & 0.021 & 0.5 & 0.37 & 0.049 & 0.02 & 0.5 & 0.36 & 0.049 \\ 
\hline
\end{tabular}
}

\label{tab:mod1_cpl}

\end{table}

\begin{table}[b]
\centering
\renewcommand{\arraystretch}{1.4} 
\caption{MOD1 | GP fit}
\begin{tabular}{cc|c|c|c|c|c|}
\hline
\multicolumn{1}{|c|}{\multirow{2}{*}{\textbf{Data set}}}    
& \multicolumn{3}{c|}{\textbf{Fiducial}}      
& \multicolumn{3}{c|}{\textbf{Extended}}  \\ \cline{2-7}

\multicolumn{1}{|c|}{}    
& $\boldsymbol{\frac{\Delta H_\mathrm{0}}{H_\mathrm{0}}}$ & $\boldsymbol{\frac{\Delta \Omega_\mathrm{m}}{\Omega_\mathrm{m}}}$ &$\boldsymbol{\tilde{\epsilon}}$ 
& $\boldsymbol{\frac{\Delta H_\mathrm{0}}{H_\mathrm{0}}}$ & $\boldsymbol{\frac{\Delta \Omega_\mathrm{m}}{\Omega_\mathrm{m}}}$ &$\boldsymbol{\tilde{\epsilon}}$                        \\ \hline
\multicolumn{1}{|c|}{\panp $\&$ LVK(O5)} & 0.044 & 0.4 & 0.067 & 0.041 & 0.4 & 0.064 \\ 
\multicolumn{1}{|c|}{\panp $\&$ ET $\Delta$} & 0.0097 & 0.4 & 0.03 & 0.0089 & 0.38 & 0.029 \\ 
\multicolumn{1}{|c|}{\panp $\&$ ET $\Delta$ + CE} & 0.0082 & 0.32 & 0.025 & 0.0077 & 0.32 & 0.024 \\ 
\multicolumn{1}{|c|}{\panp $\&$ ET 2L} & 0.009 & 0.41 & 0.031 & 0.0083 & 0.37 & 0.027 \\ 
\multicolumn{1}{|c|}{\panp $\&$ ET 2L + CE} & 0.008 & 0.35 & 0.026 & 0.0074 & 0.31 & 0.024 \\ 
\multicolumn{1}{|c|}{LVK(O5)} & 0.055 & 1.8 & 0.15 & 0.057 & 1.8 & 0.14 \\ 
\multicolumn{1}{|c|}{LVKI A$^\#$} & 0.044 & 1.3 & 0.11 & 0.043 & 1.3 & 0.1 \\ 
\multicolumn{1}{|c|}{ET $\Delta$} & 0.015 & 0.58 & 0.046 & 0.014 & 0.52 & 0.042 \\ 
\multicolumn{1}{|c|}{ET $\Delta$ + CE} & 0.012 & 0.45 & 0.035 & 0.01 & 0.37 & 0.031 \\ 
\multicolumn{1}{|c|}{ET 2L} & 0.014 & 0.53 & 0.042 & 0.012 & 0.45 & 0.036 \\ 
\multicolumn{1}{|c|}{ET 2L + CE} & 0.011 & 0.43 & 0.034 & 0.0098 & 0.38 & 0.029 \\ 
\hline
\end{tabular}
\label{tab:mod1_gp}
\end{table}

\begin{table}[b]
\centering
\renewcommand{\arraystretch}{1.4} 

\caption{MOD2 | PEDE fit}

\scriptsize{
\begin{tabular}{cc|c|c|c|c|c|c|c|c|}
\hline
\multicolumn{1}{|c|}{\multirow{2}{*}{\textbf{Data set}}}    
& \multicolumn{3}{c|}{\textbf{Fiducial}}      
& \multicolumn{3}{c|}{\textbf{Extended}}  
& \multicolumn{3}{c|}{\textbf{Very extended}} \\ \cline{2-10}

\multicolumn{1}{|c|}{}    
& $\boldsymbol{\frac{\Delta H_\mathrm{0}}{H_\mathrm{0}}}$ & $\boldsymbol{\frac{\Delta \Omega_\mathrm{m}}{\Omega_\mathrm{m}}}$ &$\boldsymbol{\tilde{\epsilon}}$ 
& $\boldsymbol{\frac{\Delta H_\mathrm{0}}{H_\mathrm{0}}}$ & $\boldsymbol{\frac{\Delta \Omega_\mathrm{m}}{\Omega_\mathrm{m}}}$&$\boldsymbol{\tilde{\epsilon}}$ 
& $\boldsymbol{\frac{\Delta H_\mathrm{0}}{H_\mathrm{0}}}$ & $\boldsymbol{\frac{\Delta \Omega_\mathrm{m}}{\Omega_\mathrm{m}}}$ &$\boldsymbol{\tilde{\epsilon}}$                        \\ \hline
\multicolumn{1}{|c|}{\panp $\&$ BAO} & 0.019 & 0.04 & 0.012 & -- & -- & -- & -- & -- & -- \\ 
\multicolumn{1}{|c|}{\panp $\&$ SH$_0$ES} & 0.032 & 0.1 & 0.028 & -- & -- & -- & -- & -- & -- \\ 
\multicolumn{1}{|c|}{\panp $\&$ LVK (O5)} & 0.044 & 0.1 & 0.033 & -- & -- & -- & -- & -- & -- \\ 
\multicolumn{1}{|c|}{\panp $\&$ ET$\Delta$} & 0.009 & 0.09 & 0.013 & 0.0081 & 0.088 & 0.0087 & -- & -- & -- \\ 
\multicolumn{1}{|c|}{\panp $\&$ ET$\Delta$+CE} & 0.0075 & 0.078 & 0.011 & 0.0066 & 0.072 & 0.0073 & -- & -- & -- \\ 
\multicolumn{1}{|c|}{LVK (O5)} & 0.057 & 1.8 & 0.14 & 0.057 & 1.8 & 0.14 & 0.057 & 1.8 & 0.14 \\ 
\multicolumn{1}{|c|}{LVKI A$^\#$} & 0.044 & 0.98 & 0.079 & 0.043 & 0.94 & 0.075 & 0.042 & 0.92 & 0.073 \\ 
\multicolumn{1}{|c|}{ET HF $\Delta$} & 0.016 & 0.28 & 0.026 & 0.015 & 0.25 & 0.023 & 0.015 & 0.24 & 0.023 \\ 
\multicolumn{1}{|c|}{ET$\Delta$} & 0.013 & 0.2 & 0.02 & 0.011 & 0.17 & 0.017 & 0.011 & 0.16 & 0.016 \\ 
\multicolumn{1}{|c|}{ET$\Delta$ + CE} & 0.0095 & 0.12 & 0.013 & 0.0079 & 0.1 & 0.011 & 0.0078 & 0.097 & 0.01 \\ 
\multicolumn{1}{|c|}{ET HF 2L} & 0.013 & 0.22 & 0.021 & 0.012 & 0.19 & 0.018 & 0.011 & 0.17 & 0.018 \\ 
\multicolumn{1}{|c|}{ET2L} & 0.011 & 0.17 & 0.017 & 0.0096 & 0.14 & 0.014 & 0.0093 & 0.13 & 0.014 \\ 
\multicolumn{1}{|c|}{ET2L + CE} & 0.0088 & 0.11 & 0.012 & 0.0074 & 0.096 & 0.01 & 0.0072 & 0.092 & 0.0097 \\ 
\hline
\end{tabular}
}

\label{tab:mod2_pede}
\end{table}

\begin{table}[b]
\centering
\renewcommand{\arraystretch}{1.4} 

\caption{MOD2 | CPL fit}

\scriptsize{
\begin{tabular}{cc|c|c|c|c|c|c|c|c|c|c|c|}
\hline
\multicolumn{1}{|c|}{\multirow{2}{*}{\textbf{Data set}}}    
& \multicolumn{4}{c|}{\textbf{Fiducial}}      
& \multicolumn{4}{c|}{\textbf{Extended}}  
& \multicolumn{4}{c|}{\textbf{Very extended}} \\ \cline{2-13}

\multicolumn{1}{|c|}{}    
& $\boldsymbol{\frac{\Delta H_\mathrm{0}}{H_\mathrm{0}}}$ & $\boldsymbol{\frac{\Delta \Omega_\mathrm{m}}{\Omega_\mathrm{m}}}$& $\boldsymbol{\frac{\Delta w_\mathrm{0}}{w_\mathrm{0}}}$ &$\boldsymbol{\tilde{\epsilon}}$ 
& $\boldsymbol{\frac{\Delta H_\mathrm{0}}{H_\mathrm{0}}}$ & $\boldsymbol{\frac{\Delta \Omega_\mathrm{m}}{\Omega_\mathrm{m}}}$& $\boldsymbol{\frac{\Delta w_\mathrm{0}}{w_\mathrm{0}}}$ &$\boldsymbol{\tilde{\epsilon}}$ 
& $\boldsymbol{\frac{\Delta H_\mathrm{0}}{H_\mathrm{0}}}$ & $\boldsymbol{\frac{\Delta \Omega_\mathrm{m}}{\Omega_\mathrm{m}}}$& $\boldsymbol{\frac{\Delta w_\mathrm{0}}{w_\mathrm{0}}}$ &$\boldsymbol{\tilde{\epsilon}}$ 
                                                \\ \hline
\multicolumn{1}{|c|}{\panp $\&$ LVK (O5)} & 0.045 & 0.45 & 0.28 & 0.071 & -- & -- & -- & -- & -- & -- & -- & -- \\ 
\multicolumn{1}{|c|}{\panp $\&$ ET$\Delta$} & 0.016 & 0.42 & 0.26 & 0.039 & 0.015 & 0.43 & 0.26 & 0.039 & -- & -- & -- & -- \\ 
\multicolumn{1}{|c|}{\panp $\&$ ET$\Delta$+CE} & 0.014 & 0.43 & 0.24 & 0.038 & 0.014 & 0.4 & 0.22 & 0.035 & -- & -- & -- & -- \\ 
\multicolumn{1}{|c|}{LVK (O5)} & 0.064 & 1.8 & 1.1 & 0.16 & 0.064 & 1.8 & 1.1 & 0.16 & 0.064 & 1.8 & 1.1 & 0.16 \\ 
\multicolumn{1}{|c|}{LVKI A$^\#$} & 0.051 & 1.1 & 0.97 & 0.12 & 0.05 & 1.1 & 0.96 & 0.11 & 0.05 & 1.1 & 0.95 & 0.11 \\ 
\multicolumn{1}{|c|}{ET HF $\Delta$} & 0.03 & 0.65 & 0.76 & 0.069 & 0.029 & 0.62 & 0.73 & 0.068 & 0.029 & 0.62 & 0.72 & 0.067 \\ 
\multicolumn{1}{|c|}{ET$\Delta$} & 0.027 & 0.56 & 0.64 & 0.061 & 0.027 & 0.52 & 0.58 & 0.059 & 0.027 & 0.5 & 0.55 & 0.058 \\ 
\multicolumn{1}{|c|}{ET$\Delta$ + CE} & 0.023 & 0.48 & 0.44 & 0.052 & 0.023 & 0.44 & 0.38 & 0.049 & 0.023 & 0.41 & 0.37 & 0.046 \\ 
\multicolumn{1}{|c|}{ET HF 2L} & 0.028 & 0.58 & 0.68 & 0.063 & 0.028 & 0.52 & 0.61 & 0.06 & 0.027 & 0.51 & 0.59 & 0.059 \\ 
\multicolumn{1}{|c|}{ET2L} & 0.026 & 0.53 & 0.58 & 0.058 & 0.025 & 0.48 & 0.51 & 0.055 & 0.025 & 0.48 & 0.48 & 0.054 \\ 
\multicolumn{1}{|c|}{ET2L + CE} & 0.023 & 0.48 & 0.41 & 0.051 & 0.022 & 0.44 & 0.36 & 0.047 & 0.022 & 0.44 & 0.35 & 0.047 \\ 
\hline
\end{tabular}
}
\label{tab:mod2_cpl}
\end{table}

\begin{table}[b]
\centering
\renewcommand{\arraystretch}{1.4} 

\caption{MOD2 | GP fit}
\begin{tabular}{cc|c|c|c|c|c|}
\hline
\multicolumn{1}{|c|}{\multirow{2}{*}{\textbf{Data set}}}    
& \multicolumn{3}{c|}{\textbf{Fiducial}}      
& \multicolumn{3}{c|}{\textbf{Extended}}  \\ \cline{2-7}

\multicolumn{1}{|c|}{}    
& $\boldsymbol{\frac{\Delta H_\mathrm{0}}{H_\mathrm{0}}}$ & $\boldsymbol{\frac{\Delta \Omega_\mathrm{m}}{\Omega_\mathrm{m}}}$ &$\boldsymbol{\tilde{\epsilon}}$ 
& $\boldsymbol{\frac{\Delta H_\mathrm{0}}{H_\mathrm{0}}}$ & $\boldsymbol{\frac{\Delta \Omega_\mathrm{m}}{\Omega_\mathrm{m}}}$ &$\boldsymbol{\tilde{\epsilon}}$                        \\ \hline
\multicolumn{1}{|c|}{\panp $\&$ ET$\Delta$} & 0.01 & 0.4 & 0.031 & 0.0095 & 0.4 & 0.03 \\ 
\multicolumn{1}{|c|}{\panp $\&$ ET$\Delta$ + CE} & 0.0088 & 0.37 & 0.028 & 0.0081 & 0.34 & 0.026 \\ 
\multicolumn{1}{|c|}{\panp $\&$ ET2L} & 0.0095 & 0.4 & 0.031 & 0.0088 & 0.4 & 0.028 \\ 
\multicolumn{1}{|c|}{\panp $\&$ ET2L + CE} & 0.0087 & 0.35 & 0.027 & 0.0078 & 0.35 & 0.025 \\ 
\multicolumn{1}{|c|}{ET$\Delta$} & 0.015 & 0.55 & 0.045 & 0.014 & 0.5 & 0.04 \\ 
\multicolumn{1}{|c|}{ET$\Delta$ + CE} & 0.012 & 0.43 & 0.035 & 0.011 & 0.33 & 0.03 \\ 
\multicolumn{1}{|c|}{ET2L} & 0.014 & 0.5 & 0.041 & 0.013 & 0.45 & 0.036 \\ 
\multicolumn{1}{|c|}{ET2L + CE} & 0.012 & 0.45 & 0.037 & 0.0099 & 0.31 & 0.027 \\ 
\hline
\end{tabular}
\label{tab:mod2_gp}
\end{table}

\clearpage
\bibliography{main}
\bibliographystyle{JHEP}

\end{document}